\pdfoutput=1
\documentclass[12pt]{article}
\usepackage[letterpaper,margin=1in]{geometry}
\usepackage{amsmath,amssymb,mathtools}
\usepackage{graphicx}
\usepackage{booktabs}
\usepackage{longtable}
\usepackage{threeparttable}
\usepackage{array}
\usepackage{float}
\usepackage{needspace}
\usepackage{setspace}
\usepackage[round]{natbib}
\usepackage{xurl}
\usepackage{pdflscape}
\usepackage[skip=6pt]{caption}
\usepackage{adjustbox}
\usepackage{xcolor}
\definecolor{linkcol}{rgb}{0.10,0.25,0.50}

\newif\ifanon
\ifdefined\ANON\anontrue\else\anonfalse\fi
\usepackage{hyperref}
\ifanon
  \hypersetup{colorlinks=true,linkcolor=linkcol,citecolor=linkcol,urlcolor=linkcol,
    pdftitle={The Rural Margin of Territorial Peace: Within-Municipality Evidence from Colombia},
    pdfauthor={},pdfsubject={},pdfkeywords={},pdfcreator={},pdfproducer={}}
\else
  \hypersetup{colorlinks=true,linkcolor=linkcol,citecolor=linkcol,urlcolor=linkcol,
    pdftitle={The Rural Margin of Territorial Peace: Within-Municipality Evidence from Colombia},pdfauthor={Hector Galindo-Silva}}
\fi

\makeatletter
\let\primitiveinput\@@input
\renewcommand\paragraph{\@startsection{paragraph}{4}{\z@}%
  {1.2ex \@plus0.3ex \@minus.2ex}{-1em}{\normalfont\normalsize\itshape}}
\makeatother
\newcolumntype{L}[1]{>{\raggedright\arraybackslash\hspace{0pt}}p{#1}}
\newcolumntype{C}[1]{>{\centering\arraybackslash\hspace{0pt}}p{#1}}
\newcommand{\tablebodyfont}{\setstretch{1}\fontsize{9pt}{10.75pt}\selectfont}
\newcommand{\tablenotefont}{\setstretch{1}\fontsize{8pt}{9.5pt}\selectfont}
\newcommand{\tablenotesgap}{\vspace{-10pt}}
\newsavebox{\exhibittablebox}
\newlength{\exhibittablewidth}
\newenvironment{tablecontent}{\begin{lrbox}{\exhibittablebox}\tablebodyfont\ignorespaces}{%
  \unskip\end{lrbox}\global\exhibittablewidth=\wd\exhibittablebox
  \begin{center}\usebox{\exhibittablebox}\end{center}}

\title{\textbf{The Rural Margin of Territorial Peace:\\ Within-Municipality Evidence from Colombia}}
\ifanon
\author{}
\else
\author{
 Hector Galindo-Silva\\
  \texttt{galindoh@javeriana.edu.co}\thanks{Department of Economics, Pontificia Universidad Javeriana, Bogot\'a, Colombia. Any remaining errors are my own.}
}
\fi
 \newcommand{\BarcasCrossRawN}{503}
\newcommand{\BarcasCrossRawPct}{15.8}
\newcommand{\BarcasCrossFamilyPct}{33.5}
\newcommand{\BarcasAgeDistinctSeventeen}{6}
\newcommand{\BarcasAgeDistinctTwentyThree}{26.5}
\newcommand{\BarcasMainB}{0.582}
\newcommand{\BarcasMainP}{0.005}
\newcommand{\BarcasMainLo}{0.187}
\newcommand{\BarcasMainHi}{0.976}
\newcommand{\BarcasMainPreFive}{0.062}
\newcommand{\BarcasMainPreEleven}{-0.015}
\newcommand{\BarcasMainPreP}{0.921}
\newcommand{\BarcasMainRegionB}{0.436}
\newcommand{\BarcasMainRegionP}{0.004}
\newcommand{\BarcasMainCleanB}{0.625}
\newcommand{\BarcasMainCleanP}{0.008}
\newcommand{\BarcasMainCleanPreP}{0.686}
\newcommand{\BarcasMainCleanPreFive}{0.081}
\newcommand{\BarcasMainCleanWildP}{0.006}

\newcommand{\BarcasMainPoolB}{0.587}

\newcommand{\BarcasMainDoseB}{0.186}
\newcommand{\BarcasMainDoseP}{0.057}
\newcommand{\BarcasMainDoseWildP}{0.248}

\newcommand{\BarcasCoordB}{0.445}
\newcommand{\BarcasCoordP}{0.007}

\newcommand{\BarcasCoordCleanB}{0.582}
\newcommand{\BarcasCoordCleanP}{0.003}

\newcommand{\BarcasCivicB}{0.328}
\newcommand{\BarcasCivicP}{0.070}

\newcommand{\BarcasCivicCleanB}{0.253}
\newcommand{\BarcasCivicCleanP}{0.193}

\newcommand{\BarcasPartyB}{0.446}
\newcommand{\BarcasPartyP}{0.027}

\newcommand{\BarcasPartyCleanB}{0.474}
\newcommand{\BarcasPartyCleanP}{0.044}

\newcommand{\BarcasPartyDoseB}{0.260}

\newcommand{\BarcasPartyActivityCleanP}{0.082}
\newcommand{\BarcasPalancaCleanB}{-0.048}
\newcommand{\BarcasPalancaCleanP}{0.727}
\newcommand{\BarcasOnceB}{0.590}

\newcommand{\BarcasOnceCleanB}{0.624}

\newcommand{\BarcasAllocB}{0.600}
\newcommand{\BarcasAllocP}{0.018}
\newcommand{\BarcasAllocUnweightedB}{0.642}

\newcommand{\BarcasStrictB}{0.442}
\newcommand{\BarcasStrictP}{0.016}

\newcommand{\BarcasMocoaB}{0.605}

\newcommand{\BarcasHorseFullB}{0.779}
\newcommand{\BarcasHorseFullP}{0.040}
\newcommand{\BarcasHorseSameB}{0.574}

\newcommand{\BarcasWaveNinetySevenB}{0.585}

\newcommand{\BarcasPeerB}{0.322}
\newcommand{\BarcasPeerP}{0.041}
\newcommand{\BarcasConflictB}{0.368}
\newcommand{\BarcasConflictP}{0.016}
\newcommand{\BarcasPartyCompositeB}{0.446}
\newcommand{\BarcasPartyCompositeP}{0.028}
\newcommand{\BarcasPartyWorkB}{0.326}
\newcommand{\BarcasPartyWorkP}{0.030}
\newcommand{\BarcasScale}{0.698}
\newcommand{\BarcasRawB}{0.406}
\newcommand{\BarcasSelectionUnit}{0.052}
\newcommand{\BarcasSelectionBlock}{0.057}
\newcommand{\BarcasCellsCleanTr}{0.396}
\newcommand{\BarcasCellsCleanTu}{0.164}

\newcommand{\BarcasCellsCleanDdd}{0.425}

\newcommand{\BarcasCellsPooledDdd}{0.456}
\newcommand{\BarcasLooCovMin}{0.538}
\newcommand{\BarcasLooCovMax}{0.611}
\newcommand{\BarcasLooCovP}{0.008}
\newcommand{\BarcasLooCompMin}{0.448}
\newcommand{\BarcasLooCompMax}{0.651}
\newcommand{\BarcasLooCompP}{0.009}
\newcommand{\BarcasLooDeptMin}{0.449}
\newcommand{\BarcasLooDeptMax}{0.652}
\newcommand{\BarcasLooDeptP}{0.019} 
 \newcommand{\BarcasTrimSupportB}{0.541}

\newcommand{\BarcasTrimOneB}{0.446}

\newcommand{\BarcasTrimTwoB}{0.434}

\newcommand{\BarcasAgeSexB}{0.582}
\newcommand{\BarcasAgeAdjustedB}{0.564}
\newcommand{\BarcasEtcrBaseB}{0.510}
\newcommand{\BarcasEtcrBaseP}{0.014}
\newcommand{\BarcasEtcrEarlyB}{0.565}
\newcommand{\BarcasEtcrEarlyP}{0.030}
\newcommand{\BarcasCocaBaseB}{0.581}
\newcommand{\BarcasCocaBaseP}{0.019}
\newcommand{\BarcasCocaEarlyB}{0.694}
\newcommand{\BarcasCocaEarlyP}{0.005}
\newcommand{\BarcasGapComparisonB}{-0.341}
\newcommand{\BarcasGapComparisonP}{0.076}
\newcommand{\BarcasGapCoveredB}{0.284}
\newcommand{\BarcasGapCoveredP}{0.026} 
 \newcommand{\BarcasDoseSizePrimaryB}{0.188}

\newcommand{\BarcasDoseSizePrimaryWildP}{0.204}
\newcommand{\BarcasDoseSizePartyB}{0.257}

\newcommand{\BarcasDoseSizePartyWildP}{0.215}
\newcommand{\BarcasDoseSizeN}{3,213}
\newcommand{\BarcasDoseSizeClusters}{22} 
 \newcommand{\BarcasDimCivicB}{0.401}
\newcommand{\BarcasDimCivicP}{0.169}
\newcommand{\BarcasDimPoliticalB}{0.142}
\newcommand{\BarcasDimPoliticalP}{0.196}
\newcommand{\BarcasDimHorizontalB}{0.070}
\newcommand{\BarcasDimHorizontalP}{0.687}
\newcommand{\BarcasDimPrimaryMin}{0.539}
\newcommand{\BarcasDimPrimaryMax}{0.614}

\newcommand{\BarcasAggregateEqualElevenB}{0.533}
\newcommand{\BarcasAggregateEqualElevenP}{0.008}
\newcommand{\BarcasAggregateEqualElevenCommonB}{0.541}
\newcommand{\BarcasAggregateEqualFourteenB}{0.347}
\newcommand{\BarcasAggregateEqualFourteenP}{0.068}
\newcommand{\BarcasAggregateEqualFourteenCommonB}{0.350}
\newcommand{\BarcasAggregateOriginalTrioB}{0.282}
\newcommand{\BarcasAggregateOriginalTrioP}{0.137}
\newcommand{\BarcasAggregateOriginalTrioCommonB}{0.284}
\newcommand{\BarcasOmitFamilyMin}{0.508}
\newcommand{\BarcasOmitFamilyMax}{0.747}

\newcommand{\BarcasOmitComponentMin}{0.560}
\newcommand{\BarcasOmitComponentMax}{0.607}
 
\date{October 2026}

\newcommand{\BarcasSourceExcludedB}{0.581638}
\newcommand{\BarcasAgeFullP}{0.006}
\newcommand{\BarcasAgeFullPreP}{0.957}
\newcommand{\BarcasDedupB}{0.186}
\newcommand{\BarcasDedupP}{0.057}

\newcommand{\BarcasCoordLooMin}{0.400}
\newcommand{\BarcasCoordLooMax}{0.480}
\newcommand{\BarcasCoordLooP}{0.013}

\newcommand{\NovJointBasePP}{16.1}
\newcommand{\NovJointBaseLo}{0.7}
\newcommand{\NovJointBaseHi}{31.5}

\newcommand{\NovPartyBasePP}{9.6}
\newcommand{\NovPartyBaseLo}{1.0}
\newcommand{\NovPartyBaseHi}{18.2}

\newcommand{\NovMentionsBase}{-0.006}

\newcommand{\NovMentionsBaseP}{0.940}

\newcommand{\NovConflictItemBase}{0.368}

\newcommand{\NovConflictObservedBase}{0.350}

\newcommand{\NovMentionsEarlier}{-0.094}

\newcommand{\NovMentionsEarlierP}{0.320}

\newcommand{\NovConflictItemEarlier}{0.540}

\newcommand{\NovConflictObservedEarlier}{0.522}

\newcommand{\NovTrioBaseMedian}{0.344}

\newcommand{\NovTrioEarlierMedian}{0.322}

\newcommand{\NovZomacMunicipalities}{6}
\newcommand{\NovZomacEarlierEndpoints}{5}
\newcommand{\NovZomacBaselineEndpoints}{4}
\newcommand{\NovZomacEndpoints}{\NovZomacEarlierEndpoints}
\newcommand{\NovCommonMunicipalities}{36}
\newcommand{\NovCommonCovered}{17}
\newcommand{\NovCommonBaseline}{0.481}
\newcommand{\NovCommonBaselineP}{0.006}
\newcommand{\NovCommonEarlier}{0.475}
\newcommand{\NovCommonEarlierP}{0.031}
\newcommand{\NovRuralCompDeclining}{10}
\newcommand{\NovRuralCompMunicipalities}{16}
\newcommand{\NovRuralCompMedian}{-0.222}
\newcommand{\NovRuralCompNonthin}{-0.188}
\newcommand{\NovRuralCompNonthinN}{11}
\newcommand{\NovRuralCompLooMin}{-0.363}
\newcommand{\NovRuralCompLooMax}{-0.200}

\begin{document}

\addtocontents{toc}{\protect\setcounter{tocdepth}{-1}}
\maketitle
\vspace{-2.4em}

\begin{abstract}
\singlespacing
\noindent
Colombia's peace agreement introduced participatory development programs and congressional constituencies reserved for conflict victims with a rural electorate. I use repeated waves of the Social Capital Barometer and a triple-difference event study to compare changes in rural--urban engagement gaps in municipalities covered by both reforms with those elsewhere. Rural engagement evolved more favorably relative to urban engagement in the covered municipalities. The index captures coordination around shared problems, civic participation and party linkage, with coordination providing the clearest evidence. The result persists when the reference period precedes the agreement and the baseline wave is excluded. Program activity and organizational membership patterns are consistent with participatory investment and political mobilization. The findings document a change in engagement beyond voting under the joint territorial reforms.

\end{abstract}


\newpage

\section{Introduction}\label{sec:intro}

Residents of dispersed rural settlements face substantial costs of organizing, limited access to public authorities, and weak incentives for political actors to respond to their demands. These barriers can be especially severe in rural areas of developing countries where public institutions have limited capacity to respond to collective demands and can be further weakened by armed conflict \citep{worldbank2011}. Community organizations can help residents coordinate around shared problems, but this capacity does not necessarily translate into influence over public decisions. When that connection is weak, residents may have little reason to bear the costs of sustained collective engagement \citep{mansurirao2013}.

Can reforms that give rural residents a stronger political voice and a say in public investment encourage broader collective engagement? Dedicated representation may increase the returns to organizing by making collective demands electorally consequential, while development programs built around community assemblies can create opportunities to deliberate, coordinate and interact with public authorities. Yet participation may remain confined to program activities or depend on established intermediaries, without extending to wider civic and political engagement. Evidence from community-driven development shows that improved local public goods need not be accompanied by sustained changes in collective action or local institutions \citep{casey2012,humphreys2019,olken2010}.

Colombia's peace agreement provides a setting in which to examine this question. It introduced participatory rural development programs that involve communities in setting public investment priorities: the Development Programs with a Territorial Approach (PDET). It also established additional congressional seats reserved for conflict victims and elected by rural voters: the Special Transitional Peace Constituencies (CITREP). I examine whether the combined reforms strengthened rural engagement relative to urban engagement. The analysis compares changes in rural--urban engagement gaps between municipalities included in CITREP, all of which are also covered by PDET, and municipalities outside the constituencies.

I use repeated waves of Colombia's Social Capital Barometer (BARCAS) to track these changes. Here, \emph{public-collective engagement} refers to attitudes and reported activities concerning collective problems, civic affairs and political organizations, distinguished from the wider networks, trust and reciprocity commonly studied as social capital. I estimate the change in rural--urban engagement gaps associated with joint coverage using a triple-difference event study. I first compare engagement changes in the countryside with those in its own urban center, removing the component shared by both groups. I then compare these rural--urban changes between municipalities covered by both PDET and CITREP and comparison municipalities outside the peace constituencies. Accounting for municipality-wide changes matters because coverage was concentrated in places affected by conflict, poverty and weak institutions. Comparison municipalities provide a benchmark for the evolution of the internal gaps. The specification allows each municipality its own persistent rural--urban difference. A causal interpretation requires that the gaps would have evolved similarly across the two groups without the reforms; earlier survey waves provide evidence on this assumption.

To capture engagement in collective and public affairs, I construct a primary index combining coordination around collective problems, civic participation and party linkage. Its estimated differential is \BarcasMainB\ standard deviations (municipality-clustered $p=\BarcasMainP$). This positive estimate means that rural engagement evolves more favorably relative to urban engagement in municipalities covered by both PDET and CITREP than in municipalities outside the peace constituencies. The design attributes this relative change to coverage if rural--urban gaps would otherwise have evolved in parallel.

Among the index's constituents, coordination provides the clearest evidence across reference periods. Party linkage is positive in the baseline, but individual party activity is less precise with an earlier reference; evidence for civic participation and broader collective activity is also less conclusive. Comparisons within a fixed municipal sample help explain this favorable differential in rural--urban engagement gaps: it can reflect both improvement in covered rural areas and rural engagement falling behind urban engagement in comparison municipalities.

The primary result remains positive after allowing rural trajectories to vary with pre-existing conflict and development characteristics or regional changes. It also survives restrictions holding municipal coverage fixed, alternative weights, reweighting municipal sample allocation, and the exclusion of individual municipalities or departments. Checks excluding early reincorporation-site municipalities and adjusting for recorded pre-agreement coca exposure address specific concurrent rural changes, while geographic clustering and bootstrap inference assess precision. Using 2011 as the reference and excluding 2017 gives a similar estimate. A direct average of the same underlying measures, without first summarizing each domain, also retains the result.

What might explain this relative rural gain in engagement? PDET assemblies provide opportunities to discuss shared priorities and coordinate proposals, while CITREP gives social organizations a role in nominating candidates and a reason to mobilize rural residents. I examine these channels through program activity and organizational membership. Within covered municipalities, rural--urban engagement gaps evolve more favorably where more PDET initiatives were adopted, with the strongest association in party linkage. Membership in organization types explicitly named as CITREP candidate nominators also shows a positive relative rural change, particularly for ethnic organizations. These complementary associations are consistent with participatory investment and organizational mobilization contributing to the relative rural gain under the combined reforms.

\medskip

This paper contributes evidence on collective engagement under the territorial reforms that followed Colombia's peace agreement. The analysis estimates the differential change in engagement where participatory development and dedicated rural representation were introduced together. Their overlapping coverage leaves their separate effects and complementarity unidentified. Studies of community-driven development distinguish gains in public goods from changes in participation and organizational capacity \citep{casey2012,mansurirao2013,humphreys2019}. In post-conflict Liberia, \citet{fearonhumphreysweinstein2015} measure collective action through contributions in a field experiment. Here the survey captures coordination, civic participation and party linkage. These measures extend work on turnout under Colombia's territorial programs \citep{gelvez2023,giraldo2019apatia}.

This institutional perspective also connects the paper to research on how reserved seats affect policy and citizen participation \citep{pande2003,chattopadhyay2004,beaman2009,iyer2012,dunningnilekani2013}. \citet{poertner2023} links representation to turnout and institution-directed attitudes through organizationally embedded representatives, and \citet{goyal2024} examines mobilization through recruitment to canvassing. The analysis extends this organizational perspective to engagement beyond voting when political representation expands alongside participatory rural development.

Finally, the interpretation draws on research on political linkage, which informs the distinction between engagement and dependence on intermediaries \citep{kitschelt2007,keefer2008,stokes2013}. Research on conflict and cooperation shows that exposure to violence can shape subsequent participation and cooperation \citep{blattman2009,bellowsmiguel2009,bauer2016}. This makes conflict conditions and other consequences of the peace process relevant to assessing the rural comparison.

\medskip

Section~\ref{sec:theory} presents the institutional context and conceptual framework. Sections~\ref{sec:data} and~\ref{sec:strategy} describe the data and empirical strategy. Section~\ref{sec:results} reports the results, followed by identification and robustness checks in Section~\ref{sec:evidence} and evidence on the proposed mechanisms in Section~\ref{sec:attribution}. Section~\ref{sec:conclusion} concludes.

\section{Rural Engagement and the Institutional Reform}\label{sec:theory}

\subsection{Institutional background}\label{sec:instruments}\label{sec:election}

In 2016, the Colombian government and the Revolutionary Armed Forces of Colombia (FARC), then the country's largest guerrilla organization, signed a peace agreement to end their armed confrontation. This confrontation had lasted more than five decades and was concentrated in rural peripheries \citep{gmh2013basta}. In these territories, state presence was uneven and governance relied on regional power holders \citep{gonzalez2014poder,gutierrez2019clientelistic,acemoglu2013}. Rural residents reached public authorities through brokers and local bosses rather than organizations able to aggregate their demands \citep{robinson2013colombia,escobar2002clientelism,fergusson2022weak}. Two institutional reforms arising from the agreement are central to the analysis \citep{acuerdo2016}.

The first reform introduced a participatory framework for rural development. Decree Law 893 of 2017 created sixteen Development Programs with a Territorial Approach (PDET) in 170 municipalities prioritized mainly for poverty, conflict incidence, institutional weakness and illicit crops, confining them to the rural zone in eight larger municipalities \citep{decreto893}. Coordinated by the Agency for Territorial Renewal (ART), PDET links community priorities to public investment under the agreement's rural reform. Its areas of intervention include land, infrastructure, health, education, housing, productive activities, food security and reconciliation \citep{dnp2020pdet}.

Each PDET has a subregional action plan, initially negotiated through local consultations between late 2017 and early 2019. Rural residents and community organizations identified needs and proposed initiatives at the village level; these proposals fed into municipal agreements and then into the subregional plans \citep{dnp2020pdet}. Adopted initiatives are proposals incorporated into the plans to address territorial needs. ART coordinates their development into investment projects with national and territorial authorities \citep{decreto893}. This process gives rural communities a formal route for bringing collective demands into development planning.

The second reform expanded political representation through sixteen Special Transitional Peace Constituencies (CITREP), each electing one additional member of the Chamber of Representatives for 2022--2026 and 2026--2030 \citep{actoleg2021}.\footnote{Shelved in 2017 over a disputed Senate majority count, the reform became Legislative Act 02 of 2021 after the Constitutional Court declared it approved \citep{corte2021}.} Three rules set it apart. Parties with legal status or seats in Congress, the FARC's successor included, could not nominate: lists could be registered only by victims', peasant and other social organizations, significant groups of citizens and, in ethnic territories, ethnic authorities.\footnote{For the first election, social, peasant and women's organizations had to demonstrate five years of legal existence or local activity; ethnic authorities followed separate registration requirements \citep[arts.\ 2 and 4]{decreto1207}. Thus the rules could encourage membership and mobilization within established organizations rather than the creation of new endorsing bodies.} Each list paired two conflict victims rooted in the territory, one of each gender. And only citizens registered at rural polling places could vote: the urban center (\emph{cabecera}) was excluded, voters received a separate ballot in addition to their ordinary Chamber vote, and new rural polling places were mandated. All 167 CITREP municipalities are also covered by PDET, bringing participatory development and dedicated political representation into the same rural territories \citep{actoleg2021,art_pdetmun}.\footnote{Only three PDET municipalities fall outside CITREP: Manaure Balc\'on del Cesar and San Diego in Cesar, and Mocoa in Putumayo.}

The first constituency election took place on 13 March 2022, alongside the ordinary legislative elections.\footnote{Of roughly 1.25 million eligible voters, 533,664 cast a CITREP ballot, a turnout of 42.8 percent; 403 candidates had registered on 203 lists \citep{fip2022que,registraduria2021inscripcion}. Of the sixteen winners, three were women; five had been endorsed by victims' organizations, four by other social organizations, three by Afro-Colombian community councils, two by peasant associations, one by a women's organization and one by a significant group of citizens \citep{fip2022que}.} Post-election assessments describe gains in representation for grassroots organizations alongside reports of links between some elected candidates and established political networks \citep{fip2022que}.

\subsection{Conceptual framework and empirical predictions}\label{sec:predictions}

How could reforms combining participatory investment with dedicated representation strengthen rural engagement beyond voting or program assemblies? The combination offers two possible channels: greater political returns to organizing and more opportunities to participate in investment decisions. Both could encourage public-collective engagement by connecting residents' collective demands to public decision-making. These channels are most plausible where local organizations can mobilize residents, community proposals influence public priorities and representatives have incentives to respond.

The first mechanism is political salience. Rural residents previously voted for the Chamber through departmental constituencies alongside urban electorates. The additional rural constituencies create a dedicated electoral arena, with their own ballot and candidate-endorsement rules. Eligible organizations gain a reason to recruit and articulate demands; candidates gain a reason to campaign in rural settlements (\emph{veredas}); and residents may perceive a greater return to attending to public affairs. Research on expanded electoral inclusion shows how changing whose votes matter can redirect political attention and resources \citep{fujiwara2015,casciowashington2014}. Here the response depends on whether organizations and effective mobilization bring those opportunities to residents.

The additional rural campaign could also mobilize networks connected to ordinary parties even though those parties could not nominate CITREP candidates. The direction of party linkage is ambiguous because dedicated representation can strengthen those networks or make them less central to organizing. The 2022 national campaigns could independently change party linkage in places where rural and urban voters responded differently.

The second mechanism is participatory rural investment. Participation may become more attractive when residents expect collective demands to yield tangible improvements. Working together on investment proposals may also strengthen coordination beyond the planning process. Place-based investment can change local opportunities and incentives \citep{klinemoretti2014,neumarksimpson2015}. A broader engagement response depends on whether these experiences extend beyond project-specific participation, a distinction emphasized by the mixed evidence on civic outcomes from community-driven development \citep{casey2012,humphreys2019}. Implementation failures can also disappoint participatory expectations and weaken this connection \citep{bachmann2025}.

Both mechanisms suggest that the reforms could strengthen engagement more among rural residents than among residents of the same municipality's urban center. Colombia's rural--urban divide combines dispersed settlement with persistent gaps in access to public services and infrastructure, including between the urban centers and outlying settlements of rural municipalities \citep{dnp2015campo}. These conditions can make coordinating demands and reaching public authorities more costly. By bringing investment decisions into local assemblies and giving rural voters a dedicated electoral arena, the reforms could reduce these barriers and increase the returns to organizing more in the countryside. Urban residents may also respond to investment and political mobilization, so the prediction concerns a relative rural gain.{\interfootnotelinepenalty=10000\footnote{Other pathways could reinforce these mechanisms. Repeated participation could build civic skills and trust, while more responsive public authorities could strengthen residents' belief that participation matters. Brokerage could also mobilize residents without increasing their autonomy. Changes in security or livelihoods following demobilization offer explanations beyond the two reforms; conflict exposure itself can shape cooperation and participation \citep{bauer2016}. These possibilities motivate the broader outcomes and identification checks below.}} The prediction centers on three substantive margins---coordination around shared problems, civic participation and party linkage---grouped into families for the study's primary public-collective engagement index. Formal collective action and organizational membership provide related outcomes, defined in the next section.

These mechanisms need not affect all social-capital domains alike. Assemblies could strengthen social relations, mobilization could increase participation in ordinary elections, and contact with public authorities could change institutional trust or reliance on brokers. These secondary outcomes assess whether the response extends beyond the primary public-collective core. The mechanisms also suggest more specific comparisons. The investment channel motivates asking whether rural--urban engagement gaps evolve more favorably where more PDET initiatives are adopted; the representation channel motivates comparing membership changes in organization types explicitly named as constituency nominators with those in other secular types. The latter could also nominate as generic social organizations, so this is not an eligible--ineligible comparison. Because the reforms overlap geographically and program activity may reflect existing organizational capacity, these comparisons provide complementary evidence on the channels behind the broader response.

This prediction motivates a triple-difference event study comparing changes in rural--urban engagement gaps between municipalities covered by both PDET and CITREP and comparison municipalities. The design focuses on the relative rural response while accounting for municipality-wide changes. The next sections describe the data and empirical strategy.

\section{Data}\label{sec:data}\label{sec:wave1997}\label{sec:outcomes}

The main data source is the Social Capital Barometer (BARCAS), a face-to-face survey of social capital and civic attitudes fielded in Colombia in 1997, 2005, 2011, 2017 and 2023 under a stratified multi-stage design \citep{sudarsky2022methodological,barcas2023}.\footnote{The survey was produced by the Corporaci\'on para el Control Social (Contrial), which holds the rights to the microdata. The Centro de Datos CEDE, Universidad de los Andes, serves as the repository and facilitated access under a confidentiality agreement.} The survey re-samples municipalities rather than individuals, so the data are repeated cross-sections with a municipal panel structure.\footnote{The analysis uses the distributed national releases. The producer also documents a dedicated CITREP sample in 2023 that partly overlaps the national sample \citep{barcascontexto}. The available files do not identify membership in these overlapping samples or establish continuity of rural sampling points. Appendix~\ref{app:sampling2023} discusses this distinction and reports the sample counts. The estimates concern the surveyed municipalities; representativeness for all 167 constituency municipalities is not established.} The producer does not distribute sampling weights. The unweighted regression estimates a contrast for the observed survey sample under the model's implicit weights. With heterogeneous effects, this can differ from the average for the rural population. I distinguish residents of the urban center (\emph{cabecera}) from those in rural settlements, reflecting the reforms' rural focus. Municipal identifiers, rural--urban coding and outcome measures are harmonized across waves.\footnote{Municipal identifiers and the collective-problems module change format. Rural residence is coded as \texttt{area}=1 before 2023 and \texttt{area}=2 in 2023, although all dictionaries label 2 as rural. Bogot\'a's predominantly urban sample and municipal rurality and poverty indicators support this correction. Appendix~\ref{app:data} documents these checks and the harmonization.}

The baseline compares 2017 with 2023 to focus on the shorter implementation window. The published 2017 survey contains unusually many repeated response profiles and limited variation within municipalities, and the fieldwork month is unknown. Using 2011 and excluding 2017 avoids relying on that release and predates PDET activities, but spans more of the peace process and leaves only one pre-agreement change, from 2005 to 2011. The two windows therefore differ in survey comparability and exposure to other changes during the peace process. Appendix~\ref{app:sourceaudit} documents the 2017 patterns.

The~estimation sample includes 12,466 respondents in 66 municipalities across all four waves.\footnote{BARCAS also provides data for 1997, but that wave includes only eight covered municipalities and 149 rural respondents in those municipalities, compared with nineteen and 395 in 2005. The main analysis therefore begins in 2005; including 1997 leaves the primary estimate similar (Section~\ref{sec:evidence}).} Twenty-two municipalities are covered by both PDET and CITREP; all are observed in 2017 and 2023, while each of the two earlier waves includes nineteen. The comparison focuses on changes in rural--urban engagement gaps between these jointly covered municipalities and municipalities outside the peace constituencies. Mocoa is the only comparison municipality covered by PDET. Excluding it gives a primary estimate of \BarcasMocoaB{} (Section~\ref{sec:evidence}). Both rural and urban respondents are observed in 195 of the 238 municipality--wave cells, including 80 of the 82 cells in jointly covered municipalities. Table~\ref{tab:coverage} summarizes coverage, and Figure~\ref{fig:map} in Appendix~\ref{app:map} locates the municipalities.

The principal outcomes capture residents' engagement in collective problem-solving, civic affairs and political organizations. For this study, selected measures are grouped by substantive content into nine outcome families, using harmonized versions across waves. The primary engagement index combines three families---horizontal coordination, civic participation and party linkage---because these capture complementary margins through which participatory investment and dedicated representation may encourage engagement beyond voting and program assemblies. Horizontal coordination captures reliance on joint action and openness to resolving conflicts; civic participation covers local meetings, committees, petitions, volunteering, media engagement and attention to public affairs; and party linkage combines membership, trust in parties and reported party activity. Within civic participation, attention and political education capture engagement with public affairs rather than completed acts. These families therefore combine attitudes with reported behavior rather than measuring realized collective action alone.\footnote{Among rural respondents in 2011, pooling covered and comparison municipalities, 56.9 percent mentioned joint action as a way collective problems are solved, 12.9 percent reported having signed a petition, and 7.2 percent reported having worked for a political party. The joint-action item records either mention, not necessarily the first. These are unweighted shares among respondents with nonmissing answers. Table~\ref{tab:outcomedescriptives} provides descriptive statistics by coverage and residence.}

The remaining six families distinguish related collective outcomes from wider social and political responses. Organizational membership covers affiliation with parties, religious organizations, guilds and unions; formal collective action adds petition-signing dispositions and membership in organizations that support respondents' interests. Broader affiliation is examined separately because membership need not entail civic involvement. These are overlapping substantive domains, not estimated factors: petitions also enter civic participation. Electoral participation, brokerage and clientelism, diffuse social relations, and institutional trust assess whether any response extends beyond the public-collective domain. Brokerage is distinct because party attachment need not involve personal intermediaries. These are contrast domains, not outcomes assumed to be unaffected.

\Needspace{3\baselineskip}
These outcome families draw on a pool of 44 composite indices distributed by BARCAS in all five waves, supplemented with individual participation and membership items. Selecting and harmonizing these measures yields a catalogue of 54 measures, including alternative codings retained for measurement diagnostics. Thirty-five distinct measures enter the nine families; the remaining nineteen are examined individually, with subsets also used in complementary engagement and organization-type analyses. Family indices average component $z$-scores to avoid scale-driven weights, and are re-standardized on the 2011 rural comparison distribution \citep{kling2007,anderson2008}. The primary index gives equal weight to the three standardized families, regardless of component counts, and is re-standardized on the same group.\footnote{Tables~\ref{tab:dictionary} and~\ref{tab:additionalmeasures} list the measures and their uses. Equal weights apply to constructed components within families, not to underlying survey questions: some composites overlap, particularly in party linkage. Counting each underlying party item once gives similar primary estimates (Section~\ref{sec:robustness}). Constructed components are complete in the estimation sample; this does not imply complete raw responses. Appendix~\ref{app:outcomes} explains inclusion and boundary choices, construction and missing-data treatment. The 2011 scaling reference is distinct from the event study's 2017 reference wave.} The three families index coordination around shared problems, participation in public affairs and connections to political organizations. The broader BARCAS dimensions include activities such as voting and mutual aid that extend beyond that construct. Section~\ref{sec:multiplicity} examines sensitivity to the choice of families.\footnote{Temporal harmonization also holds fixed the content of political activity and trust in media, legal institutions and control institutions. Nine simple composites are reconstructed from their questionnaire items; excluding five source records that violate these identities leaves the primary estimate unchanged to three decimals. Appendix~\ref{app:data} documents these checks.}

A complementary index, outside the nine-family set, combines membership in communal action boards, peasant associations and cooperatives with community meetings, local committees, volunteering and petition-signing dispositions. Three separate sensitivity checks omit petitions, count only having signed, or restrict the index to completed activities without memberships. Separate organization-type indices compare membership in the statute's nominating categories with other secular organizations; these groups are distinct from the categories used in the organizational-membership family.

Municipal characteristics and PDET records complement the survey in the identification checks and mechanism analyses. The CEDE municipal panel \citep{cedepanel2019} supplies population, rurality, unsatisfied basic needs, distance to the departmental capital, homicides and forced displacement. The trajectory controls in Section~\ref{sec:robchar} use 2015 values, before the peace agreement and PDET, with basic needs observed in 2011. Historical residence coding is validated using 2017 rurality and 2011 poverty values. ART's public registries identify PDET municipalities and count initiatives adopted in each municipal action plan, overall and by pillar \citep{art_pdetmun,art2022}. These establish coverage and support the analysis of program activity in Section~\ref{sec:attribution}.\footnote{CEDE's \href{https://datoscede.uniandes.edu.co/catalogo-de-datos/}{catalogue} provides access to the General Characteristics and Conflict and Violence files. ART's \href{https://www.datos.gov.co/d/idrk-ba8y}{Municipios PDET} and \href{https://www.datos.gov.co/d/gmvf-t63e}{Iniciativas PDET} are openly downloadable.}

\section{Empirical Strategy}\label{sec:strategy}

\subsection{Baseline specification}\label{sec:spec}

The empirical strategy compares changes in rural--urban engagement gaps between municipalities covered by the peace constituencies and comparison municipalities. The baseline anchors the comparison on 2017 to focus on the shorter 2017--2023 implementation-period window; a complementary specification uses 2011 and excludes 2017 entirely to assess robustness to the reference period and wave inclusion.\footnote{The PDET programs and CITREP constituencies have different start dates. Decree Law 893 was issued in May 2017, assemblies began in late 2017 and plans were signed in 2018--2019; constituency representatives were first elected in March 2022. The producer does not publish the 2017 fieldwork month, so that wave cannot be certified as preceding every program activity. The earlier-reference estimates also address this timing ambiguity and remain positive.} I estimate the contrast using a triple-difference event study:
\begin{equation}\label{eq:esddd}
\begin{split}
Y_{imt}={} & \alpha_{mt}+\sum_{\tau\neq 2017}\pi_{\tau}\bigl(R_i\times T_m\times\mathbf 1\{t=\tau\}\bigr)+\sum_{\tau\neq 2017}\rho_\tau\bigl(R_i\times\mathbf 1\{t=\tau\}\bigr)\\
&+\gamma_m R_i+X_i'\theta+\varepsilon_{imt}.
\end{split}
\end{equation}
Here, $Y_{imt}$ is the engagement outcome for respondent $i$ in municipality $m$ and survey wave $t$. $T_m$ equals one for constituency-covered municipalities, and $R_i$ equals one for respondents living in rural settlements rather than the urban center. The indicator $\mathbf 1\{t=\tau\}$ equals one in wave $\tau$ and zero otherwise, so each interaction applies only to that wave.

The municipality-by-wave effects $\alpha_{mt}$ absorb everything that is constant within a municipality and wave, including coverage and its interaction with any wave, and the component of spending, security, price or weather shocks that affects countryside and town equally, but not any differential rural exposure. The term $\gamma_m R_i$ gives every municipality its own persistent rural--urban gap, which is equivalent here to municipality-by-residence effects. Relative to 2017, $\pi_\tau$ measures the additional change in the rural--urban gap in covered municipalities compared with comparison municipalities, while $\rho_\tau$ measures the change in that gap in comparison municipalities. $X_i$ contains respondent sex in the baseline, $\theta$ is its coefficient vector, and $\varepsilon_{imt}$ is the error term. Other specifications allow rural--urban trajectories to vary with pre-existing municipal characteristics (Section~\ref{sec:evidence}); the online appendix also adjusts for age and its interaction with rural residence over all four waves.

The coefficient of interest is $\pi_{2023}$. With gaps defined as rural minus urban outcomes, a positive value means that the gap rises more, or falls less, in covered municipalities than in comparison municipalities between 2017 and 2023. For engagement outcomes, this is a relative rural gain; a negative value indicates the reverse. A zero coefficient means no additional gap change, not unchanged engagement. For the standardized indices, magnitudes are in standard deviations of the outcome in the 2011 rural comparison group, not percentage or percentage-point changes. The contrast can reflect rural or urban movements and does not, by itself, measure an absolute change in rural outcomes.

The pre-reform coefficients $\pi_{2005}$ and $\pi_{2011}$ give the joint pre-trend test $H_0:\pi_{2005}=\pi_{2011}=0$. Because every coefficient of interest is an interaction with rural residence, only the municipality-wave cells that contain both rural and urban respondents, counted in Section~\ref{sec:data}, identify it. A pooled two-period analogue replaces the wave-specific interactions with a single post-reform indicator and compares 2023 with the pooled earlier waves rather than with 2017.\footnote{The pooled analogue is $Y_{imt}=\alpha_{mt}+\pi\,(R_i\times T_m\times\text{Post}_t)+\rho\,(R_i\times\text{Post}_t)+\gamma_m R_i+X_i'\theta+\varepsilon_{imt}$. Here $\text{Post}_t=\mathbf 1\{t=2023\}$; $\rho$ is the post-reform change in the comparison rural--urban gap and $\pi$ the additional change in covered municipalities, both relative to pooled earlier waves. The pooled primary-index estimate remains positive and close to the baseline (Section~\ref{sec:evidence}).} Standard errors are clustered by municipality, the level at which coverage varies, so the effective sample for inference is 66 municipalities rather than 12,466 respondents.

\Needspace{3\baselineskip}
A direct alternative to the triple difference is a municipal difference-in-differences, comparing engagement changes before and after the reform in covered and comparison municipalities. Separate rural and urban estimates can distinguish their changes, but municipal shocks shared by countryside and town can bias both comparisons. The triple difference removes this common component through municipality-by-wave effects, allowing covered and comparison municipalities to follow different overall trajectories. However, Appendix~\ref{app:did} reports pooled and residence-specific municipal estimates as complementary comparisons: the primary-index estimate is positive and larger among rural respondents, but several outcomes show differential pre-reform trends. The triple difference is therefore the baseline: it requires parallel counterfactual rural--urban gaps, not parallel municipal trends within each residence group.\footnote{In the simple two-period case, the triple difference is the rural DiD minus the urban DiD. Recovering the regression coefficient this way requires compatible samples, weights and model restrictions. The appendix specifications use different fixed effects, so their coefficients cannot be mechanically subtracted to recover equation~\eqref{eq:esddd}. Municipality-by-wave effects also absorb intervention effects shared by both residence groups; an absolute rural effect requires assumptions about urban exposure. Differences in coefficient magnitudes alone do not establish bias in either specification. Section~\ref{sec:sides} displays each group's movements.} 

\subsection{Identification and inference}\label{sec:identification}\label{sec:pdet}\label{sec:inference}

The estimand is the relative rural response to the combined PDET rural development programs and CITREP peace constituencies. Groups reflect residence, not polling-place registration.\footnote{In 2017, 44 percent of the covered municipalities' population lived in the urban center, but 61 percent of their electoral roll was registered there \citep{moe2017ponencia}. These figures predate the 2021 registration period and the new rural polling places.} A causal interpretation of $\pi_{2023}$ in equation~\eqref{eq:esddd} requires that, absent these reforms, rural--urban engagement gaps would have evolved in parallel in covered and comparison municipalities. This is the parallel-trends restriction for the gap \citep{oldenmoen2022}. After absorbing common municipal shocks and persistent rural--urban differences, the remaining threat is a change in the gap correlated with coverage for reasons other than the intervention. Section~\ref{sec:evidence} reports the detailed results of the identification checks and inference exercises outlined below.

The primary index's pre-reform differentials are very small in magnitude and jointly statistically indistinguishable from zero. Using 2011 as the reference and excluding 2017 also retains a positive estimate and no detectable pre-reform differential. \citeauthor{rambachanroth2023}'s (\citeyear{rambachanroth2023}) sensitivity interval for the primary index remains positive under departures from parallel trends as large as the largest pre-reform wave-to-wave change in the differential. The relative rural gain therefore survives this relaxation of the identifying assumption.

Concurrent rural changes following demobilization could confound the comparison \citep{blattman2009,bellowsmiguel2009,bauer2016,shenk2023,prem2022selective}. I allow rural paths to vary with pre-existing conflict and development characteristics and across macro-regions, exclude early reincorporation-site municipalities, and adjust for recorded pre-agreement coca exposure. The primary estimate remains positive throughout, reducing concern about these alternatives. Changes in sample composition could also generate the differential. Fixed-coverage samples and reweighting respondent allocation across municipalities preserve a positive primary estimate, reducing concern about observed geographic composition. These checks do not, however, rule out bias from household or settlement turnover within municipalities.\footnote{Appendix~\ref{app:sampling2023} discusses the dedicated CITREP sample and sampling continuity within municipalities.}

Finally, inference separately addresses geographic dependence and outcome selection. Municipality clustering is complemented by broader clustering, wild cluster bootstrap tests \citep{cameron2008,roodman2019} and a cluster jackknife. Within-region coverage permutations support placebo tests, Romano--Wolf multiple-testing corrections \citep{romanowolf2005,clarkeromano2020} and index-selection sensitivity, assuming exchangeability, not randomized coverage.

\section{Results}\label{sec:results}

\subsection{Baseline results}\label{sec:baselineres}

Table~\ref{tab:baseline} presents the main results. The outcome of interest in column (1) is the public-collective engagement index, which combines coordination, civic participation and party linkage. In Panel A, the baseline estimate of $\pi_{2023}$ is \BarcasMainB{} standard deviations (municipality-clustered $p=\BarcasMainP$, 95 percent interval $[\BarcasMainLo,\BarcasMainHi]$).\footnote{For the unrescaled average of the three standardized families, the differential is \BarcasRawB{}; dividing by its standard deviation in the 2011 rural comparison group (\BarcasScale{}) yields the reported estimate.} The positive coefficient means that, between 2017 and 2023, rural engagement evolved more favorably relative to urban engagement in municipalities covered by both reforms than in comparison municipalities. The point estimate is just over half a standard deviation. The standard errors imply approximate minimum detectable effects of 0.55 standard deviations for Panel A and 0.64 for Panel B at 80 percent power and a two-sided 5 percent level. Effects near this range can be estimated imprecisely, and conditioning on significance can exaggerate magnitudes when the true effect is smaller \citep{gelmancarlin2014}.

The relative gain is consistent with an engagement response to the territorial reforms beyond voting. Panel B retains a similar positive estimate when 2011 is the reference and 2017 is excluded entirely: \BarcasMainCleanB{} standard deviations ($p=\BarcasMainCleanP$; wild cluster bootstrap $p=\BarcasMainCleanWildP$). The relative rural gain therefore also appears without the 2017 release. Figure~\ref{fig:primary} plots the wave-specific estimates for both references; the rural and urban movements behind the contrast are examined below. Section~\ref{sec:multiplicity} assesses sensitivity to index choice alongside conventional inference.

Columns (2) to (4) of Table~\ref{tab:baseline} report results for the three constituent families. In Panel A, the estimates are positive for all three: \BarcasCoordB{} for coordination ($p=\BarcasCoordP$), \BarcasCivicB{} for civic participation ($p=\BarcasCivicP$), and \BarcasPartyB{} for party linkage ($p=\BarcasPartyP$), before multiplicity adjustment. All three remain positive in Panel B, with coordination and party linkage individually significant at 5 percent in both panels; civic participation is less precisely estimated. The signs align with the predicted relative rural gain across these public-collective margins, though precision differs.\footnote{Figure~\ref{fig:eventstudy} in the supplement shows the families' wave-specific coefficients; Appendix~\ref{app:reference2011} discusses estimates for additional outcomes and selected individual measures using 2011 as the reference and excluding 2017.}

Table~\ref{tab:components_main} examines the orientations and reported activities underlying these family-level results. Mentioning joint resolution of collective problems and agreeing that conflicts can be resolved once aired give positive estimates of \BarcasPeerB{} ($p=\BarcasPeerP$) and \BarcasConflictB{} ($p=\BarcasConflictP$). All six civic-participation components are positive, although none is individually significant at 5 percent. The membership--trust--activity composite and reported party activity give \BarcasPartyCompositeB{} ($p=\BarcasPartyCompositeP$) and \BarcasPartyWorkB{} ($p=\BarcasPartyWorkP$).\footnote{The party composites overlap; the membership--trust composite does not measure trust alone. Table~\ref{tab:party_once} reports the underlying items separately, and Table~\ref{tab:components} gives the full component regressions.} On the original binary scales the estimates correspond to changes in the relative gap of \NovJointBasePP{} percentage points for joint resolution (95 percent interval $[\NovJointBaseLo,\NovJointBaseHi]$) and \NovPartyBasePP{} for reported party activity ($[\NovPartyBaseLo,\NovPartyBaseHi]$). These estimates compare changes in rural--urban differences between coverage groups. Reported party activity is less precisely estimated when 2017 is excluded ($p=\BarcasPartyActivityCleanP$); the next subsection examines whether the response extends to broader collective activity.

\subsection{Other outcome families}\label{sec:behavioral}\label{sec:contrast}

The conceptual framework also asks whether institutional opportunities encourage participation beyond orientations and political linkage. Table~\ref{tab:families} examines this extension through formal collective action and organizational membership in columns (1) and (2).\footnote{The membership family covers parties, religious organizations, guilds and unions, rather than the explicitly named nominating types examined below. Parties could not nominate CITREP candidates. This family therefore measures broader organizational engagement, not a direct response to the statutory nominating categories.} Their estimates are positive but imprecise, at 0.422 ($p=0.154$) and 0.294 ($p=0.357$); their joint pre-trend tests give $p=0.101$ and $0.269$. Formal action also has a significant 2005 differential, making its pre-period pattern less stable than the primary index's. The signs are compatible with a relative rural expansion of membership and formal action, but the wide intervals do not establish it. Columns (3) to (6) report electoral participation, brokerage and clientelism, diffuse social relations, and institutional trust. Figure~\ref{fig:families} compares all nine families with the primary index.

Complementary measures of reported memberships and activities also have positive but imprecise estimates. An index combining membership in communal action boards, peasant associations and cooperatives with meetings, committees, volunteering and petition-signing dispositions yields 0.341 ($p=0.157$). Alternative codings that remove hypothetical willingness to sign petitions or retain only completed acts give estimates of 0.281--0.337, but all confidence intervals include zero.\footnote{Table~\ref{tab:behavioral} reports the original index and three variants: omitting petitions, counting only having signed alongside memberships and other activities, and retaining only the four completed activities without memberships.} The party-activity result above provides evidence beyond attitudes, but these broader measures leave an expansion of collective activity unresolved. Differences in significance do not establish differences in effects.

The contrast domains assess the wider social and political responses considered in the conceptual framework. All four intervals include zero, leaving wider changes in social capital unresolved rather than establishing that these domains were unaffected.\footnote{Table~\ref{tab:intervals} reports the intervals and compares them with an illustrative benchmark of $\pm0.33$ standard deviations, not a prespecified smallest effect of interest or a formal equivalence margin. The intervals are conventional and pointwise; they do not adjust for selection or simultaneous coverage.}

\Needspace{3\baselineskip}
For reliance on intermediaries, the estimate for mentioning a \emph{palanca}, or personal connection, when facing a collective problem is close to zero and imprecise when 2017 is excluded ($\BarcasPalancaCleanB$, $p=\BarcasPalancaCleanP$). This leaves the predicted response in autonomy from intermediaries unresolved.\footnote{The harmonized any-mention indicator also enters the brokerage family's opportunism component. Table~\ref{tab:clean2011} reports the earlier-reference estimate; Table~\ref{tab:speccurve} provides estimates for all 54 catalogue measures.}

\subsection{Which side of the gap moves}\label{sec:sides}

Figure~\ref{fig:cells} plots the primary-index trajectories for rural and urban residents in covered and comparison municipalities. It keeps the same 33 municipalities---17 covered and 16 comparison---with rural and urban respondents in every wave, weighting municipalities equally within each coverage group. The figure displays 2005, 2011 and 2023; the baseline event study continues to use 2017 as its reference.\footnote{Figure~\ref{fig:cells_allwaves} displays all four waves with the same municipalities, weights and outcome scale. Table~\ref{tab:cells} reports changes against 2017 and the pooled earlier waves. Table~\ref{tab:clean_cells} reports changes against 2011 and the mean of 2005 and 2011, excluding 2017 throughout.}

Between 2011 and 2023, the primary index rises by \BarcasCellsCleanTr{} in covered rural areas and \BarcasCellsCleanTu{} in covered urban centers. In comparison municipalities, it falls by 0.283 in rural areas and 0.090 in urban centers. Thus rural engagement gains relative to urban engagement in covered municipalities, while losing ground in comparison municipalities. The resulting descriptive triple difference is \BarcasCellsCleanDdd{}. It differs from the regression estimate because the latter uses regression weights, sex adjustment and broader municipal coverage. Urban improvements do not contradict the conceptual prediction, which concerns a relative rural advantage rather than an absence of urban responses. The corresponding regression decomposition also gives a favorable change in the covered rural--urban gap and an unfavorable change in the comparison gap; neither component alone is an absolute causal effect.\footnote{Table~\ref{tab:regressiongaps} reports changes from the regressions underlying Table~\ref{tab:baseline}. The covered and comparison gap changes are 0.113 and $-0.468$ with reference in 2017, and 0.284 and $-0.341$ when the reference is 2011 and 2017 is excluded.}

Compared instead with the mean of 2005 and 2011, covered rural engagement is near its earlier level, while comparison rural engagement is lower. The descriptive differential remains positive, at \BarcasCellsPooledDdd{}. The rural advantage can therefore involve sustaining engagement against decline, not simply increasing rural levels.

Figure~\ref{fig:municipalchanges} shows how each municipality contributed to the changes in Figure~\ref{fig:cells}. Rural engagement falls in \NovRuralCompDeclining{} of the \NovRuralCompMunicipalities{} comparison municipalities, with a median change of \NovRuralCompMedian{} standard deviations. The average rural decline persists when municipalities are omitted one at a time or when each endpoint must have more than five rural respondents (Appendix~\ref{app:novsupport}).

\section{Identification and Robustness}\label{sec:evidence}

This section examines the evidence for parallel counterfactual gaps and comparable sampled populations, then assesses stability across specifications, samples and inference methods.

\subsection{Evidence on the identifying assumption}\label{sec:sensitivity}\label{sec:anchors}\label{sec:robchar}\label{sec:composition}

The primary index's pre-reform coefficients are very small in magnitude and jointly statistically indistinguishable from zero: \BarcasMainPreFive{} in 2005 and $\BarcasMainPreEleven$ in 2011, with joint $p=\BarcasMainPreP$ (Panel A of Table~\ref{tab:baseline} and Figure~\ref{fig:primary}). The constituent families likewise show no statistically detectable pre-period differentials. These patterns are consistent with parallel counterfactual gaps.\footnote{Appendix~\ref{app:sourceaudit} discusses survey diagnostics relevant to interpreting these tests. Appendix~\ref{app:did} reports the municipal DiD comparisons.} Table~\ref{tab:honestdid} assesses bounded departures using \citet{rambachanroth2023}. For the fixed primary index, the 95 percent interval remains $[0.086,1.166]$ at $\bar M=1$; zero first enters at $\bar M=1.5$ on the tested grid. The relative-magnitude restriction uses the observed pre-period changes in the gap, with limited information about their variability across time. This aggregate robustness does not extend uniformly to the constituents: their intervals first include zero at $\bar M=1$ for coordination, $0.5$ for party linkage and $0$ for civic participation.

Panel B of Table~\ref{tab:baseline} assesses dependence on 2017 by excluding that release and using 2011 as the reference. Restricting both regressions to the same \NovCommonMunicipalities{} municipalities with both residence groups in 2005, 2011 and 2023 gives estimates of \NovCommonBaseline{} and \NovCommonEarlier{}, respectively (Appendix~\ref{app:novsupport}). Alongside the positive post-reform estimate discussed above, the 2005 coefficient relative to 2011 is \BarcasMainCleanPreFive{} ($p=\BarcasMainCleanPreP$), providing separate evidence consistent with parallel gaps. Pooling 2005 and 2011 instead gives \BarcasMainPoolB{}. Appendix~\ref{app:reference2011} reports these alternative time comparisons and the constituent outcomes: coordination is consistently positive and precisely estimated, while party linkage varies more and civic participation is less conclusive.\footnote{The 2011--2023 comparison covers a longer period of the peace process and has one pre-reform interval. The appendix reports group trajectories, item-level results and parallel-trends sensitivity without 2017. Its sensitivity interval is positive at $\bar M=1$ but includes zero at $\bar M=1.5$; the bounds compare observed-wave changes over a six-year pre-period and a twelve-year post-period. Including the sparsely sampled 1997 wave instead gives \BarcasWaveNinetySevenB{}; Appendix~\ref{app:1997} documents that extension.}

Allowing rural trajectories to vary with pre-existing conflict and development characteristics also preserves the sign (Table~\ref{tab:horserace}). On the complete-covariate sample, the unadjusted estimate is \BarcasHorseSameB{}; joint adjustment gives \BarcasHorseFullB{} ($p=\BarcasHorseFullP$). Table~\ref{tab:balance} documents the large baseline differences between coverage groups, and Table~\ref{tab:support} reports restrictions to common covariate support, non-capital municipalities and municipalities with substantial rural populations. All retain positive primary estimates of similar magnitude, although the rurality restriction is less precise. These checks support robustness to alternative municipal comparison sets. A comparison restricted to ZOMAC municipalities outside PDET has only \NovZomacEndpoints{} controls with both residence groups in 2011 and 2023 (Appendix~\ref{app:novsupport}).

Two further checks address exposure to concurrent rural changes (Appendix~\ref{app:submissionchecks}). Excluding the six sampled municipalities hosting early territorial reincorporation spaces (ETCR) gives \BarcasEtcrBaseB{} for the primary index ($p=\BarcasEtcrBaseP$); allowing rural trajectories to vary with recorded coca hectares in 2015 and their reporting status gives \BarcasCocaBaseB{} ($p=\BarcasCocaBaseP$). Both remain positive and conventionally significant when 2017 is excluded. Appendix~\ref{app:novpolitics} examines whether the results are sensitive to differential rural trajectories associated with the geography of presidential voting.

Reweighting addresses changes in the geographic composition of respondents. Matching the 2023 municipal allocation to 2011, separately within coverage and residence groups, gives \BarcasAllocB{} ($p=\BarcasAllocP$), compared with \BarcasAllocUnweightedB{} on the identical unweighted sample (Table~\ref{tab:allocation}). Together with fixed-coverage checks, this supports stability to observed geographic composition.\footnote{Reweighting uses only 2011 and 2023 and has no pre-trend test. The rural-share confidence interval permits substantial change (Table~\ref{tab:composition}); the four-wave age adjustment retains a similar positive primary estimate (Appendix~\ref{app:age}), and Appendix~\ref{app:sampling2023} discusses sampling within municipalities.}

\subsection{Robustness}\label{sec:robustness}\label{sec:robsamples}\label{sec:robinference}\label{sec:multiplicity}

Panel~A of Table~\ref{tab:specs} shows that allowing separate rural paths for every macro-region and wave preserves the positive primary-index estimate (\BarcasMainRegionB{}, $p=\BarcasMainRegionP$). These controls absorb rural--urban changes shared within macro-regions, whether substantive or due to survey measurement. Civic participation and party linkage remain positive but are not individually significant at 5 percent under this adjustment (Table~\ref{tab:specsc}). Coordination continues to provide the clearest family-level evidence.

Holding municipal coverage more tightly fixed also preserves the result. In Panel~A, requiring rural and urban respondents in all four waves leaves 33 municipalities (17 covered) and gives \BarcasStrictB{} ($p=\BarcasStrictP$). The estimate also remains positive when municipalities must have rural respondents in both 2011 and 2023; this endpoint restriction is not a balanced four-wave panel. Panel~B of Table~\ref{tab:specs} reports positive estimates after excluding Mocoa or Santa Marta and under alternative weights. Dropping each covered or comparison municipality, or each department, preserves the positive sign and conventional significance at 5 percent. Finally, counting each underlying party item once, rather than through overlapping published composites, gives a primary-index estimate of \BarcasOnceB{} and \BarcasOnceCleanB{} when 2017 is excluded (Table~\ref{tab:party_once}). Overlap does not drive the finding.

The total number of alternatives reported in the collective-problems module changes by \NovMentionsBase{} in the relative rural gap ($p=\NovMentionsBaseP$), and by \NovMentionsEarlier{} when 2017 is excluded ($p=\NovMentionsEarlierP$). These estimates do not indicate a general increase in mentions. Appendix~\ref{app:novmeasurement} compares response patterns across common survey items and discusses changes in category-specific reporting. The primary finding also persists after excluding municipalities with the most extreme gap changes. On the common endpoint sample, trimming one or two municipalities from each tail of the primary-index gap change within each coverage group gives \BarcasTrimOneB{} and \BarcasTrimTwoB{}, compared with \BarcasTrimSupportB{} without trimming. Appendix~\ref{app:trimming} reports these diagnostics on the unchanged outcome scale; because selection uses observed changes, they assess magnitude sensitivity rather than identify a preferred effect.

The primary-index result remains statistically significant under broader geographic clustering, the cluster jackknife, wild cluster bootstrap tests and fixed-index placebo permutations (Table~\ref{tab:inference}). These comparisons are relevant when policy coverage is concentrated in a limited number of clusters \citep{conleytaber2011,mackinnonwebb2018}.

Figure~\ref{fig:indexspec} shows the estimates for all 84 averages of three families. Every estimate is positive under both reference choices, with medians of \NovTrioBaseMedian{} and \NovTrioEarlierMedian{} standard deviations of the respective indices. The indices share measures and their estimates are correlated. The primary index ranks first by absolute $t$-statistic in the baseline and second with 2017 excluded. Under the baseline, the reference $p$-values that allow choice among the 84 trios are \BarcasSelectionUnit{} for municipal reassignment and \BarcasSelectionBlock{} for block reassignment (Appendix~\ref{app:selection}). Broader indices that retain the primary trio also yield positive estimates (Table~\ref{tab:altindices}). Romano--Wolf adjustment over all nine families leaves all nine above the 5 percent threshold. Coordination has the smallest adjusted $p$-values, 0.051 and 0.055 (Table~\ref{tab:multiplicity}).

Appendix~\ref{app:barcasdimensions} separates changes in weights from changes in content. On a common scale, giving each of the eleven measures equal weight yields 0.541, compared with 0.582 for the primary index. Using fourteen measures from the closest BARCAS dimensions gives 0.350 with equal measure weights and 0.284 with equal dimension weights. The wider measures capture additional constructs and yield smaller, less precise differentials.

\section{Complementary Evidence on the Mechanisms}\label{sec:attribution}

PDET assemblies provide opportunities to coordinate investment proposals. Within covered municipalities, the primary gap changes more favorably where more initiatives were adopted (Table~\ref{tab:dose}). The gradient is \BarcasMainDoseB{} per standard deviation of initiative counts, with conventional $p=\BarcasMainDoseP$ and wild cluster bootstrap $p=\BarcasMainDoseWildP$. Existing organizational capacity could contribute to this association. The appendix reports pillar comparisons, sensitivity to counting and the checks for municipal influence and size.\footnote{See Tables~\ref{tab:dosesize} and~\ref{tab:dose_audit} and Appendix~\ref{app:attribution}.}

CITREP endorsement rules give named organizations an opportunity to recruit and mobilize members. The gap changes by 0.672 in membership of measured nominating types and by 0.206 in other secular types (Table~\ref{tab:nominating}). Their direct contrast is 0.466 ($p=0.099$), with the largest change in membership of ethnic organizations. Other secular organizations may also nominate under the general rules, and ethnic organizations also participate in PDET assemblies. The named-types advantage is less pronounced with 2017 excluded.\footnote{Table~\ref{tab:membershipcoding} examines alternative codings of inactive membership. Appendix~\ref{app:attribution} reports the disaggregated results.} These associations are consistent with mobilization through both reforms. Their overlap leaves the channels empirically intertwined.

\section{Conclusion}\label{sec:conclusion}

Colombia's peace agreement expanded the opportunities for rural residents to participate in local investment and elect dedicated representatives. The engagement gap between rural and urban residents moved in favor of rural residents in municipalities covered by both reforms, relative to the change in comparison municipalities. The contrast persists when the reference is moved to 2011 and the 2017 release is excluded. The group trajectories show that a more favorable gap can reflect both changes in covered areas and deterioration elsewhere.

Coordination around collective problems provides the clearest evidence across reference periods. Results from reported civic activities and mobilization are more variable, and the associations with PDET initiatives and organizational membership suggest possible channels. The common geographic coverage prevents attribution of the relative change to either program alone or to their interaction.

The evidence places coordination and civic engagement alongside electoral participation as outcomes of interest in territorial peace reforms. Additional survey waves, fieldwork documentation and administrative records could establish whether the changes persisted and clarify how residents used the opportunities opened by the agreement.

\section*{Data and code availability}

The Social Capital Barometer (BARCAS) was produced by the Corporaci\'on para el Control Social (Contrial), which holds the rights to the microdata. The Centro de Datos CEDE, Universidad de los Andes, serves as the repository and provided access under a confidentiality agreement. Replication requires the five BARCAS releases and the General Characteristics and Conflict and Violence files from the 2019 CEDE Municipal Panel. On 19 November 2026, CEDE confirmed that external evaluators may request the versions used in this study and their documentation at \texttt{datoscede@uniandes.edu.co}, subject to its access procedures and conditions. The source manifest identifies the exact files and their SHA-256 checksums.

The replication archive includes five Stata/Mata programs, permitted public inputs, source documentation, aggregate reference values and automated verification routines. It excludes the BARCAS microdata, respondent-level derivatives and the two CEDE Municipal Panel source files. The archive will be supplied for editorial and referee review. The accuracy and reliability of the results and their interpretation are the sole responsibility of the author.

The online supplement documents survey harmonization and validation in Appendix~\ref{app:data} and provides the supporting estimates, outcome definitions and sensitivity analyses referenced throughout the paper.

\section*{Declarations}

\noindent\textit{Competing interests.} The author declares no known competing financial interests or personal relationships that could have appeared to influence the work reported in this paper.

\medskip
\noindent\textit{Funding.} This research received no specific grant from any funding agency in the public, commercial or not-for-profit sectors.

\medskip
\Needspace{4\baselineskip}
\noindent\textit{Ethics.} The study uses de-identified secondary microdata from a survey conducted and distributed by a third party. No new data were collected from human subjects and no individual can be identified in the files used.

\medskip
\noindent\textit{Generative AI.} During the preparation of this work the author used a large language model to assist with code development and review, editorial revision, and checking numerical consistency between the text and the estimation logs. The author reviewed and edited the output and takes full responsibility for the content of the article.
\clearpage
\bibliographystyle{apalike}
\bibliography{references}
\clearpage
\renewcommand{\thetable}{\Roman{table}}
\setcounter{table}{0}\setcounter{figure}{0}
\section*{Figures and Tables}

\begin{figure}[H]\centering
\captionsetup{font={normalsize,md}}
\caption{The Primary Index under Two Reference Choices}\label{fig:primary}
\includegraphics[width=0.98\textwidth]{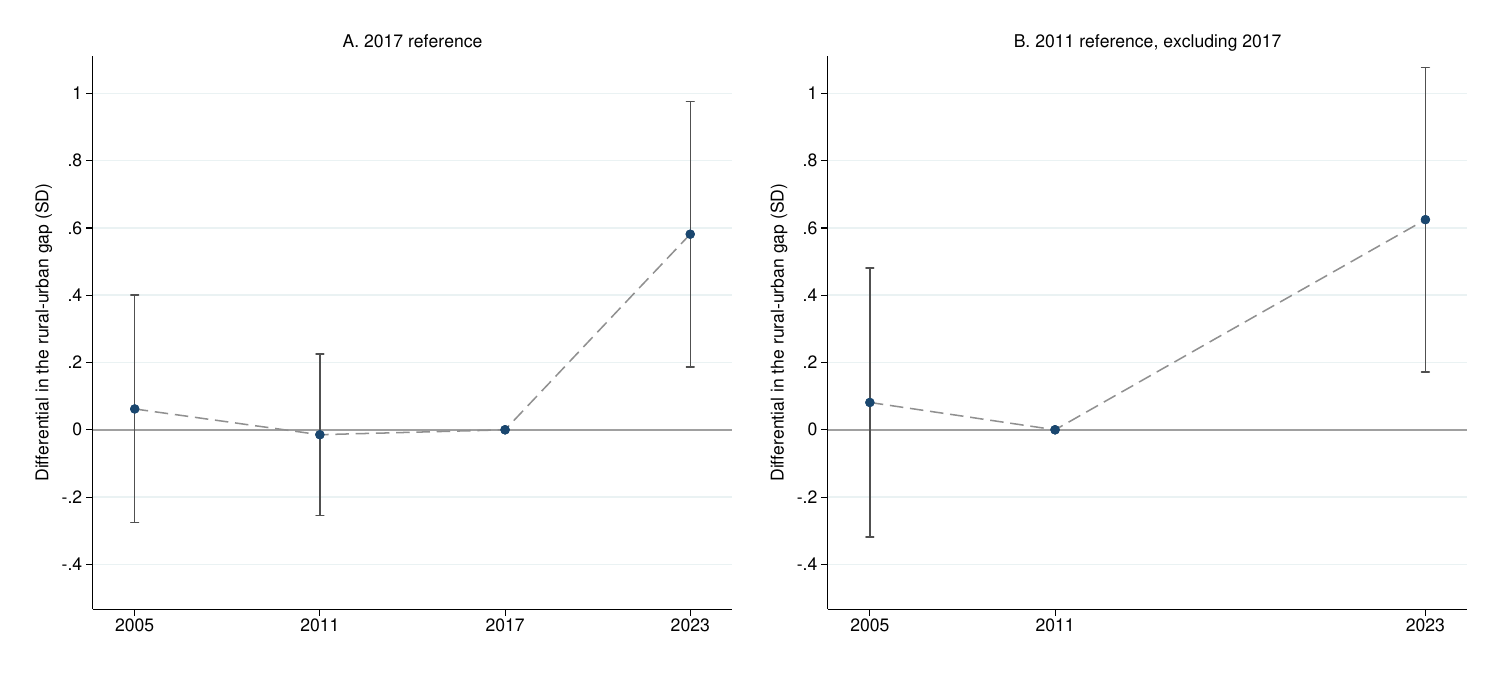}
\begin{minipage}{0.9\textwidth}\vspace{0.15cm}
\tablenotefont\textit{Note:} Each point measures the change since the panel's reference wave in the rural--urban engagement gap, comparing covered and comparison municipalities. The left panel uses all four waves and 2017 as reference. The right panel uses 2005, 2011 and 2023, with 2011 as reference. Reference waves are normalized to zero and have no interval. Both panels use the same outcome scale and municipality-by-wave and municipality-by-residence fixed effects. Bars are conventional 95\% municipality-clustered intervals.
\end{minipage}\end{figure}
\begin{figure}[H]\centering
\captionsetup{font={normalsize,md}}
\caption{The Primary Index and All Nine Families}\label{fig:families}
\includegraphics[width=0.92\textwidth]{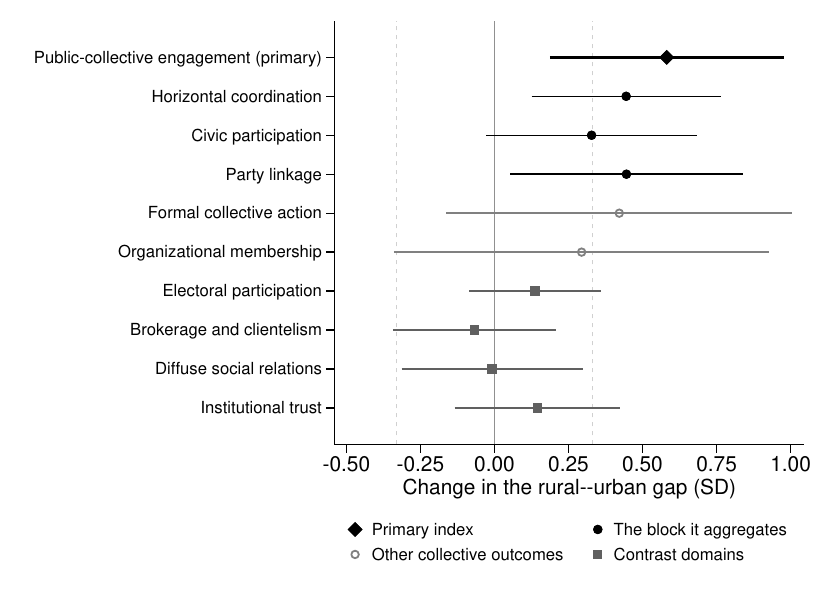}
\begin{minipage}{0.92\textwidth}\vspace{0.15cm}
\tablenotefont\textit{Note:} Coefficients and conventional 95\% confidence intervals from equation~\eqref{eq:esddd}, with both sets of fixed effects and municipality clustering. The dashed lines mark $-0.33$ and $+0.33$, the common illustrative benchmark of Table~\ref{tab:intervals}. Marker shapes distinguish the primary index, its constituents, other collective outcomes and contrast domains.
\end{minipage}\end{figure}

\begin{figure}[H]\centering
\captionsetup{font={normalsize,md}}
\caption{Four Group Trajectories on Fixed Longitudinal Support}\label{fig:cells}
\includegraphics[width=0.95\textwidth]{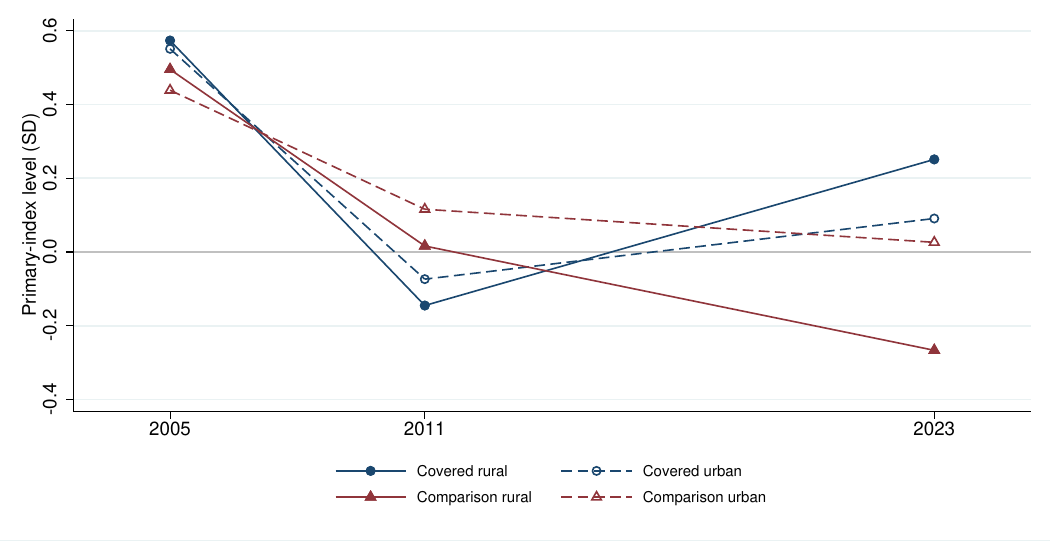}
\begin{minipage}{0.95\textwidth}\vspace{0.15cm}\tablenotefont
\textit{Note:} Means for the same 33 municipalities (17 covered, 16 comparison), each with rural and urban observations in all four waves. Municipalities have equal weight within their coverage group in every year. Scaling uses the full-sample 2011 rural comparison cell. The figure omits 2017; Figure~\ref{fig:cells_allwaves} displays the complete series with identical municipalities, weights and scale. The baseline event study retains 2017 as its reference. Table~\ref{tab:clean_cells} reports changes excluding that wave. These are descriptive levels, not separately identified treatment effects.
\end{minipage}
\end{figure}

\begin{figure}[H]\centering
\captionsetup{font={normalsize,md}}
\caption{Municipal Changes in Engagement from 2011 to 2023}\label{fig:municipalchanges}
\includegraphics[width=0.98\textwidth]{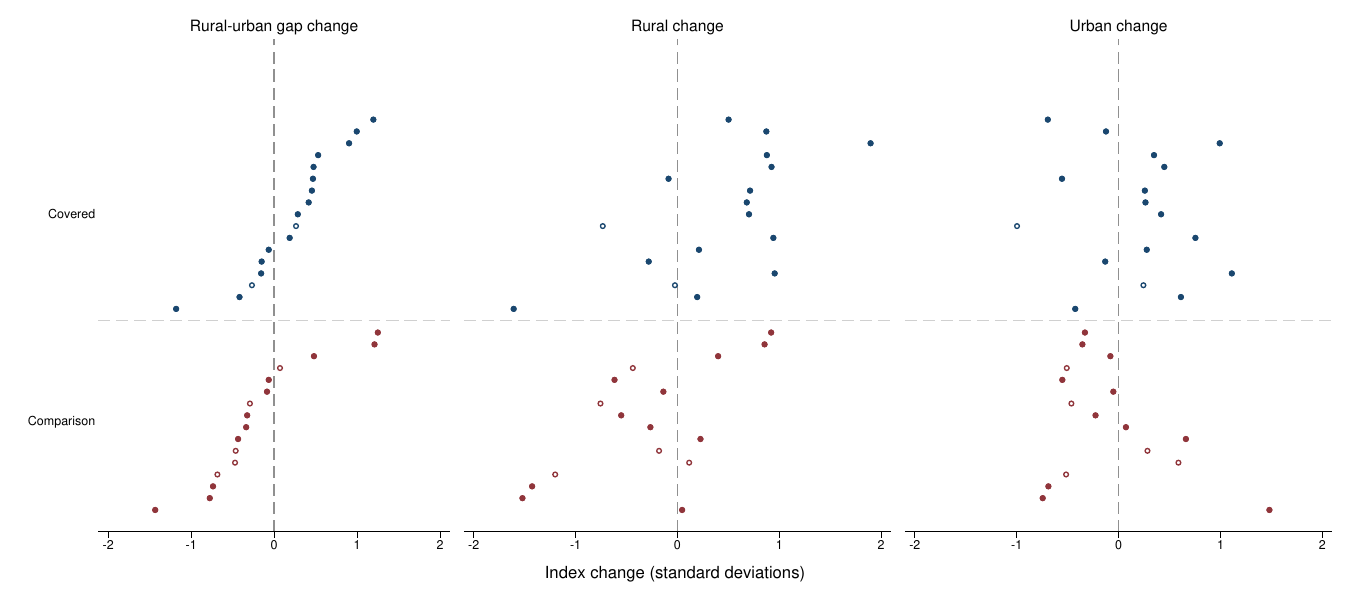}
\begin{minipage}{0.95\textwidth}\vspace{0.15cm}\tablenotefont
\textit{Note:} Each point represents one municipality with rural and urban respondents at both endpoints. Changes use the primary index in its 2011 rural-comparison scale. Hollow markers identify municipalities with at most five rural respondents at either endpoint.
\end{minipage}
\end{figure}

\begin{figure}[H]\centering
\captionsetup{font={normalsize,md}}
\caption{Engagement Differentials across Index Definitions}\label{fig:indexspec}
\includegraphics[width=0.98\textwidth]{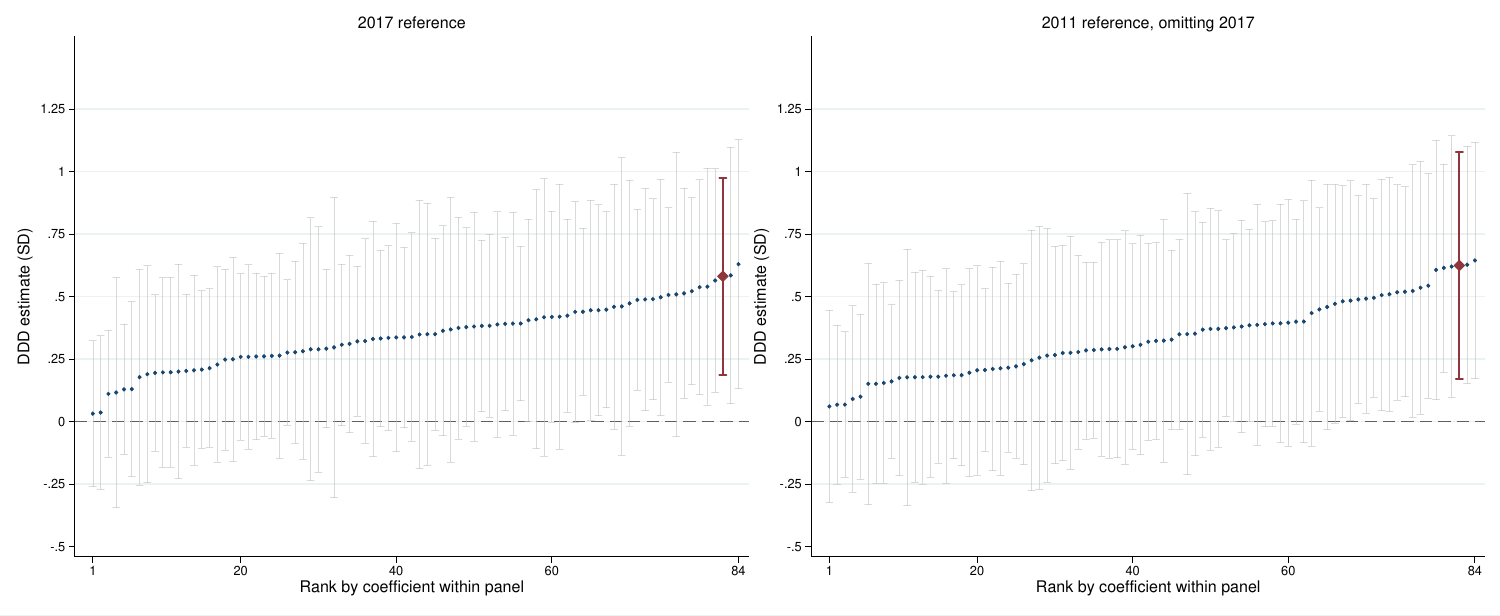}
\begin{minipage}{0.95\textwidth}\vspace{0.15cm}\tablenotefont
\textit{Note:} Each panel ranks the estimates for all 84 equally weighted combinations of three of the nine families. The maroon diamond marks the primary index. Every aggregate is expressed in its own 2011 rural comparison standard deviation. Intervals are pointwise with municipality clustering.
\end{minipage}
\end{figure}

\begin{table}[H]{\renewcommand{\arraystretch}{0.9}\setlength{\tabcolsep}{2pt}
\captionsetup{font={normalsize,md}}
\caption{Rural and Urban Coverage by Wave}\label{tab:coverage}\vspace{-0.3cm}
\begin{tablecontent}\begin{adjustbox}{max width=0.98\linewidth, max totalheight=0.86\textheight}
\begin{tabular}{L{2.1cm}*{8}{C{1.5cm}}}

\hline\hline \addlinespace[0.15cm]
 & (1) & (2) & (3) & (4) & (5) & (6) & (7) & (8) \\
\addlinespace[0.10cm]
 & \multicolumn{4}{c}{Covered municipalities} & \multicolumn{4}{c}{Comparison} \\
\cmidrule(lr){2-5}\cmidrule(lr){6-9}
Wave & Rural & Urban & Total N & Munic. & Rural & Urban & Total N & Munic. \\
\addlinespace[0.15cm]\hline\addlinespace[0.15cm]
2005 & 395 & 358 & 753 & 19 & 433 & 1839 & 2272 & 38 \\
2011 & 356 & 396 & 752 & 19 & 357 & 1919 & 2276 & 38 \\
2017 & 397 & 455 & 852 & 22 & 390 & 1946 & 2336 & 40 \\
2023 & 338 & 518 & 856 & 22 & 332 & 2037 & 2369 & 40 \\
\addlinespace[0.15cm]\hline\hline

\end{tabular}\end{adjustbox}\end{tablecontent}\tablenotesgap
\centering\begin{minipage}{\exhibittablewidth}\tablenotefont\textbf{Notes:} Numbers of interviewed rural and urban respondents by wave and coverage group; Total N is their sum, and Munic. counts distinct municipalities. Municipalities with both residence groups number 17, 19, 22 and 22 among covered units, and 32, 32, 24 and 27 among comparison units, respectively. The two covered municipality--wave cells without rural respondents occur in 2005. The 1997 wave is excluded; Section~\ref{sec:wave1997} reports its coverage.\end{minipage}\par}\end{table}

\begin{table}[H]{\renewcommand{\arraystretch}{0.95}\setlength{\tabcolsep}{2pt}
\captionsetup{font={normalsize,md}}
\caption{The Primary Index and Its Constituents: Baseline and Earlier Reference}\label{tab:baseline}\vspace{-0.3cm}
\begin{tablecontent}\begin{adjustbox}{max width=0.98\linewidth, max totalheight=0.86\textheight}
\begin{tabular}{L{5.3cm}*{4}{C{2.4cm}}}

\hline\hline \addlinespace[0.15cm]
 & (1) & (2) & (3) & (4) \\
 & Primary index & Horizontal coordination & Civic participation & Party linkage \\
\addlinespace[0.10cm]\hline\addlinespace[0.15cm]
\emph{Panel A:} & \multicolumn{4}{c}{Baseline: 2017 reference} \\
\cmidrule(lr){2-5}
Rural $\times$ Covered $\times$ 2023 & 0.582*** & 0.445*** & 0.328* & 0.446** \\
 & (0.198) & (0.160) & (0.178) & (0.197) \\
 & [0.187, 0.976] & [0.126, 0.764] & [-0.027, 0.683] & [0.053, 0.838] \\
Conventional $p$ & 0.005 & 0.007 & 0.070 & 0.027 \\
Wild cluster bootstrap $p$ & 0.003 & 0.006 & 0.084 & 0.030 \\
Rural $\times$ Covered $\times$ 2011 & -0.015 & -0.102 & 0.067 & 0.005 \\
 & (0.120) & (0.119) & (0.126) & (0.119) \\
Rural $\times$ Covered $\times$ 2005 & 0.062 & -0.114 & 0.124 & 0.119 \\
 & (0.169) & (0.135) & (0.180) & (0.171) \\
Pre-trend $p$ & 0.921 & 0.555 & 0.752 & 0.775 \\
R-squared & 0.171 & 0.127 & 0.151 & 0.129 \\
Observations & 12,466 & 12,466 & 12,466 & 12,466 \\
Municipalities (clusters) & 66 & 66 & 66 & 66 \\
\addlinespace[0.18cm]
\emph{Panel B:} & \multicolumn{4}{c}{2011 reference, excluding 2017} \\
\cmidrule(lr){2-5}
Rural $\times$ Covered $\times$ 2023 & 0.625*** & 0.582*** & 0.253 & 0.474** \\
 & (0.227) & (0.188) & (0.192) & (0.230) \\
 & [0.172, 1.078] & [0.208, 0.957] & [-0.131, 0.636] & [0.014, 0.935] \\
Conventional $p$ & 0.008 & 0.003 & 0.193 & 0.044 \\
Wild cluster bootstrap $p$ & 0.006 & 0.002 & 0.199 & 0.046 \\
Rural $\times$ Covered $\times$ 2005 & 0.081 & 0.001 & 0.055 & 0.114 \\
 & (0.200) & (0.164) & (0.184) & (0.186) \\
Pre-trend $p$ & 0.686 & 0.994 & 0.764 & 0.543 \\
R-squared & 0.158 & 0.117 & 0.173 & 0.102 \\
Observations & 9,278 & 9,278 & 9,278 & 9,278 \\
Municipalities (clusters) & 64 & 64 & 64 & 64 \\
\addlinespace[0.10cm]\hline\addlinespace[0.15cm]
Municipality $\times$ wave FE & Y & Y & Y & Y \\
Municipality $\times$ residence FE & Y & Y & Y & Y \\
Sex controls & Y & Y & Y & Y \\
\addlinespace[0.15cm]\hline\hline

\end{tabular}\end{adjustbox}\end{tablecontent}\tablenotesgap
\centering\begin{minipage}{\exhibittablewidth}\tablenotefont\textbf{Notes:} Panel A estimates equation~\eqref{eq:esddd} on waves 2005, 2011, 2017 and 2023, with 2017 as the reference. Panel B uses waves 2005, 2011 and 2023, with 2011 as the reference; all 2017 observations are excluded. Each coefficient is the change from its panel's reference in the rural-minus-urban gap, in covered relative to comparison municipalities. Both panels include the displayed fixed effects and respondent sex. Municipality-clustered standard errors appear in parentheses and conventional 95\% intervals in brackets. The outcome scale is identical across panels: the 2011 rural comparison distribution. Column (1) averages the families in columns (2)--(4) and is re-standardized. Pre-trend $p$ tests the 2005 and 2011 coefficients jointly in Panel A and the 2005 coefficient in Panel B. Both panels report conventional cluster-$t$ and wild cluster bootstrap $p$-values for the 2023 coefficient; the bootstrap uses 9,999 replications. Table~\ref{tab:inference} reports placebo permutations. Stars use conventional $p$-values: *, **, *** denote 10\%, 5\% and 1\% significance. Section~\ref{sec:multiplicity} reports selection-sensitive inference.\end{minipage}\par}\end{table}

\begin{table}[H]{\renewcommand{\arraystretch}{0.95}\setlength{\tabcolsep}{2pt}
\captionsetup{font={normalsize,md}}
\caption{Components of the Primary Engagement Index}\label{tab:components_main}\vspace{-0.3cm}
\begin{tablecontent}\begin{adjustbox}{max width=0.98\linewidth, max totalheight=0.86\textheight}
\begin{tabular}{L{7.4cm}C{1.9cm}C{3.0cm}C{2.1cm}}

\hline\hline\addlinespace[0.15cm]
Outcome & $\pi_{2023}$ & 95\% interval & Pre-trend $p$ \\
\addlinespace[0.10cm]\hline\addlinespace[0.15cm]
\multicolumn{4}{l}{\textit{Horizontal coordination}} \\
Joint resolution mentioned & 0.322** & [0.014, 0.630] & 0.642 \\
Conflicts resolvable when aired & 0.368** & [0.070, 0.666] & 0.471 \\
\multicolumn{4}{l}{\textit{Civic participation}} \\
Civic activity & 0.225 & [-0.070, 0.519] & 0.596 \\
Local activity & 0.220 & [-0.144, 0.583] & 0.796 \\
Media activity & 0.319 & [-0.128, 0.765] & 0.835 \\
Volunteer work & 0.101 & [-0.247, 0.449] & 0.051 \\
Citizen politics & 0.062 & [-0.258, 0.383] & 0.986 \\
Political education & 0.122 & [-0.087, 0.332] & 0.220 \\
\multicolumn{4}{l}{\textit{Party linkage}} \\
Membership, trust and activity & 0.446** & [0.050, 0.841] & 0.745 \\
Membership and trust & 0.353* & [-0.021, 0.726] & 0.427 \\
Party activity & 0.326** & [0.033, 0.619] & 0.881 \\
Observations (each regression) & \multicolumn{3}{c}{12,466} \\
\hline\hline

\end{tabular}\end{adjustbox}\end{tablecontent}\tablenotesgap
\centering\begin{minipage}{\exhibittablewidth}\tablenotefont\textbf{Notes:} Estimates of $\pi_{2023}$ from equation~\eqref{eq:esddd}, relative to 2017. All eleven components of the three families are retained. Regressions include municipality-by-wave and municipality-by-residence effects and respondent sex. Outcomes use the 2011 rural comparison standardization. The 95\% intervals use municipality-clustered standard errors. Pre-trend $p$ tests $\pi_{2005}=\pi_{2011}=0$. The party composites have overlapping content; the membership--trust composite does not measure trust alone. Table~\ref{tab:components} reports the full regressions, including standard errors and pre-reform coefficients. Stars use conventional pointwise tests: *, **, *** denote 10\%, 5\% and 1\% significance.\end{minipage}\par}\end{table}

\begin{table}[H]{\renewcommand{\arraystretch}{0.95}\setlength{\tabcolsep}{2pt}
\captionsetup{font={normalsize,md}}
\caption{Baseline Results: Other Outcome Families}\label{tab:families}\vspace{-0.3cm}
\begin{tablecontent}\begin{adjustbox}{max width=0.98\linewidth, max totalheight=0.86\textheight}
\begin{tabular}{L{4.8cm}*{6}{C{1.67cm}}}

\hline\hline \addlinespace[0.15cm]
 & \multicolumn{2}{c}{Other collective outcomes} & \multicolumn{4}{c}{Contrast domains} \\
\cmidrule(lr){2-3}\cmidrule(lr){4-7}
 & (1) & (2) & (3) & (4) & (5) & (6) \\
 & Formal coll. action & Org. membership & Electoral particip. & Brokerage & Diffuse relations & Inst. trust \\
\addlinespace[0.10cm]\hline\addlinespace[0.15cm]
Rural $\times$ Covered $\times$ 2023 & 0.422 & 0.294 & 0.137 & -0.067 & -0.008 & 0.145 \\
 & (0.292) & (0.318) & (0.111) & (0.138) & (0.153) & (0.140) \\
 & [-0.163, 1.006] & [-0.340, 0.929] & [-0.085, 0.359] & [-0.343, 0.209] & [-0.313, 0.297] & [-0.135, 0.424] \\
Rural $\times$ Covered $\times$ 2011 & 0.100 & 0.018 & -0.037 & -0.080 & 0.067 & -0.007 \\
 & (0.147) & (0.142) & (0.148) & (0.142) & (0.124) & (0.129) \\
Rural $\times$ Covered $\times$ 2005 & 0.372** & 0.259 & -0.019 & -0.175 & 0.119 & 0.066 \\
 & (0.174) & (0.184) & (0.133) & (0.154) & (0.166) & (0.124) \\
Pre-trend $p$ & 0.101 & 0.269 & 0.956 & 0.481 & 0.731 & 0.861 \\
Permutation $p$ & 0.094 & 0.300 & 0.315 & 0.586 & 0.955 & 0.320 \\
R-squared & 0.093 & 0.084 & 0.196 & 0.079 & 0.205 & 0.236 \\
Observations & 12,466 & 12,466 & 12,466 & 12,466 & 12,466 & 12,466 \\
\addlinespace[0.10cm]\hline\addlinespace[0.15cm]
Municipality $\times$ wave FE & Y & Y & Y & Y & Y & Y \\
Municipality $\times$ residence FE & Y & Y & Y & Y & Y & Y \\
Municipalities (clusters) & 66 & 66 & 66 & 66 & 66 & 66 \\
\addlinespace[0.15cm]\hline\hline

\end{tabular}\end{adjustbox}\end{tablecontent}\tablenotesgap
\centering\begin{minipage}{\exhibittablewidth}\tablenotefont\textbf{Notes:} Estimates of equation~\eqref{eq:esddd} under the baseline specification of Table~\ref{tab:baseline}, normalized on 2017, waves 2005--2023. Standard errors clustered by municipality in parentheses and 95\% confidence intervals in brackets. Outcomes are standardized on the 2011 rural comparison distribution. $X_i$ contains respondent sex. Pre-trend $p$ tests $\pi_{2005}=\pi_{2011}=0$. Columns (1) and (2) are the other collective outcomes; columns (3) to (6) are contrast domains, for which no sign is imposed. Permutation $p$ uses 1,999 within-region municipal reassignments (Table~\ref{tab:inference}). Table~\ref{tab:intervals} compares the confidence intervals with an illustrative benchmark. *, **, *** denote significance at 10\%, 5\% and 1\%.\end{minipage}\par}\end{table}

\begin{table}[H]{\renewcommand{\arraystretch}{0.95}\setlength{\tabcolsep}{2pt}
\captionsetup{font={normalsize,md}}
\caption{Pre-Reform Differentials: Sensitivity to Violations of Parallel Trends}\label{tab:honestdid}\vspace{-0.3cm}
\begin{tablecontent}\begin{adjustbox}{max width=0.98\linewidth, max totalheight=0.86\textheight}
\begin{tabular}{L{5.3cm}*{4}{C{2.4cm}}}

\hline\hline \addlinespace[0.15cm]
 & (1) & (2) & (3) & (4) \\
 & Primary index & Coordination & Civic particip. & Party linkage \\
\addlinespace[0.10cm]\hline\addlinespace[0.15cm]
Rural $\times$ Covered $\times$ 2023 & 0.582 & 0.445 & 0.328 & 0.446 \\
 & (0.198) & (0.160) & (0.178) & (0.197) \\
\addlinespace[0.10cm]
 & \multicolumn{4}{c}{95\% sensitivity intervals, relative-magnitudes bound $\bar M$} \\
\cmidrule(lr){2-5}
Original interval & [0.194, 0.969] & [0.132, 0.758] & [-0.020, 0.676] & [0.061, 0.831] \\
$\bar M=0.00$ & [0.190, 0.973] & [0.128, 0.761] & [-0.024, 0.680] & [0.056, 0.835] \\
$\bar M=0.25$ & [0.180, 0.999] & [0.100, 0.777] & [-0.026, 0.703] & [0.030, 0.854] \\
$\bar M=0.50$ & [0.170, 1.041] & [0.071, 0.792] & [-0.042, 0.741] & [-0.005, 0.889] \\
$\bar M=0.75$ & [0.136, 1.091] & [0.029, 0.821] & [-0.065, 0.786] & [-0.057, 0.932] \\
$\bar M=1.00$ & [0.086, 1.166] & [-0.020, 0.864] & [-0.118, 0.854] & [-0.133, 1.001] \\
$\bar M=1.50$ & [-0.066, 1.335] & [-0.141, 0.971] & [-0.254, 1.005] & [-0.306, 1.164] \\
$\bar M=2.00$ & [-0.244, 1.523] & [-0.273, 1.094] & [-0.414, 1.174] & [-0.491, 1.349] \\
\addlinespace[0.10cm]
Breakdown $\bar M$ & 1.50 & 1.00 & 0.00 & 0.50 \\
\addlinespace[0.15cm]\hline\hline

\end{tabular}\end{adjustbox}\end{tablecontent}\tablenotesgap
\centering\begin{minipage}{\exhibittablewidth}\tablenotefont\textbf{Notes:} 95\% sensitivity intervals for $\pi_{2023}$ under the relative-magnitudes restriction of \citet{rambachanroth2023}, bounding post-reform departures from parallel trends by a multiple $\bar M$ of the largest pre-reform wave-to-wave movement. Inference accounts for uncertainty in the pre-reform coefficients. $\bar M=0$ imposes exact post-reform parallel trends. The original interval uses the asymptotic Gaussian critical value returned by \texttt{honestdid}, rather than the cluster-$t$ critical value in Table~\ref{tab:baseline}. Breakdown is the first evaluated grid value whose interval contains zero. C-LF method, 5\% level. These intervals condition on outcome definitions and do not jointly adjust for index selection. Table~\ref{tab:allsensitivity} reports the same procedure for all families.\end{minipage}\par}\end{table}

\begin{table}[H]{\renewcommand{\arraystretch}{0.95}\setlength{\tabcolsep}{1pt}
\captionsetup{font={normalsize,md}}
\caption{Evidence on the Identifying Assumption: Pre-Existing Municipal Characteristics}\label{tab:horserace}\vspace{-0.3cm}
\begin{tablecontent}\begin{adjustbox}{max width=0.98\linewidth, max totalheight=0.86\textheight}
\begin{tabular}{L{4.3cm}*{9}{C{1.24cm}}}

\hline\hline \addlinespace[0.15cm]
 & (1) & (2) & (3) & (4) & (5) & (6) & (7) & (8) & (9) \\
 & Full sample & Same sample & Homicides & Displacement & Basic needs & Rurality & Population & Dist. to capital & All six \\
\addlinespace[0.10cm]\hline\addlinespace[0.15cm]
 & \multicolumn{9}{c}{Dependent variable: Primary index} \\
\cmidrule(lr){2-10}
Rural $\times$ Covered $\times$ 2023 & 0.582*** & 0.574*** & 0.566*** & 0.579** & 0.703** & 0.570** & 0.592*** & 0.632*** & 0.779** \\
 & (0.198) & (0.204) & (0.204) & (0.252) & (0.287) & (0.250) & (0.206) & (0.235) & (0.370) \\
Pre-trend $p$ & 0.921 & 0.943 & 0.941 & 0.892 & 0.896 & 0.981 & 0.891 & 0.886 & 0.712 \\
R-squared & 0.171 & 0.164 & 0.164 & 0.164 & 0.165 & 0.164 & 0.164 & 0.164 & 0.166 \\
Observations & 12,466 & 12,105 & 12,105 & 12,105 & 12,105 & 12,105 & 12,105 & 12,105 & 12,105 \\
\addlinespace[0.10cm]\hline\addlinespace[0.15cm]
Covariate $\times$ rural $\times$ wave & N & N & Y & Y & Y & Y & Y & Y & Y \\
Common covariate sample & N & Y & Y & Y & Y & Y & Y & Y & Y \\
\addlinespace[0.15cm]\hline\hline

\end{tabular}\end{adjustbox}\end{tablecontent}\tablenotesgap
\centering\begin{minipage}{\exhibittablewidth}\tablenotefont\textbf{Notes:} Estimates of equation~\eqref{eq:esddd} for the primary index. Only the 2023 coefficient is displayed; the 2005 and 2011 interactions are included. Column (1) is the full-sample baseline; column (2) is the unadjusted baseline on the complete-covariate sample used in columns (3)--(9). Columns (3)--(8) add one standardized characteristic interacted with rural residence and wave; column (9) adds all six. Homicides, displacement, population, rurality and distance to the departmental capital are measured in 2015; unsatisfied basic needs are measured in 2011, the latest available observation before 2016. All are held fixed across survey waves. Homicides, displacement and population enter as $\log(1+x)$, distance in hundreds of kilometers, and poverty and rurality as indices. The interactions permit trajectories associated linearly with measured pre-agreement characteristics, not a bound on confounding. Table~\ref{tab:horseracec} reports the constituents. Standard errors clustered by municipality in parentheses. Outcomes are standardized on the 2011 rural comparison distribution; $X_i$ contains sex. *, **, *** denote 10\%, 5\% and 1\%.\end{minipage}\par}\end{table}

\begin{table}[H]{\renewcommand{\arraystretch}{0.95}\setlength{\tabcolsep}{2pt}
\captionsetup{font={normalsize,md}}
\caption{Robustness: Fixed Effects, Samples and Weights}\label{tab:specs}\vspace{-0.3cm}
\begin{tablecontent}\begin{adjustbox}{max width=0.98\linewidth, max totalheight=0.86\textheight}
\begin{minipage}{\dimexpr14.6cm+24pt\relax}\centering
\begin{tabular}{L{5.1cm}*{5}{C{1.9cm}}}

\hline\hline \addlinespace[0.15cm]
 &  & \multicolumn{2}{c}{Fixed effects} & \multicolumn{2}{c}{Sample} \\
\cmidrule(lr){3-4}\cmidrule(lr){5-6}
 & (1) & (2) & (3) & (4) & (5) \\
 & Baseline & No munic.\ $\times$ residence FE & Region $\times$ residence $\times$ wave FE & Strict support & Rural in 2011 and 2023 \\
\addlinespace[0.10cm]\hline\addlinespace[0.15cm]
\emph{Panel A:} & \multicolumn{5}{c}{Fixed effects and municipal coverage} \\
\cmidrule(lr){2-6}
Rural $\times$ Covered $\times$ 2023 & 0.582*** & 0.513*** & 0.436*** & 0.442** & 0.599*** \\
 & (0.198) & (0.183) & (0.146) & (0.174) & (0.220) \\
Pre-trend $p$ & 0.921 & 0.956 & 0.753 & 0.976 & 0.928 \\
R-squared & 0.171 & 0.165 & 0.174 & 0.221 & 0.215 \\
Observations & 12,466 & 12,466 & 12,466 & 5,032 & 6,312 \\
\addlinespace[0.10cm]\hline\addlinespace[0.15cm]
Municipality $\times$ residence FE & Y & N & Y & Y & Y \\
Region $\times$ residence $\times$ wave FE & N & N & Y & N & N \\

\end{tabular}
\par\vspace{0.18cm}
\begin{tabular}{L{5.1cm}*{4}{C{\dimexpr2.375cm+1pt\relax}}}

\emph{Panel B:} & \multicolumn{4}{c}{Municipality exclusions and weights} \\
\cmidrule(lr){2-5}
 & \multicolumn{2}{c}{Sample} & \multicolumn{2}{c}{Weights} \\
\cmidrule(lr){2-3}\cmidrule(lr){4-5}
 & (6) & (7) & (8) & (9) \\
 & Excl.\ Mocoa & Excl.\ Santa Marta & Equal munic.\ weights & Regional reweighting \\
\addlinespace[0.10cm]\hline\addlinespace[0.15cm]
Rural $\times$ Covered $\times$ 2023 & 0.605*** & 0.598*** & 0.569*** & 0.558*** \\
 & (0.202) & (0.199) & (0.209) & (0.193) \\
Pre-trend $p$ & 0.908 & 0.906 & 0.954 & 0.497 \\
R-squared & 0.171 & 0.171 & 0.195 & 0.171 \\
Observations & 12,406 & 12,226 & 12,466 & 12,466 \\
\addlinespace[0.10cm]\hline\addlinespace[0.15cm]
Municipality $\times$ residence FE & Y & Y & Y & Y \\
Region $\times$ residence $\times$ wave FE & N & N & N & N \\
\addlinespace[0.15cm]\hline\hline

\end{tabular}
\end{minipage}\end{adjustbox}\end{tablecontent}\tablenotesgap
\centering\begin{minipage}{\exhibittablewidth}\tablenotefont\textbf{Notes:} Both panels report estimates of equation~\eqref{eq:esddd} for the primary index. The 2005 and 2011 interactions are included but not displayed. Column (1) is the baseline. Column (2) omits municipality-by-residence effects; column (3) adds a separate rural path for every macro-region and wave. Column (4) keeps municipalities with rural and urban respondents in all four waves (33 municipalities, 17 covered); column (5) requires rural respondents in both 2011 and 2023, not every wave; columns (6) and (7) exclude Mocoa, the only program municipality outside a constituency in the sample, and Santa Marta. Column (8) weights every municipality equally. Column (9) reweights to stratum-by-region cells, treating missing stratum as a separate category; because stratum is almost entirely missing for rural respondents outside 2017, this is not a demonstrated rural-stratum balance adjustment. The separate age diagnostic appears in Table~\ref{tab:agediagnostic}. Dropping each covered municipality gives \BarcasLooCovMin{} to \BarcasLooCovMax{} (largest $p=\BarcasLooCovP$); dropping comparison municipalities gives \BarcasLooCompMin{} to \BarcasLooCompMax{} (largest $p=\BarcasLooCompP$), and dropping departments gives \BarcasLooDeptMin{} to \BarcasLooDeptMax{} (largest $p=\BarcasLooDeptP$). Table~\ref{tab:specsc} reports the constituents. Standard errors clustered by municipality in parentheses. Outcomes are standardized on the 2011 rural comparison distribution. *, **, *** denote significance at 10\%, 5\% and 1\%.\end{minipage}\par}\end{table}

\begin{table}[H]{\renewcommand{\arraystretch}{0.95}\setlength{\tabcolsep}{2pt}
\captionsetup{font={normalsize,md}}
\caption{Complementary Evidence: Program Activity}\label{tab:dose}\vspace{-0.3cm}
\begin{tablecontent}\begin{adjustbox}{max width=0.98\linewidth, max totalheight=0.86\textheight}
\begin{tabular}{L{4.2cm}*{5}{C{2.1cm}}}

\hline\hline \addlinespace[0.15cm]
 & (1) & (2) & (3) & (4) & (5) \\
 & Primary index & Coordination & Civic particip. & Party linkage & Membership, named types \\
\addlinespace[0.10cm]\hline\addlinespace[0.15cm]
\emph{Panel A:} & \multicolumn{5}{c}{Post-reform gradient} \\
\cmidrule(lr){2-6}
Rural $\times$ Post $\times$ Initiatives & 0.186 & 0.066 & 0.064 & 0.260 & 0.115 \\
 & (0.092) & (0.073) & (0.052) & (0.123) & (0.163) \\
 & [-0.006, 0.378] & [-0.087, 0.218] & [-0.044, 0.171] & [0.005, 0.516] & [-0.224, 0.453] \\
Conventional $p$ & 0.057 & 0.381 & 0.231 & 0.046 & 0.489 \\
Wild-bootstrap $p$ & 0.248 & 0.386 & 0.307 & 0.251 & 0.500 \\
Leave-one-out range & [0.099, 0.239] & [0.021, 0.105] & [0.038, 0.088] & [0.099, 0.313] & [0.006, 0.244] \\
R-squared & 0.246 & 0.160 & 0.233 & 0.195 & 0.248 \\
Observations & 3,213 & 3,213 & 3,213 & 3,213 & 3,213 \\
\addlinespace[0.20cm]
\emph{Panel B:} & \multicolumn{5}{c}{Pre-reform placebo} \\
\cmidrule(lr){2-6}
Rural $\times$ 2011 $\times$ Initiatives & 0.031 & 0.045 & 0.014 & 0.006 & -0.020 \\
 & (0.062) & (0.078) & (0.049) & (0.057) & (0.072) \\
Rural $\times$ 2005 $\times$ Initiatives & -0.125 & -0.017 & -0.142 & -0.103 & -0.353** \\
 & (0.115) & (0.076) & (0.127) & (0.102) & (0.138) \\
Joint $p$ & 0.541 & 0.668 & 0.545 & 0.601 & 0.051 \\
R-squared & 0.248 & 0.161 & 0.235 & 0.196 & 0.255 \\
Observations & 3,213 & 3,213 & 3,213 & 3,213 & 3,213 \\
\addlinespace[0.10cm]\hline\addlinespace[0.15cm]
Municipality $\times$ wave FE & Y & Y & Y & Y & Y \\
Sample & Covered & Covered & Covered & Covered & Covered \\
Municipalities (clusters) & 22 & 22 & 22 & 22 & 22 \\
\addlinespace[0.15cm]\hline\hline

\end{tabular}\end{adjustbox}\end{tablecontent}\tablenotesgap
\centering\begin{minipage}{\exhibittablewidth}\tablenotefont\textbf{Notes:} Covered municipalities only. Panel A compares 2023 with the pooled 2005--2017 waves, not with 2017 alone, and regresses each outcome on the interaction of rural residence, the post-reform wave and the number of initiatives adopted in the municipality's action plan, standardized, with a rural-by-post interaction, municipality-by-wave and municipality-by-residence effects, and $X_i$. Panel B replaces the post-reform interaction with separate interactions for 2005, 2011 and 2023; joint $p$ tests the two pre-reform gradients. Column (5) is membership in the organization types the statute names as nominators; the initiative normalization is held fixed when omitting each municipality in turn. Wild $p$ uses 9,999 municipality-cluster bootstrap replications, seed 20261006. Panel A reports conventional and bootstrap $p$-values without stars; Panel B's stars use conventional cluster-$t$ tests. Initiative counts use snapshot rows; normalization uses the mean and standard deviation among covered respondents in 2023. Deduplication sensitivities appear in Table~\ref{tab:dose_audit}. Standard errors clustered by municipality in parentheses and 95\% confidence intervals in brackets. *, **, *** denote significance at 10\%, 5\% and 1\%.\end{minipage}\par}\end{table}

\begin{table}[H]{\renewcommand{\arraystretch}{0.95}\setlength{\tabcolsep}{2pt}
\captionsetup{font={normalsize,md}}
\caption{Complementary Evidence: Membership by Organization Type}\label{tab:nominating}\vspace{-0.3cm}
\begin{tablecontent}\begin{adjustbox}{max width=0.98\linewidth, max totalheight=0.86\textheight}
\begin{tabular}{L{4.9cm}*{6}{C{1.65cm}}}

\hline\hline \addlinespace[0.15cm]
 & \multicolumn{3}{c}{Membership indices} & \multicolumn{3}{c}{Components of (1)} \\
\cmidrule(lr){2-4}\cmidrule(lr){5-7}
 & (1) & (2) & (3) & (4) & (5) & (6) \\
 & Named types & Other secular types & Difference (1)$-$(2) & Peasant associations & Women's organizations & Ethnic authorities \\
\addlinespace[0.10cm]\hline\addlinespace[0.15cm]
Rural $\times$ Covered $\times$ 2023 & 0.672** & 0.206 & 0.466* & 0.210 & 0.299 & 0.985*** \\
 & (0.307) & (0.333) & (0.278) & (0.201) & (0.218) & (0.370) \\
 & [0.059, 1.285] & [-0.458, 0.870] & [-0.090, 1.022] & [-0.192, 0.612] & [-0.136, 0.734] & [0.247, 1.724] \\
Rural $\times$ Covered $\times$ 2011 & 0.037 & -0.041 & 0.078 & -0.112 & 0.110 & 0.084 \\
 & (0.142) & (0.135) & (0.127) & (0.097) & (0.141) & (0.159) \\
Rural $\times$ Covered $\times$ 2005 & 0.414* & 0.157 & 0.257 & 0.699* & 0.224 & -0.003 \\
 & (0.215) & (0.233) & (0.163) & (0.357) & (0.191) & (0.209) \\
Pre-trend $p$ & 0.166 & 0.653 & 0.297 & 0.078 & 0.438 & 0.863 \\
Permutation $p$ & 0.032 & 0.515 & 0.113 & --- & --- & --- \\
R-squared & 0.149 & 0.108 & 0.096 & 0.219 & 0.060 & 0.175 \\
Observations & 12,466 & 12,466 & 12,466 & 12,466 & 12,466 & 12,466 \\
\addlinespace[0.10cm]\hline\addlinespace[0.15cm]
Municipality $\times$ wave FE & Y & Y & Y & Y & Y & Y \\
Municipality $\times$ residence FE & Y & Y & Y & Y & Y & Y \\
Municipalities (clusters) & 66 & 66 & 66 & 66 & 66 & 66 \\
\addlinespace[0.15cm]\hline\hline

\end{tabular}\end{adjustbox}\end{tablecontent}\tablenotesgap
\centering\begin{minipage}{\exhibittablewidth}\tablenotefont\textbf{Notes:} Estimates of equation~\eqref{eq:esddd} under the baseline specification. Column (1) averages standardized membership in the three types the statute names as nominators that the survey records in every wave; column (2) in six other secular types (sports, cultural, educational, cooperative and charitable organizations and communal action boards), which could also nominate as generic social organizations; column (3) is their difference. Columns (4) to (6) are the components of column (1). The ethnic component records membership in ethnic organizations, a proxy for mobilization around nominating authorities, which could nominate only where a constituency overlaps ethnic territories. Permutation $p$ for the three indices uses 1,999 within-region municipal reassignments (Table~\ref{tab:inference}). Standard errors clustered by municipality in parentheses and 95\% confidence intervals in brackets. Outcomes are standardized on the 2011 rural comparison distribution. $X_i$ contains respondent sex. *, **, *** denote significance at 10\%, 5\% and 1\%.\end{minipage}\par}\end{table}

\clearpage
\pdfbookmark[0]{Online Appendix}{online-appendix}
\section*{Online Appendix}

\noindent This supplement documents the supporting estimates, survey harmonization and sensitivity analyses referenced in the paper. It includes municipal difference-in-differences comparisons, additional evidence on the proposed mechanisms, outcome definitions, and inference allowing index selection or departures from parallel trends. Table and section numbers match the paper's references, and the replication package reproduces the exhibits. BARCAS was produced by the Corporaci\'on para el Control Social (Contrial), which holds the rights to the microdata. The Centro de Datos CEDE, Universidad de los Andes, serves as the repository and facilitated access under a confidentiality agreement \citep{barcas2023}. The accuracy and reliability of the results derived from them, and their interpretation, are the sole responsibility of the author.

\medskip
\noindent\textit{Guide to the supplement.} Appendix~\ref{app:data} documents survey harmonization, with source-file diagnostics in Section~\ref{app:sourceaudit} and the national and CITREP sampling discussion in Section~\ref{app:sampling2023}. Appendix~\ref{app:reference2011} expands the earlier-reference comparison in Panel B of the paper's Table~\ref{tab:baseline}, including alternative anchors, group trajectories and outcome-level estimates. Appendices~\ref{app:selection} and~\ref{app:allsensitivity} detail inference allowing index choice and departures from parallel trends, respectively. The contents below locate the remaining supporting analyses.

\clearpage
\addtocontents{toc}{\protect\setcounter{tocdepth}{2}}
{\renewcommand{\contentsname}{Contents of the Online Appendix}\singlespacing\small\tableofcontents}
\clearpage

\appendix
\renewcommand{\theHsection}{appendix.\Alph{section}}
\renewcommand{\theHtable}{appendix.\arabic{table}}
\renewcommand{\theHfigure}{appendix.\arabic{figure}}
\renewcommand{\theHequation}{appendix.\arabic{equation}}
\renewcommand{\thesection}{\Alph{section}}
\renewcommand{\thetable}{A\arabic{table}}
\setcounter{table}{0}
\renewcommand{\thefigure}{A\arabic{figure}}
\setcounter{figure}{0}
\renewcommand{\theequation}{A\arabic{equation}}
\setcounter{equation}{0}

\section{Survey Geography}\label{app:map}

\begin{figure}[H]\centering
\captionsetup{font={normalsize,md}}
\caption{Where the Constituencies Are, and Where the Survey Reaches Them}\label{fig:map}
\includegraphics[width=0.72\textwidth]{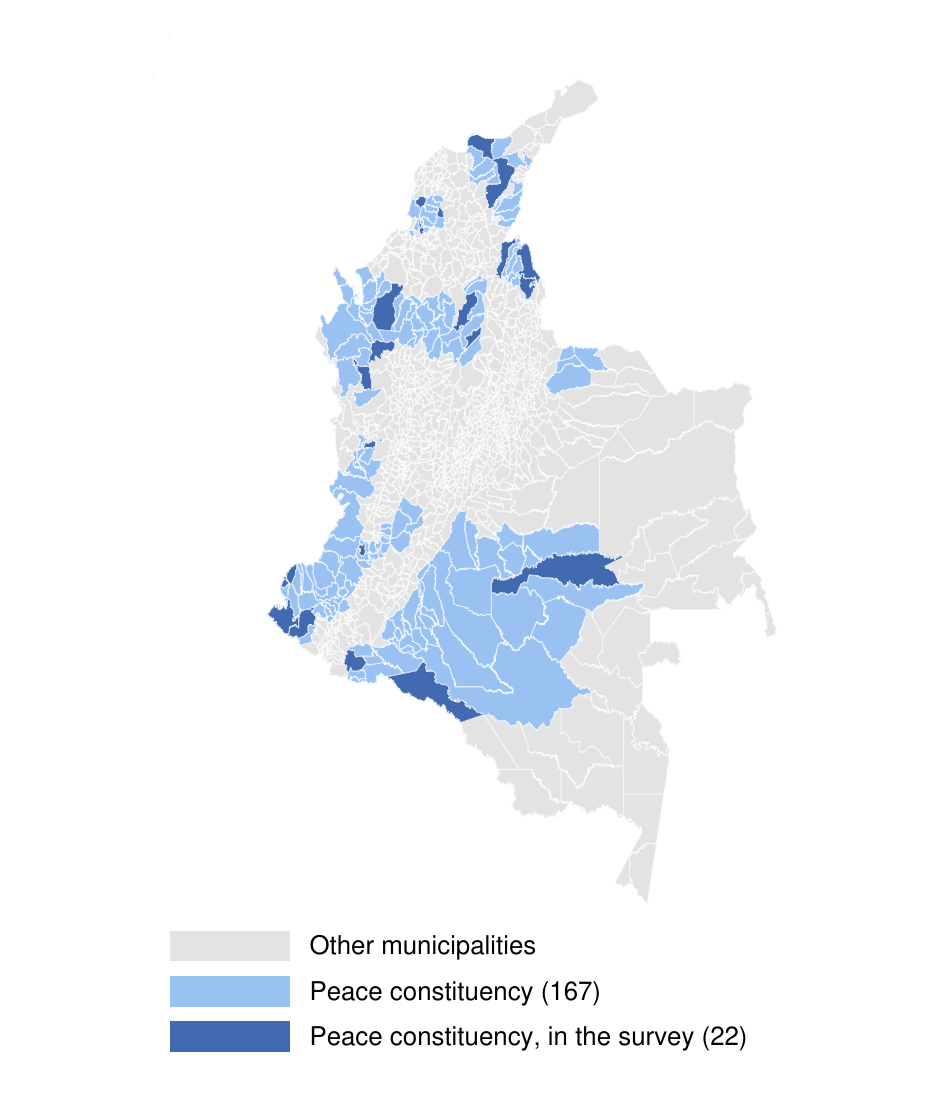}
\begin{minipage}{0.9\textwidth}\vspace{0.15cm}
\tablenotefont\textit{Note:} The 167 municipalities assigned to one of the sixteen peace constituencies by transitory Article 2 of Legislative Act 02 of 2021, and the 22 of them the Social Capital Barometer samples in both 2017 and 2023. Coverage is concentrated in the Pacific, the Caribbean periphery, Catatumbo and the Amazon--Orinoco piedmont; the survey reaches the first three more densely than the fourth, limiting extrapolation across the full reform territory. Boundaries from geoBoundaries \citep[COL ADM2, CC BY 4.0;][]{runfola2020}.
\end{minipage}\end{figure}

\clearpage
\section{Outcome Construction and Descriptive Levels}\label{app:outcomes}

Availability and harmonization determine which versions enter the comparable families; content determines assignment. The primary block covers orientations toward shared problems, civic involvement and political attachment (Section~\ref{sec:predictions}), not social capital as a whole. Ordinary voting, interpersonal support and general institutional evaluations address wider responses. This classification is overlapping, not unique.

Political attention and education capture engagement with public affairs rather than completed acts; openness to airing conflicts is an orientation toward coordination, not observed cooperation. Party trust enters alongside membership and activity as political attachment, not civic action or autonomy from brokers. Broader affiliation need not entail public involvement. Formal collective action adds membership in organizations supporting respondents' interests and petition dispositions to general affiliation; petitions also enter civic participation, so adding this family broadens coverage while reweighting shared content.

The 54-measure catalogue contains selected composites, harmonized replacements, individual items and diagnostic codings, not all 44 producer composites plus ten items. Tables~\ref{tab:dictionary} and~\ref{tab:additionalmeasures} list the 35 family measures and 19 others with their uses. All 54 are estimated in Table~\ref{tab:speccurve}; the five diagnostic versions with changing scales or mention formats are not comparable treatment outcomes.

Components are standardized on rural comparison respondents in 2011, averaged equally within families, and re-standardized on the same group. The primary index repeats these steps over the three families, weighting families equally. Reference moments stay fixed across waves and specifications; the 2017 event-study reference determines the temporal contrast, not outcome scaling. Complementary indices do not enlarge the nine-family multiple-testing and selection set.

The averaging routine uses available components and returns missing if none are observed. 
All 35 constructed components entering the nine families are observed for all 12,466 respondents in the 2005--2023 estimation sample. Every family and the primary index therefore uses its full component set for every respondent; the available-component averaging rule does not change their weights across respondents.

 Complete constructed measures do not imply complete raw responses. The rebuilt party sum assigns zero when all three underlying items are missing; opportunism treats a missing base component as zero. The party composites overlap in membership, trust and activity. Table~\ref{tab:party_once} standardizes and averages each underlying item once, with a common complete-item sample check separating weighting from nonresponse.

\clearpage

\begin{table}[H]{\renewcommand{\arraystretch}{1.08}\setlength{\tabcolsep}{3pt}
\captionsetup{font={normalsize,md}}
\caption{Descriptive Outcome Levels by Coverage and Residence}\label{tab:outcomedescriptives}\vspace{-0.3cm}
\begin{tablecontent}
\begin{tabular}{L{5.0cm}*{4}{C{2.50cm}}}

\hline\hline \addlinespace[0.15cm]
 & (1) & (2) & (3) & (4) \\
Outcome & Covered rural & Covered urban & Comparison rural & Comparison urban \\
\addlinespace[0.15cm]\hline\addlinespace[0.15cm]
\multicolumn{5}{l}{\emph{2011}} \\
Primary engagement index & -0.187 (0.876) & 0.034 (0.986) & 0.000 (1.000) & -0.015 (0.962) \\
Horizontal coordination & 0.040 (1.002) & 0.138 (0.938) & 0.000 (1.000) & 0.009 (0.941) \\
Civic participation & -0.196 (0.950) & 0.048 (1.110) & 0.000 (1.000) & 0.091 (1.005) \\
Party linkage & -0.236 (0.773) & -0.116 (0.997) & 0.000 (1.000) & -0.131 (0.989) \\
Formal collective action & -0.160 (0.735) & 0.001 (0.984) & 0.000 (1.000) & 0.130 (1.134) \\
Organizational membership & -0.094 (0.578) & 0.011 (0.874) & 0.000 (1.000) & 0.036 (1.109) \\
Electoral participation & -0.069 (1.072) & 0.153 (1.040) & 0.000 (1.000) & -0.141 (1.003) \\
Brokerage and clientelism & -0.159 (0.970) & 0.040 (1.053) & 0.000 (1.000) & 0.027 (1.001) \\
Diffuse social relations & 0.234 (0.880) & 0.203 (0.866) & 0.000 (1.000) & -0.200 (0.959) \\
Institutional trust & 0.035 (0.904) & -0.077 (0.905) & 0.000 (1.000) & -0.060 (0.858) \\
Observations, each outcome & 356 & 396 & 357 & 1,919 \\
\addlinespace[0.15cm]
\multicolumn{5}{l}{\emph{2017}} \\
Primary engagement index & -0.329 (0.634) & -0.379 (0.595) & -0.323 (0.764) & -0.252 (0.785) \\
Horizontal coordination & -0.225 (0.378) & -0.249 (0.327) & -0.205 (0.492) & -0.185 (0.743) \\
Civic participation & 0.106 (0.912) & 0.069 (0.851) & 0.054 (0.985) & 0.049 (0.909) \\
Party linkage & -0.570 (0.445) & -0.615 (0.412) & -0.526 (0.629) & -0.390 (0.738) \\
Formal collective action & 0.252 (1.091) & 0.229 (1.035) & 0.195 (1.143) & 0.234 (1.260) \\
Organizational membership & 0.297 (1.320) & 0.239 (1.255) & 0.197 (1.281) & 0.218 (1.393) \\
Electoral participation & -0.696 (0.580) & -0.659 (0.596) & -0.605 (0.694) & -0.440 (0.772) \\
Brokerage and clientelism & 0.062 (1.080) & 0.141 (1.126) & 0.009 (1.051) & -0.029 (1.113) \\
Diffuse social relations & 0.441 (0.769) & 0.494 (0.755) & 0.376 (0.864) & 0.125 (0.942) \\
Institutional trust & -0.766 (0.733) & -0.782 (0.700) & -0.701 (0.774) & -0.595 (0.835) \\
Observations, each outcome & 397 & 455 & 390 & 1,946 \\
\addlinespace[0.15cm]
\hline\hline

\end{tabular}\end{tablecontent}\tablenotesgap
\centering\begin{minipage}{\exhibittablewidth}\tablenotefont\textbf{Notes:} Unweighted respondent means with standard deviations in parentheses. Each observation count applies to all ten outcomes in its column and year. Outcomes are standardized using the 2011 rural comparison distribution, so that column has mean zero and standard deviation one by construction. The 2017 values retain the same scale. These are full-wave respondent samples, not the fixed-municipality, equally weighted sample of Figure~\ref{fig:cells}; differences may reflect sample composition and are not treatment effects. Urban denotes residence in the municipality's urban center (\emph{cabecera}). The 2017 release also contains repeated profiles documented in Appendix~\ref{app:sourceaudit}; low levels should not be taken as evidence of a genuine decline. It is the event-study reference, but its fieldwork month is unavailable, so it cannot be certified as preceding every program activity.\end{minipage}\par
}\end{table}

\begin{table}[H]{\renewcommand{\arraystretch}{0.92}\setlength{\tabcolsep}{2pt}
\captionsetup{font={normalsize,md}}
\caption{Outcome Dictionary: Family Components}\label{tab:dictionary}\vspace{-0.3cm}
\begin{tablecontent}\begin{adjustbox}{max width=0.98\linewidth, max totalheight=0.82\textheight, keepaspectratio}
\begin{tabular}{L{4.3cm}L{8.1cm}L{2.3cm}}

\hline\hline \addlinespace[0.15cm]
 & (1) & (2) \\
Family and components & What the respondent was asked & Rebuilt \\
\addlinespace[0.15cm]\hline\addlinespace[0.15cm]
\emph{Horizontal coordination} & & \\
~~clhoriz\_f & Mentions joint action by everyone as a way a collective problem gets solved & any mention \\
~~conflic & Agrees that conflicts are easier to resolve once brought into the open & rebuilt from items \\
\emph{Civic participation} & & \\
~~actvciv & Petition-signing: has done, might do, or would never do & rebuilt from items \\
~~actvloc & Attended a community or school meeting; served on a local committee & rebuilt from items \\
~~actvmed & Called a radio or television programme; wrote to a newspaper & rebuilt from items \\
~~travolum & Does unpaid voluntary work & rebuilt from items \\
~~politciud & Politicisation: politics as part of everyday responsibility & no \\
~~edupolit & Receives and processes public and political information & no \\
\emph{Party linkage} & & \\
~~ptido\_c & Sum of party membership, trust in parties and reported party activity & common items \\
~~hmeptid & Party membership and trust in parties & rebuilt from items \\
~~actptid & Activities carried out within a party & no \\
\emph{Formal collective action} & & \\
~~ovgptido, ovreli, ovgremi, ovsindi & Member of a party, religious, guild or union organisation & no \\
~~oripert & Belongs to an organisation that supports one's interests & no \\
~~actvciv & Petition-signing disposition (also in civic participation) & rebuilt from items \\
\emph{Organizational membership} & The four membership items above, without petition or interest organisation & no \\
\emph{Electoral participation} & & \\
~~vutotal & Voted in mayoral, gubernatorial, presidential, council, assembly, Chamber and Senate races & no \\
~~actpol & Political discussion, demonstrations and seeking or holding elected office & common items \\
~~eslabon, ejeslab & Linkage to legislators and to the executive & no \\
\emph{Brokerage and clientelism} & & \\
~~oportun\_f & Opportunism, and mentioning a broker as a route to a collective problem & any mention \\
~~cljquia\_f & Mentions a representative as a route to a collective problem & any mention \\
~~particul, client & Particularism; clientelist exchange in political decisions & no \\
\emph{Diffuse social relations} & & \\
~~solidar, recipro, ayhortz & Solidarity, reciprocity, horizontal help among neighbours & no \\
~~actraba, activecin & Relations at work and with neighbours & no \\
\emph{Institutional trust} & & \\
~~cfvpolt, cfvlega, cfvcentre & Trust in parties, legal and control institutions & fixed content \\
~~cfvgrup, cfvmovi, cfvmedi & Trust in interest groups, movements, and press/television & fixed media content \\
\addlinespace[0.15cm]\hline\hline

\end{tabular}\end{adjustbox}\end{tablecontent}\tablenotesgap
\centering\begin{minipage}{\exhibittablewidth}\tablenotefont\textbf{Notes:} The nine families use 35 distinct measures. Organizational membership is nested in formal collective action, and the petition item also enters civic participation; the families are not disjoint sets. The party components also overlap in their underlying membership, trust and activity content. ``Any mention'' indicates a collective-problems indicator harmonized across available mention fields; ``common items'' indicates an index restricted to items fielded throughout. Appendix~\ref{app:data} documents harmonization. Table~\ref{tab:additionalmeasures} completes the dictionary with the 19 measures outside these families, including their uses and comparability limitations. Appendix~\ref{app:outcomes} explains inclusion and boundary choices, aggregation and component availability.\end{minipage}\par
}\end{table}

\begin{table}[H]{\renewcommand{\arraystretch}{1.05}\setlength{\tabcolsep}{3pt}
\captionsetup{font={normalsize,md}}
\caption{Outcome Dictionary: Additional Measures}\label{tab:additionalmeasures}\vspace{-0.3cm}
\begin{tablecontent}
\begin{tabular}{L{2.3cm}L{5.1cm}L{7.7cm}}

\hline\hline \addlinespace[0.15cm]
 & (1) & (2) \\
Measure & Question content & Use and reason for exclusion from the nine families \\
\addlinespace[0.15cm]\hline\addlinespace[0.15cm]
\texttt{actedu} & Attendance at a community meeting & Complementary index; individual item underlying local participation. \\
\texttt{actlocal} & Service on a local committee & Complementary index; individual item underlying local participation. \\
\texttt{travolu} & Unpaid voluntary work & Complementary index; item counterpart of the volunteering composite. \\
\texttt{apetici} & Has signed, might sign or would never sign a petition & Complementary index; item counterpart of the petition composite. \\
\texttt{ovagrar} & Peasant, farmer and rancher associations & Nominating-type and complementary indices; outside the four membership categories in the family. \\
\texttt{ovmujer} & Women's organizations & Nominating-type index; outside the family membership categories. \\
\texttt{ovetnia} & Ethnic authorities and councils & Nominating-type index; outside the family membership categories. \\
\texttt{ovdepor} & Sports organizations & Other-secular-type index; outside the family membership categories. \\
\texttt{ovcultu} & Cultural organizations & Other-secular-type index; outside the family membership categories. \\
\texttt{oveduca} & Educational organizations & Other-secular-type index; outside the family membership categories. \\
\texttt{ovcoop} & Cooperatives & Other-secular-type and complementary indices; outside the family membership categories. \\
\texttt{ovcari} & Charitable organizations & Other-secular-type index; outside the family membership categories. \\
\texttt{ovcomba} & Communal action boards & Other-secular-type and complementary indices; outside the family membership categories. \\
\texttt{clpalan\_f} & Mentions a \emph{palanca} as a route to solving a collective problem & Separate any-mention outcome; also enters opportunism, so it is not added again to that family. \\
\texttt{mecatot} & Participation mechanisms & Diagnostic only: scale changes across waves; omitted from comparable civic participation. \\
\texttt{habpol} & Political skills & Diagnostic only: scale changes across waves; omitted from comparable civic participation. \\
\texttt{clhoriz\_c} & Joint problem solving, legacy mixed-order coding & Diagnostic only: combines any mention in early waves with first mention later; the family uses either mention throughout. \\
\texttt{oportun\_c} & Opportunism, legacy mixed-order coding & Diagnostic only: the broker component mixes mention orders; the family uses either mention throughout. \\
\texttt{cljquia} & Vertical problem solving, distributed version & Diagnostic only: mention format changes; the family uses either mention throughout. \\
\addlinespace[0.15cm]\hline\hline

\end{tabular}\end{tablecontent}\tablenotesgap
\centering\begin{minipage}{\exhibittablewidth}\tablenotefont\textbf{Notes:} These 19 measures are outside the nine outcome-family aggregates but remain in the 54-measure catalogue. The first four participation items and memberships in peasant associations, cooperatives and communal action boards form the complementary index. Nine membership items form the nominating-type and other-secular-type indices; they are not subdivisions of the organizational-membership family in Table~\ref{tab:dictionary}. The five diagnostic versions are retained for transparency about measurement choices, not as comparable outcomes supporting causal interpretation. The completed-petition recoding used in Table~\ref{tab:behavioral} is a sensitivity transformation of \texttt{apetici}, not an additional member of the 54-measure catalogue.\end{minipage}\par
}\end{table}

\clearpage
\section{Event-Study Estimates}\label{app:eventstudy}

\begin{figure}[H]\centering
\captionsetup{font={normalsize,md}}
\caption{The Rural--Urban Gap, Before and After the Reform}\label{fig:eventstudy}
\includegraphics[width=0.95\textwidth]{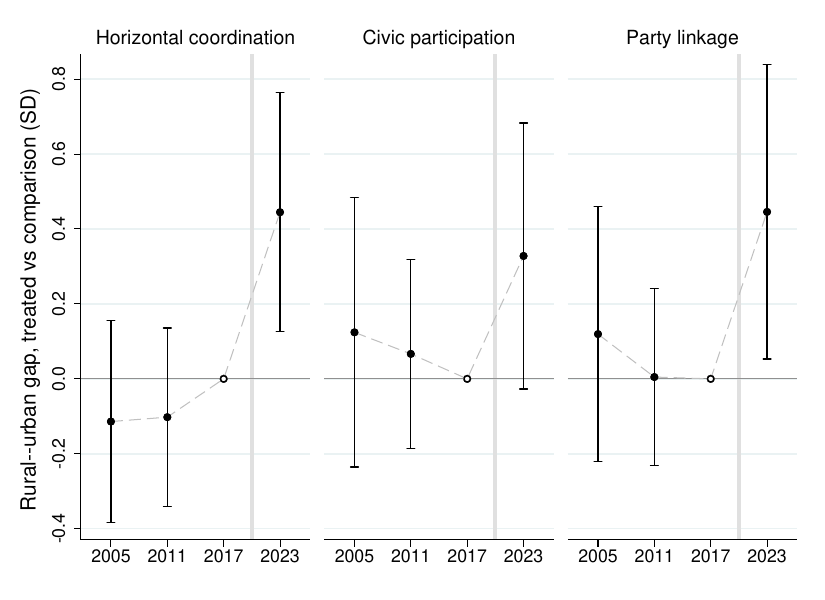}
\begin{minipage}{0.92\textwidth}\vspace{0.15cm}
\tablenotefont\textit{Note:} Coefficients $\pi_\tau$ from equation~\eqref{eq:esddd}, normalized on 2017 (hollow marker), for the three constructs both mechanisms implicate. Each point measures the change since 2017 in the rural--urban gap in covered municipalities relative to the corresponding change in comparison municipalities. Municipality-by-wave effects absorb common municipal shocks, and municipality-by-residence effects absorb persistent rural--urban differences. Bars are 95\% confidence intervals from standard errors clustered by municipality. The 1997 wave is excluded from the estimation sample; Table~\ref{tab:anchors} reports the estimates including it.
\end{minipage}\end{figure}

\clearpage
\section{Additional Tables}\label{app:tables}

\begin{table}[H]{\renewcommand{\arraystretch}{0.95}\setlength{\tabcolsep}{2pt}
\captionsetup{font={normalsize,md}}
\caption{Counting Each Underlying Party Item Once}\label{tab:party_once}\vspace{-0.3cm}
\begin{tablecontent}\begin{adjustbox}{max width=0.98\linewidth, max totalheight=0.78\textheight}
\begin{tabular}{L{6.0cm}*{3}{C{2.8cm}}}

\hline\hline \addlinespace[0.15cm]
 & (1) & (2) & (3) \\
 & Main construction & Each party item once & Observations \\
\addlinespace[0.10cm]\hline\addlinespace[0.15cm]
\multicolumn{4}{l}{\emph{2017 reference, all available observations}} \\
Primary index & 0.582*** & 0.590*** & 12,466 \\
 & (0.198) & (0.207) &  \\
\multicolumn{4}{l}{\emph{2017 reference, complete party-item sample}} \\
Primary index & 0.575*** & 0.587*** & 12,356 \\
 & (0.199) & (0.209) &  \\
\multicolumn{4}{l}{\emph{2011 reference, excluding 2017}} \\
Primary index & 0.625*** & 0.624*** & 9,278 \\
 & (0.227) & (0.233) &  \\
\emph{Underlying items, excluding 2017} & Coefficient & SE & Observations \\
Party membership (2011 reference) & 0.274 & (0.271) & 9,278 \\
Trust in parties (2011 reference) & 0.327* & (0.185) & 9,168 \\
Party activity (2011 reference) & 0.308* & (0.175) & 9,278 \\
\addlinespace[0.15cm]\hline\hline

\end{tabular}\end{adjustbox}\end{tablecontent}\tablenotesgap
\centering\begin{minipage}{\exhibittablewidth}\tablenotefont\textbf{Notes:} The first three panels compare the main primary index with one replacing its party family by an equal-weight standardized average of raw party membership, trust and activity. Both indices use the same observations within a panel. The complete-item panel requires all three party items. The final panel estimates the raw items separately, each standardized, using 2005, 2011 and 2023; its columns are coefficient, standard error and observations. All regressions include municipality-by-wave and municipality-by-residence fixed effects and respondent sex. Outcomes retain their 2011 rural-comparison standardization. Standard errors are clustered by municipality. *, **, *** denote conventional significance at 10\%, 5\% and 1\%; these tests do not jointly adjust for outcome selection.\end{minipage}\par}\end{table}

\clearpage

\begin{landscape}
\begin{table}[H]{\renewcommand{\arraystretch}{0.95}\setlength{\tabcolsep}{2pt}
\captionsetup{font={normalsize,md}}
\caption{Components of the Three Constructs}\label{tab:components}\vspace{-0.3cm}
\begin{tablecontent}\begin{adjustbox}{max width=0.98\linewidth, max totalheight=0.86\textheight}
\begin{tabular}{L{3.8cm}*{11}{C{1.30cm}}}

\hline\hline \addlinespace[0.15cm]
 & \multicolumn{2}{c}{Coordination} & \multicolumn{6}{c}{Civic participation} & \multicolumn{3}{c}{Party linkage} \\
\cmidrule(lr){2-3}\cmidrule(lr){4-9}\cmidrule(lr){10-12}
 & (1) & (2) & (3) & (4) & (5) & (6) & (7) & (8) & (9) & (10) & (11) \\
 & Peer resolution mentioned & Resolvable when aired & Civic activity & Local activity & Media activity & Volunteer work & Citizen politics & Political education & Membership, trust and activity & Membership and trust & Party activity \\
\addlinespace[0.10cm]\hline\addlinespace[0.15cm]
Rural $\times$ Covered $\times$ 2023 & 0.322** & 0.368** & 0.225 & 0.220 & 0.319 & 0.101 & 0.062 & 0.122 & 0.446** & 0.353* & 0.326** \\
 & (0.154) & (0.149) & (0.148) & (0.182) & (0.224) & (0.174) & (0.160) & (0.105) & (0.198) & (0.187) & (0.147) \\
 & [0.014, 0.630] & [0.070, 0.666] & [-0.070, 0.519] & [-0.144, 0.583] & [-0.128, 0.765] & [-0.247, 0.449] & [-0.258, 0.383] & [-0.087, 0.332] & [0.050, 0.841] & [-0.021, 0.726] & [0.033, 0.619] \\
Rural $\times$ Covered $\times$ 2011 & -0.011 & -0.148 & 0.055 & 0.070 & 0.064 & -0.219 & 0.014 & 0.229* & 0.002 & -0.036 & 0.046 \\
 & (0.129) & (0.126) & (0.110) & (0.118) & (0.197) & (0.149) & (0.154) & (0.136) & (0.118) & (0.107) & (0.121) \\
Rural $\times$ Covered $\times$ 2005 & -0.121 & -0.056 & 0.132 & 0.084 & -0.082 & 0.279 & 0.025 & -0.042 & 0.128 & 0.203 & -0.030 \\
 & (0.128) & (0.139) & (0.134) & (0.163) & (0.206) & (0.240) & (0.175) & (0.142) & (0.171) & (0.165) & (0.157) \\
Pre-trend $p$ & 0.642 & 0.471 & 0.596 & 0.796 & 0.835 & 0.051 & 0.986 & 0.220 & 0.745 & 0.427 & 0.881 \\
R-squared & 0.231 & 0.140 & 0.123 & 0.171 & 0.057 & 0.138 & 0.114 & 0.177 & 0.130 & 0.117 & 0.099 \\
Observations & 12,466 & 12,466 & 12,466 & 12,466 & 12,466 & 12,466 & 12,466 & 12,466 & 12,466 & 12,466 & 12,466 \\
\addlinespace[0.15cm]\hline\hline

\end{tabular}\end{adjustbox}\end{tablecontent}\tablenotesgap
\centering\begin{minipage}{\exhibittablewidth}\tablenotefont\textbf{Notes:} Full regressions underlying the compact component summary in Table~\ref{tab:components_main} of the paper. Estimates of equation~\eqref{eq:esddd} under the baseline specification on each standardized component of the three constructs. Columns (1) and (2): whether the respondent mentions joint resolution among ways collective problems are addressed (orientation) and agrees that conflicts are easier to resolve once aired (attitudinal). Columns (3) to (8): civic, local and media activity, volunteer work, citizen politics and political education. Columns (9) to (11): a membership--trust--activity composite, a membership--trust composite, and reported party activity; their content overlaps. Standard errors clustered by municipality in parentheses and 95\% confidence intervals in brackets. Outcomes are standardized on the 2011 rural comparison distribution. $X_i$ contains respondent sex. Pre-trend $p$ tests $\pi_{2005}=\pi_{2011}=0$. *, **, *** denote significance at 10\%, 5\% and 1\%.\end{minipage}\par}\end{table}
\end{landscape}

\clearpage

\begin{table}[H]{\renewcommand{\arraystretch}{0.95}\setlength{\tabcolsep}{2pt}
\captionsetup{font={normalsize,md}}
\caption{Reported Memberships and Completed Activities}\label{tab:behavioral}\vspace{-0.3cm}
\begin{tablecontent}\begin{adjustbox}{max width=0.98\linewidth, max totalheight=0.86\textheight}
\begin{tabular}{L{4.6cm}*{4}{C{2.48cm}}}

\hline\hline \addlinespace[0.15cm]
 & (1) & (2) & (3) & (4) \\
 & Original index & Excluding petitions & Completed acts and memberships & Completed acts only \\
\addlinespace[0.10cm]\hline\addlinespace[0.15cm]
Rural $\times$ Covered $\times$ 2023 & 0.341 & 0.307 & 0.337 & 0.281 \\
 & (0.238) & (0.245) & (0.241) & (0.191) \\
 & [-0.135, 0.817] & [-0.182, 0.795] & [-0.145, 0.819] & [-0.100, 0.662] \\
Rural $\times$ Covered $\times$ 2011 & -0.050 & -0.069 & -0.034 & -0.002 \\
 & (0.104) & (0.103) & (0.108) & (0.126) \\
Rural $\times$ Covered $\times$ 2005 & 0.449* & 0.449 & 0.464* & 0.250 \\
 & (0.255) & (0.269) & (0.254) & (0.175) \\
Pre-trend $p$ & 0.148 & 0.136 & 0.152 & 0.306 \\
R-squared & 0.198 & 0.208 & 0.201 & 0.175 \\
Observations & 12,466 & 12,466 & 12,466 & 12,466 \\
\addlinespace[0.10cm]\hline\addlinespace[0.15cm]
Organization-membership items & 3 & 3 & 3 & 0 \\
Completed-activity items & 3 & 3 & 4 & 4 \\
Petition willingness included & Y & N & N & N \\
Municipality $\times$ wave FE & Y & Y & Y & Y \\
Municipality $\times$ residence FE & Y & Y & Y & Y \\
Sex controls & Y & Y & Y & Y \\
\addlinespace[0.15cm]\hline\hline

\end{tabular}\end{adjustbox}\end{tablecontent}\tablenotesgap
\centering\begin{minipage}{\exhibittablewidth}\tablenotefont\textbf{Notes:} Estimates of equation~\eqref{eq:esddd}, with municipality-by-wave and municipality-by-residence fixed effects and sex controls. Column (1) retains the original seven-item index: memberships in communal action boards, peasant associations and cooperatives; attendance at community meetings; service on local committees; volunteering; and the three-valued petition item. Column (2) omits petitions. Column (3) keeps seven items but counts only having signed a petition, not willingness. Column (4) includes only meetings, committees, volunteering and having signed. All components are self-reports. Each component and the resulting average are standardized on the 2011 rural comparison distribution. Standard errors clustered by municipality in parentheses and 95\% intervals in brackets. Pre-trend $p$ tests the 2005 and 2011 coefficients jointly. Conventional inference is pointwise, not adjusted for selection or multiple comparisons. *, **, *** denote 10\%, 5\% and 1\%.\end{minipage}\par}\end{table}

\begin{table}[H]{\renewcommand{\arraystretch}{0.95}\setlength{\tabcolsep}{2pt}
\captionsetup{font={normalsize,md}}
\caption{Confidence Intervals and Illustrative Effect Sizes}\label{tab:intervals}\vspace{-0.3cm}
\begin{tablecontent}\begin{adjustbox}{max width=0.98\linewidth, max totalheight=0.86\textheight}
\begin{tabular}{L{4.1cm}*{9}{C{1.25cm}}}

\hline\hline \addlinespace[0.15cm]
 & \multicolumn{3}{c}{Constructs} & \multicolumn{2}{c}{Other collective} & \multicolumn{4}{c}{Contrast domains} \\
\cmidrule(lr){2-4}\cmidrule(lr){5-6}\cmidrule(lr){7-10}
 & (1) & (2) & (3) & (4) & (5) & (6) & (7) & (8) & (9) \\
 & Coordination & Civic particip. & Party linkage & Formal coll. action & Org. membership & Electoral particip. & Brokerage & Diffuse relations & Inst. trust \\
\addlinespace[0.10cm]\hline\addlinespace[0.15cm]
Rural $\times$ Covered $\times$ 2023 & 0.445*** & 0.328* & 0.446** & 0.422 & 0.294 & 0.137 & -0.067 & -0.008 & 0.145 \\
 & (0.160) & (0.178) & (0.197) & (0.292) & (0.318) & (0.111) & (0.138) & (0.153) & (0.140) \\
 & [0.126, 0.764] & [-0.027, 0.683] & [0.053, 0.838] & [-0.163, 1.006] & [-0.340, 0.929] & [-0.085, 0.359] & [-0.343, 0.209] & [-0.313, 0.297] & [-0.135, 0.424] \\
Minimum detectable effect & 0.45 & 0.50 & 0.55 & 0.82 & 0.89 & 0.31 & 0.39 & 0.43 & 0.39 \\
Interval excludes $+0.33$ & No & No & No & No & No & No & Yes & Yes & No \\
Interval excludes $-0.33$ & Yes & Yes & Yes & Yes & No & Yes & No & Yes & Yes \\
\addlinespace[0.15cm]\hline\hline

\end{tabular}\end{adjustbox}\end{tablecontent}\tablenotesgap
\centering\begin{minipage}{\exhibittablewidth}\tablenotefont\textbf{Notes:} Estimates of equation~\eqref{eq:esddd} under the baseline specification, as in Tables~\ref{tab:baseline} and \ref{tab:families}. Standard errors clustered by municipality in parentheses and 95\% confidence intervals in brackets. The minimum detectable effect is 2.80 times the standard error (80\% power, 5\% size). The last two rows compare each interval with the illustrative benchmark of $\pm0.33$ standard deviations, which is not a prespecified smallest effect of interest. *, **, *** denote significance at 10\%, 5\% and 1\%.\end{minipage}\par}\end{table}

\begin{table}[H]{\renewcommand{\arraystretch}{0.95}\setlength{\tabcolsep}{2pt}
\captionsetup{font={normalsize,md}}
\caption{Outcomes with a 2011 Reference, Excluding 2017}\label{tab:clean2011}\vspace{-0.3cm}
\begin{tablecontent}\begin{adjustbox}{max width=0.98\linewidth, max totalheight=0.78\textheight}
\begin{tabular}{L{5.3cm}*{6}{C{1.55cm}}}

\hline\hline \addlinespace[0.15cm]
 & (1) & (2) & (3) & (4) & (5) & (6) \\
 & $\pi_{2023}$ & SE & $p$ & Wild $p$ & $\pi_{2005}$ & Pre-period $p$ \\
\addlinespace[0.10cm]\hline\addlinespace[0.15cm]
Primary index & 0.625*** & 0.227 & 0.008 & 0.006 & 0.081 & 0.686 \\
Horizontal coordination & 0.582*** & 0.188 & 0.003 & 0.002 & 0.001 & 0.994 \\
Civic participation & 0.253 & 0.192 & 0.193 & 0.199 & 0.055 & 0.764 \\
Party linkage & 0.474** & 0.230 & 0.044 & 0.046 & 0.114 & 0.543 \\
Formal collective action & 0.235 & 0.274 & 0.393 & --- & 0.295 & 0.091 \\
Organizational membership & 0.173 & 0.276 & 0.533 & --- & 0.261 & 0.108 \\
Electoral participation & 0.190 & 0.176 & 0.285 & --- & 0.025 & 0.907 \\
Brokerage and clientelism & -0.021 & 0.163 & 0.898 & --- & -0.065 & 0.744 \\
Diffuse social relations & -0.037 & 0.163 & 0.819 & --- & 0.071 & 0.684 \\
Institutional trust & 0.189 & 0.178 & 0.291 & --- & 0.064 & 0.710 \\
Membership, types the statute names & 0.498* & 0.295 & 0.096 & --- & 0.353 & 0.156 \\
Other secular types & 0.155 & 0.283 & 0.586 & --- & 0.214 & 0.320 \\
Named minus other types & 0.343 & 0.263 & 0.196 & --- & 0.139 & 0.436 \\
Reported party activity & 0.308* & 0.175 & 0.082 & --- & -0.072 & 0.669 \\
Mentions a personal connection & -0.048 & 0.136 & 0.727 & --- & 0.223 & 0.168 \\
\addlinespace[0.15cm]\hline\hline

\end{tabular}\end{adjustbox}\end{tablecontent}\tablenotesgap
\centering\begin{minipage}{\exhibittablewidth}\tablenotefont\textbf{Notes:} Estimates use only 2005, 2011 and 2023. Wild $p$ uses 9,999 municipality-cluster bootstrap replications, seed 20261006, for the primary index and its three constituents; a dash denotes a test not computed. The pre-period $p$ tests the 2005 coefficient relative to 2011. Named minus other types is the difference between their separately standardized membership indices. All regressions include municipality-by-wave and municipality-by-residence fixed effects and respondent sex. Outcomes retain their 2011 rural-comparison standardization. Standard errors are clustered by municipality. *, **, *** denote conventional significance at 10\%, 5\% and 1\%; these tests do not jointly adjust for outcome selection.\end{minipage}\par}\end{table}

\clearpage

{\tablebodyfont\setlength{\tabcolsep}{2pt}\renewcommand{\arraystretch}{1.08}
\setlength{\exhibittablewidth}{\dimexpr15cm+4pt\relax}
\setlength{\LTpost}{3pt}
\captionsetup{font={normalsize,md}}
\begin{longtable}{L{3.0cm}L{4.0cm}*{4}{C{1.5cm}}}
\caption{Individual Estimates for the 54 Measures}\label{tab:speccurve}\\
\endfirsthead
\multicolumn{6}{l}{\textit{Table \thetable\ continued}}\\
Measure & Outcome family & $\pi_{2023}$ & SE & $p$ & Pre-trend $p$\\
\hline
\endhead

\hline\hline \addlinespace[0.15cm]
 & (1) & (2) & (3) & (4) & (5) \\
Measure & Outcome family & $\pi_{2023}$ & SE & $p$ & Pre-trend $p$ \\
\addlinespace[0.15cm]\hline\addlinespace[0.15cm]
\texttt{z\_actedu} & \emph{(no family)} &  0.108 & (0.159) & 0.499 & 0.582 \\
\texttt{z\_actlocal} & \emph{(no family)} &  0.297 & (0.193) & 0.130 & 0.520 \\
\texttt{z\_travolu} & \emph{(no family)} &  0.101 & (0.174) & 0.563 & 0.051 \\
\texttt{z\_apetici} & \emph{(no family)} &  0.225 & (0.148) & 0.133 & 0.596 \\
\texttt{z\_clhoriz\_f} & Horizontal coordination &  0.322** & (0.154) & 0.041 & 0.642 \\
\texttt{z\_cljquia\_f} & Brokerage and clientelism &  0.080 & (0.133) & 0.552 & 0.606 \\
\texttt{z\_clpalan\_f} & \emph{(no family)} & -0.190 & (0.132) & 0.155 & 0.415 \\
\texttt{z\_oportun\_f} & Brokerage and clientelism & -0.044 & (0.124) & 0.722 & 0.735 \\
\texttt{z\_ovgptido} & Formal collective action; Organizational membership &  0.358 & (0.265) & 0.182 & 0.671 \\
\texttt{z\_ovreli} & Formal collective action; Organizational membership &  0.256 & (0.184) & 0.169 & 0.013 \\
\texttt{z\_ovgremi} & Formal collective action; Organizational membership &  0.173 & (0.261) & 0.511 & 0.672 \\
\texttt{z\_ovsindi} & Formal collective action; Organizational membership & -0.013 & (0.309) & 0.968 & 0.384 \\
\texttt{z\_oripert} & Formal collective action &  0.286* & (0.147) & 0.056 & 0.049 \\
\texttt{z\_actvciv} & Civic participation; Formal collective action &  0.225 & (0.148) & 0.133 & 0.596 \\
\texttt{z\_ptido\_c} & Party linkage &  0.446** & (0.198) & 0.028 & 0.745 \\
\texttt{z\_hmeptid} & Party linkage &  0.353* & (0.187) & 0.064 & 0.427 \\
\texttt{z\_actptid} & Party linkage &  0.326** & (0.147) & 0.030 & 0.881 \\
\texttt{z\_actvloc} & Civic participation &  0.220 & (0.182) & 0.232 & 0.796 \\
\texttt{z\_actvmed} & Civic participation &  0.319 & (0.224) & 0.159 & 0.835 \\
\texttt{z\_mecatot}$^{\dagger}$ & \emph{(no family)} &  0.393 & (0.319) & 0.223 & 0.973 \\
\texttt{z\_travolum} & Civic participation &  0.101 & (0.174) & 0.563 & 0.051 \\
\texttt{z\_politciud} & Civic participation &  0.062 & (0.160) & 0.700 & 0.986 \\
\texttt{z\_habpol}$^{\dagger}$ & \emph{(no family)} &  0.220 & (0.166) & 0.189 & 0.766 \\
\texttt{z\_edupolit} & Civic participation &  0.122 & (0.105) & 0.247 & 0.220 \\
\texttt{z\_vutotal} & Electoral participation &  0.044 & (0.133) & 0.743 & 0.592 \\
\texttt{z\_actpol} & Electoral participation &  0.215 & (0.177) & 0.231 & 0.551 \\
\texttt{z\_eslabon} & Electoral participation &  0.190* & (0.106) & 0.077 & 0.466 \\
\texttt{z\_ejeslab} & Electoral participation & -0.027 & (0.121) & 0.821 & 0.176 \\
\texttt{z\_clhoriz\_c}$^{\dagger}$ & \emph{(no family)} &  0.228 & (0.146) & 0.122 & 0.501 \\
\texttt{z\_conflic} & Horizontal coordination &  0.368** & (0.149) & 0.016 & 0.471 \\
\texttt{z\_oportun\_c}$^{\dagger}$ & \emph{(no family)} & -0.077 & (0.108) & 0.482 & 0.556 \\
\texttt{z\_particul} & Brokerage and clientelism &  0.037 & (0.147) & 0.800 & 0.442 \\
\texttt{z\_client} & Brokerage and clientelism & -0.196 & (0.120) & 0.107 & 0.846 \\
\texttt{z\_cljquia}$^{\dagger}$ & \emph{(no family)} &  0.123 & (0.103) & 0.235 & 0.575 \\
\texttt{z\_solidar} & Diffuse social relations &  0.058 & (0.156) & 0.708 & 0.699 \\
\texttt{z\_ayhortz} & Diffuse social relations & -0.074 & (0.148) & 0.617 & 0.520 \\
\texttt{z\_recipro} & Diffuse social relations & -0.048 & (0.107) & 0.658 & 0.802 \\
\texttt{z\_actraba} & Diffuse social relations &  0.049 & (0.155) & 0.754 & 0.230 \\
\texttt{z\_activecin} & Diffuse social relations & -0.006 & (0.180) & 0.973 & 0.645 \\
\texttt{z\_cfvpolt} & Institutional trust &  0.147 & (0.160) & 0.364 & 0.343 \\
\texttt{z\_cfvcentre} & Institutional trust &  0.140 & (0.131) & 0.290 & 0.722 \\
\texttt{z\_cfvlega} & Institutional trust &  0.189* & (0.113) & 0.099 & 0.443 \\
\texttt{z\_cfvgrup} & Institutional trust &  0.032 & (0.128) & 0.807 & 0.890 \\
\texttt{z\_cfvmovi} & Institutional trust &  0.104 & (0.167) & 0.535 & 0.551 \\
\texttt{z\_cfvmedi} & Institutional trust &  0.081 & (0.132) & 0.540 & 0.954 \\
\texttt{z\_ovagrar} & Nominating types &  0.210 & (0.201) & 0.302 & 0.078 \\
\texttt{z\_ovmujer} & Nominating types &  0.299 & (0.218) & 0.174 & 0.438 \\
\texttt{z\_ovetnia} & Nominating types &  0.985*** & (0.370) & 0.010 & 0.863 \\
\texttt{z\_ovdepor} & Other secular types & -0.119 & (0.196) & 0.548 & 0.664 \\
\texttt{z\_ovcultu} & Other secular types &  0.117 & (0.287) & 0.685 & 0.635 \\
\texttt{z\_oveduca} & Other secular types &  0.373* & (0.217) & 0.091 & 0.186 \\
\texttt{z\_ovcoop} & Other secular types & -0.005 & (0.241) & 0.983 & 0.306 \\
\texttt{z\_ovcari} & Other secular types &  0.006 & (0.218) & 0.978 & 0.344 \\
\texttt{z\_ovcomba} & Other secular types &  0.408 & (0.329) & 0.219 & 0.497 \\
\addlinespace[0.10cm]
\multicolumn{6}{p{15.0cm}}{\scriptsize Of 54 measures, 5 have $p < 0.05$ on $\pi_{2023}$. Under the global null one would expect about 2.7 rejections at the 5\% level. Measures marked \emph{(no family)} are standardized by the build but enter none of the nine families, and may serve additional analyses or measurement diagnostics. The count includes diagnostic codings and duplicated content; it is not an omnibus test.} \\
\addlinespace[0.15cm]\hline\hline

\end{longtable}
\noindent\centering\begin{minipage}{\exhibittablewidth}\tablenotefont\textbf{Notes:} Equation~\eqref{eq:esddd} estimated separately on the 54 catalogue measures, including individual items and diagnostic codings. A dagger marks the five versions excluded from the comparable families because their scale or mention format changes across waves; their coefficients are measurement diagnostics, not comparable treatment outcomes. Tables~\ref{tab:dictionary} and~\ref{tab:additionalmeasures} document all 54 measures and their uses. ``Nominating types'' and ``Other secular types'' identify the nine membership items used to construct the two indices of Table~\ref{tab:nominating}; these items enter none of the nine outcome families and are not a partition of the organizational-membership family. Measures marked \emph{(no family)} also enter none of the nine, although some form the complementary index. The catalogue includes duplicated content and is not a set of 54 independent constructs; the rejection count includes the diagnostic versions and is purely descriptive. Standard errors clustered by municipality. *, **, *** denote 10\%, 5\% and 1\%.\end{minipage}\par
}

\clearpage

\section{Four-Group Trajectories with All Four Waves}\label{app:cellstrajectories}

\begin{figure}[H]\centering
\captionsetup{font={normalsize,md}}
\caption{Four Group Trajectories Including the 2017 Wave}\label{fig:cells_allwaves}
\includegraphics[width=0.95\textwidth]{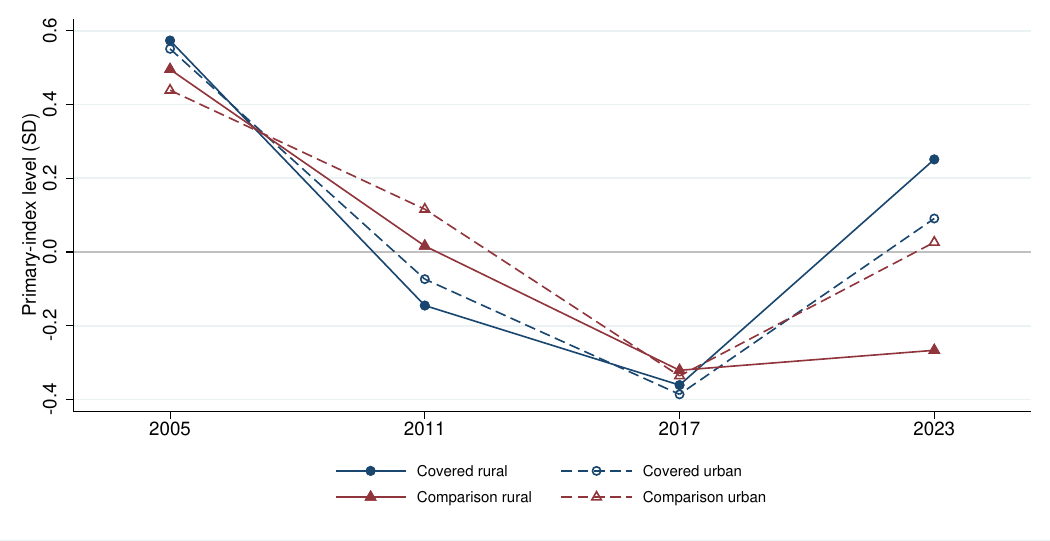}
\begin{minipage}{0.95\textwidth}\vspace{0.15cm}\tablenotefont
\textit{Note:} Means for the same 33 municipalities (17 covered, 16 comparison), each with rural and urban observations in all four waves. Municipalities have equal weight within their coverage group in every year. Scaling uses the full-sample 2011 rural comparison cell. This figure displays all four waves; Figure~\ref{fig:cells} in the paper omits only 2017, retaining identical municipalities, weights and scale. The baseline event study uses 2017 as its reference. These are descriptive levels, not separately identified treatment effects. Table~\ref{tab:cells} reports changes against 2017 and the pooled earlier waves; Table~\ref{tab:clean_cells} reports changes excluding 2017.
\end{minipage}
\end{figure}

\clearpage
\section{Further Supporting Tables}\label{app:furthertables}

\begin{table}[H]{\renewcommand{\arraystretch}{0.9}\setlength{\tabcolsep}{2pt}
\captionsetup{font={normalsize,md}}
\caption{The Four Cells: Where the Gap Comes From}\label{tab:cells}\vspace{-0.3cm}
\begin{tablecontent}\begin{adjustbox}{max width=0.98\linewidth, max totalheight=0.86\textheight}
\begin{tabular}{L{4.8cm}*{6}{C{1.65cm}}}

\hline\hline \addlinespace[0.15cm]
 & (1) & (2) & (3) & (4) & (5) & (6) \\
 & \multicolumn{2}{c}{Covered} & \multicolumn{2}{c}{Comparison} & & \\
\cmidrule(lr){2-3}\cmidrule(lr){4-5}
Outcome & Rural & Urban & Rural & Urban & DDD & Reg. \\
\addlinespace[0.15cm]\hline\addlinespace[0.15cm]
\multicolumn{7}{l}{\emph{Panel A. 2017--2023, fixed longitudinal support}} \\
\emph{Primary index} &  0.612 &  0.477 &  0.054 &  0.362 &  0.444 &  0.442** \\
\addlinespace[0.10cm]
Horizontal coordination &  0.365 &  0.222 & -0.138 &  0.197 &  0.478 &  0.486*** \\
Civic participation &  0.179 &  0.115 & -0.127 & -0.032 &  0.159 &  0.132 \\
Party linkage &  0.739 &  0.663 &  0.379 &  0.595 &  0.292 &  0.308 \\
Formal collective action & -0.058 & -0.057 & -0.134 &  0.000 &  0.133 &  0.156 \\
Organizational membership & -0.190 & -0.141 & -0.090 &  0.033 &  0.074 &  0.107 \\
Electoral participation &  0.819 &  0.706 &  0.472 &  0.523 &  0.164 &  0.130 \\
Brokerage and clientelism & -0.392 & -0.266 & -0.148 & -0.125 & -0.103 & -0.122 \\
Diffuse social relations & -0.171 & -0.202 & -0.324 & -0.410 & -0.055 & -0.078 \\
Institutional trust &  0.548 &  0.500 &  0.297 &  0.456 &  0.207 &  0.178 \\
\addlinespace[0.10cm]\multicolumn{7}{l}{\emph{Panel B. Mean of 2005, 2011 and 2017 to 2023, same support}} \\
\emph{Primary index} &  0.229 &  0.060 & -0.330 & -0.047 &  0.452 &  0.434** \\
\addlinespace[0.10cm]
Horizontal coordination &  0.113 & -0.088 & -0.375 & -0.004 &  0.572 &  0.574*** \\
Civic participation &  0.043 & -0.021 & -0.286 & -0.248 &  0.102 &  0.084 \\
Party linkage &  0.323 &  0.236 & -0.030 &  0.155 &  0.273 &  0.251 \\
Formal collective action &  0.056 &  0.062 &  0.011 &  0.054 &  0.037 &  0.067 \\
Organizational membership & -0.019 & -0.001 &  0.132 &  0.196 &  0.046 &  0.073 \\
Electoral participation &  0.312 &  0.156 &  0.003 &  0.012 &  0.164 &  0.128 \\
Brokerage and clientelism & -0.326 & -0.203 & -0.175 & -0.070 & -0.018 & -0.014 \\
Diffuse social relations & -0.158 & -0.044 & -0.158 & -0.131 & -0.088 & -0.115 \\
Institutional trust & -0.072 & -0.061 & -0.265 & -0.095 &  0.159 &  0.144 \\
\addlinespace[0.15cm]\hline\hline

\end{tabular}\end{adjustbox}\end{tablecontent}\tablenotesgap
\centering\begin{minipage}{\exhibittablewidth}\tablenotefont\textbf{Notes:} Both panels keep the same 33 municipalities (17 covered, 16 comparison) with rural and urban respondents in all four waves, totaling 5{,}032 observations. Municipalities have equal weight within each group in every year. Panel A compares 2017 and 2023; Panel B compares 2023 with the equal-wave mean of 2005, 2011 and 2017. DDD is the exact contrast of the unrounded cell changes and is descriptive. The last column is the event-study regression in Panel A and its pooled analogue in Panel B, on the same municipality support, with both sets of fixed effects and sex controls. Regression and descriptive weights differ. Stars apply only to conventional regression tests: *, **, *** denote 10\%, 5\% and 1\%.\end{minipage}\par}\end{table}

\begin{table}[H]{\renewcommand{\arraystretch}{0.95}\setlength{\tabcolsep}{2pt}
\captionsetup{font={normalsize,md}}
\caption{Four-Group Changes without the 2017 Reference}\label{tab:clean_cells}\vspace{-0.3cm}
\begin{tablecontent}\begin{adjustbox}{max width=0.98\linewidth, max totalheight=0.78\textheight}
\begin{tabular}{L{5.1cm}*{5}{C{1.9cm}}}

\hline\hline \addlinespace[0.15cm]
 & (1) & (2) & (3) & (4) & (5) \\
 & Covered rural & Covered urban & Comparison rural & Comparison urban & DDD \\
\addlinespace[0.10cm]\hline\addlinespace[0.15cm]
\emph{Panel A:} & \multicolumn{5}{c}{2011 to 2023} \\
\cmidrule(lr){2-6}
Primary index & 0.396 & 0.164 & -0.283 & -0.090 & 0.425 \\
Horizontal coordination & 0.035 & -0.141 & -0.449 & 0.042 & 0.666 \\
Civic participation & 0.421 & 0.282 & 0.000 & -0.160 & -0.020 \\
Party linkage & 0.374 & 0.204 & -0.144 & -0.070 & 0.245 \\
Formal collective action & 0.331 & 0.230 & 0.041 & 0.014 & 0.074 \\
Organizational membership & 0.218 & 0.098 & 0.171 & 0.185 & 0.134 \\
Electoral participation & 0.116 & -0.116 & -0.124 & -0.206 & 0.150 \\
Brokerage and clientelism & -0.193 & -0.147 & -0.098 & -0.117 & -0.066 \\
Diffuse social relations & 0.070 & 0.175 & 0.068 & 0.083 & -0.090 \\
Institutional trust & -0.258 & -0.285 & -0.320 & -0.221 & 0.125 \\
\emph{Panel B:} & \multicolumn{5}{c}{2005--2011 mean to 2023} \\
\cmidrule(lr){2-6}
Primary index & 0.037 & -0.148 & -0.522 & -0.251 & 0.456 \\
Horizontal coordination & -0.013 & -0.243 & -0.494 & -0.105 & 0.620 \\
Civic participation & -0.025 & -0.089 & -0.366 & -0.356 & 0.073 \\
Party linkage & 0.116 & 0.022 & -0.235 & -0.065 & 0.263 \\
Formal collective action & 0.113 & 0.122 & 0.083 & 0.080 & -0.012 \\
Organizational membership & 0.067 & 0.069 & 0.242 & 0.278 & 0.033 \\
Electoral participation & 0.058 & -0.118 & -0.231 & -0.243 & 0.164 \\
Brokerage and clientelism & -0.292 & -0.171 & -0.188 & -0.042 & 0.024 \\
Diffuse social relations & -0.152 & 0.035 & -0.075 & 0.008 & -0.104 \\
Institutional trust & -0.381 & -0.341 & -0.546 & -0.370 & 0.136 \\
\addlinespace[0.15cm]\hline\hline

\end{tabular}\end{adjustbox}\end{tablecontent}\tablenotesgap
\centering\begin{minipage}{\exhibittablewidth}\tablenotefont\textbf{Notes:} Same 33 municipalities as Figure~\ref{fig:cells}: 17 covered and 16 comparison, with both residence groups in every wave. Municipalities receive equal weight within each coverage group. Panel A subtracts 2011; Panel B subtracts the equally weighted mean of 2005 and 2011. DDD is covered rural minus covered urban minus comparison rural plus comparison urban, computed from unrounded changes. No 2017 outcome is used. These are descriptive contrasts, not separately identified treatment effects.\end{minipage}\par}\end{table}

\begin{table}[H]{\renewcommand{\arraystretch}{0.95}\setlength{\tabcolsep}{3pt}
\captionsetup{font={normalsize,md}}
\caption{Regression Decomposition of Rural--Urban Changes}\label{tab:regressiongaps}\vspace{-0.3cm}
\begin{tablecontent}\begin{adjustbox}{max width=0.98\linewidth, max totalheight=0.82\textheight}
\begin{tabular}{L{5.3cm}*{3}{C{3.0cm}}}

\hline\hline \addlinespace[0.15cm]
 & (1) & (2) & (3) \\
 & Comparison gap, $\rho$ & Covered gap, $\rho+\pi$ & Difference, $\pi$ \\
\addlinespace[0.10cm]\hline\addlinespace[0.15cm]
\emph{Panel A:} & \multicolumn{3}{c}{Primary index} \\
\cmidrule(lr){2-4}
2017 reference & -0.468** & 0.113 & 0.582*** \\
 & (0.177) & (0.088) & (0.198) \\
 & [-0.823, -0.114] & [-0.062, 0.288] & [0.187, 0.976] \\
2011 reference, no 2017 & -0.341* & 0.284** & 0.625*** \\
 & (0.189) & (0.125) & (0.227) \\
 & [-0.719, 0.037] & [0.035, 0.533] & [0.172, 1.078] \\
\emph{Panel B:} & \multicolumn{3}{c}{Horizontal coordination} \\
\cmidrule(lr){2-4}
2017 reference & -0.337*** & 0.108 & 0.445*** \\
 & (0.123) & (0.102) & (0.160) \\
 & [-0.583, -0.090] & [-0.095, 0.311] & [0.126, 0.764] \\
2011 reference, no 2017 & -0.411*** & 0.172 & 0.582*** \\
 & (0.149) & (0.114) & (0.188) \\
 & [-0.708, -0.113] & [-0.056, 0.399] & [0.208, 0.957] \\
\emph{Panel C:} & \multicolumn{3}{c}{Civic participation} \\
\cmidrule(lr){2-4}
2017 reference & -0.270 & 0.058 & 0.328* \\
 & (0.166) & (0.065) & (0.178) \\
 & [-0.601, 0.061] & [-0.071, 0.186] & [-0.027, 0.683] \\
2011 reference, no 2017 & -0.048 & 0.204** & 0.253 \\
 & (0.168) & (0.093) & (0.192) \\
 & [-0.384, 0.287] & [0.018, 0.390] & [-0.131, 0.636] \\
\emph{Panel D:} & \multicolumn{3}{c}{Party linkage} \\
\cmidrule(lr){2-4}
2017 reference & -0.374** & 0.071 & 0.446** \\
 & (0.161) & (0.112) & (0.197) \\
 & [-0.697, -0.052] & [-0.153, 0.296] & [0.053, 0.838] \\
2011 reference, no 2017 & -0.256 & 0.219 & 0.474** \\
 & (0.172) & (0.153) & (0.230) \\
 & [-0.599, 0.088] & [-0.088, 0.525] & [0.014, 0.935] \\
\addlinespace[0.15cm]\hline\hline

\end{tabular}\end{adjustbox}\end{tablecontent}\tablenotesgap
\centering\begin{minipage}{\exhibittablewidth}\tablenotefont\textbf{Notes:} Linear combinations from equation~\eqref{eq:esddd}. Each row uses the same regression for all three columns, so column (3) equals column (2) minus column (1), up to displayed rounding. The baseline uses 2005--2023; the 2011-reference regression uses 2005, 2011 and 2023 only. Municipality-by-wave and municipality-by-residence effects and respondent sex are included. Standard errors clustered by municipality in parentheses; 95\% intervals in brackets. Outcomes use the 2011 rural comparison scale. These are adjusted gap changes, not absolute rural causal effects. *, **, *** denote 10\%, 5\% and 1\%.\end{minipage}\par}\end{table}

\begin{table}[H]{\renewcommand{\arraystretch}{0.95}\setlength{\tabcolsep}{2pt}
\captionsetup{font={normalsize,md}}
\caption{Pre-Agreement Municipal Characteristics by Coverage}\label{tab:balance}\vspace{-0.3cm}
\begin{tablecontent}\begin{adjustbox}{max width=0.98\linewidth, max totalheight=0.78\textheight}
\begin{tabular}{L{5.4cm}*{5}{C{1.85cm}}}

\hline\hline \addlinespace[0.15cm]
 & (1) & (2) & (3) & (4) & (5) \\
 & Covered mean & Comparison mean & Standardized difference & Covered N & Comparison N \\
\addlinespace[0.10cm]\hline\addlinespace[0.15cm]
Unsatisfied basic needs & 57.895 & 30.760 & 1.549 & 22 & 37 \\
Rural population share & 0.483 & 0.237 & 1.083 & 22 & 37 \\
Log(1 + population) & 10.419 & 11.751 & -0.897 & 22 & 37 \\
Log(1 + homicides) & 2.321 & 3.369 & -0.638 & 22 & 37 \\
Log(1 + displacement) & 5.944 & 5.099 & 0.490 & 22 & 37 \\
Distance to capital, 100 km & 1.085 & 0.375 & 0.866 & 22 & 37 \\
\addlinespace[0.15cm]\hline\hline

\end{tabular}\end{adjustbox}\end{tablecontent}\tablenotesgap
\centering\begin{minipage}{\exhibittablewidth}\tablenotefont\textbf{Notes:} One equally weighted observation per municipality with the characteristic observed. Characteristics are measured in 2015 except unsatisfied basic needs (2011). Standardized differences divide the covered-minus-comparison mean by the square root of the average of their municipal variances. Counts are municipalities, not respondents. Source: authenticated CEDE Municipal Panel inputs.\end{minipage}\par}\end{table}

\begin{table}[H]{\renewcommand{\arraystretch}{0.95}\setlength{\tabcolsep}{2pt}
\captionsetup{font={normalsize,md}}
\caption{Sensitivity to the Municipal Comparison Set}\label{tab:support}\vspace{-0.3cm}
\begin{tablecontent}\begin{adjustbox}{max width=0.98\linewidth, max totalheight=0.78\textheight}
\begin{tabular}{L{5.3cm}*{4}{C{2.35cm}}}

\hline\hline \addlinespace[0.15cm]
 & (1) & (2) & (3) & (4) \\
 & Complete covariates & Common propensity support & Exclude capitals & Rurality at least 0.30 \\
\addlinespace[0.10cm]\hline\addlinespace[0.15cm]
\emph{Panel A:} & \multicolumn{4}{c}{Dependent variable: Primary index} \\
\cmidrule(lr){2-5}
Rural $\times$ Covered $\times$ Post & 0.574*** & 0.609** & 0.644** & 0.670* \\
 & (0.204) & (0.275) & (0.253) & (0.329) \\
Pre-trend $p$ & 0.943 & 0.725 & 0.790 & 0.644 \\
R-squared & 0.164 & 0.193 & 0.242 & 0.232 \\
Observations & 12,105 & 5,479 & 5,363 & 4,080 \\
\addlinespace[0.20cm]
\emph{Panel B:} & \multicolumn{4}{c}{Dependent variable: Horizontal coordination} \\
\cmidrule(lr){2-5}
Rural $\times$ Covered $\times$ Post & 0.474*** & 0.560** & 0.436** & 0.543** \\
 & (0.161) & (0.212) & (0.187) & (0.224) \\
Pre-trend $p$ & 0.449 & 0.424 & 0.727 & 0.594 \\
R-squared & 0.122 & 0.156 & 0.170 & 0.166 \\
Observations & 12,105 & 5,479 & 5,363 & 4,080 \\
\addlinespace[0.20cm]
\emph{Panel C:} & \multicolumn{4}{c}{Dependent variable: Civic participation} \\
\cmidrule(lr){2-5}
Rural $\times$ Covered $\times$ Post & 0.298 & 0.304 & 0.329 & 0.250 \\
 & (0.182) & (0.245) & (0.231) & (0.299) \\
Pre-trend $p$ & 0.746 & 0.931 & 0.720 & 0.344 \\
R-squared & 0.143 & 0.172 & 0.230 & 0.227 \\
Observations & 12,105 & 5,479 & 5,363 & 4,080 \\
\addlinespace[0.20cm]
\emph{Panel D:} & \multicolumn{4}{c}{Dependent variable: Party linkage} \\
\cmidrule(lr){2-5}
Rural $\times$ Covered $\times$ Post & 0.431** & 0.412 & 0.584** & 0.610* \\
 & (0.202) & (0.278) & (0.244) & (0.308) \\
Pre-trend $p$ & 0.810 & 0.800 & 0.564 & 0.484 \\
R-squared & 0.122 & 0.143 & 0.194 & 0.187 \\
Observations & 12,105 & 5,479 & 5,363 & 4,080 \\
\addlinespace[0.10cm]\hline\addlinespace[0.15cm]
\addlinespace[0.15cm]\hline\hline

\end{tabular}\end{adjustbox}\end{tablecontent}\tablenotesgap
\centering\begin{minipage}{\exhibittablewidth}\tablenotefont\textbf{Notes:} Column (1) uses municipalities with all six pre-agreement characteristics. Column (2) restricts their estimated logit coverage propensity to the intersection of the covered and comparison score ranges; the logit uses the six characteristics in Table~\ref{tab:balance}, not outcomes. Column (3) excludes departmental capitals, identified by municipal DANE codes ending in 001, from the full baseline sample. Column (4) requires observed pre-agreement rural population share of at least 0.30. These are sensitivity restrictions, not a claim of as-if-random coverage. All regressions include municipality-by-wave and municipality-by-residence fixed effects and respondent sex. Outcomes retain their 2011 rural-comparison standardization. Standard errors are clustered by municipality. *, **, *** denote conventional significance at 10\%, 5\% and 1\%; these tests do not jointly adjust for outcome selection.\end{minipage}\par}\end{table}

\begin{table}[H]{\renewcommand{\arraystretch}{0.95}\setlength{\tabcolsep}{2pt}
\captionsetup{font={normalsize,md}}
\caption{Holding Municipal Sample Allocation at Its 2011 Distribution}\label{tab:allocation}\vspace{-0.3cm}
\begin{tablecontent}\begin{adjustbox}{max width=0.98\linewidth, max totalheight=0.78\textheight}
\begin{tabular}{L{6.0cm}*{2}{C{4.2cm}}}

\hline\hline \addlinespace[0.15cm]
 & (1) & (2) \\
 & Unweighted & 2011 municipal allocation \\
\addlinespace[0.10cm]\hline\addlinespace[0.15cm]
\emph{Panel A:} & \multicolumn{2}{c}{Dependent variable: Primary index} \\
\cmidrule(lr){2-3}
Rural $\times$ Covered $\times$ Post & 0.642** & 0.600** \\
 & (0.238) & (0.243) \\
 & [0.160, 1.123] & [0.109, 1.091] \\
Pre-trend $p$ & --- & --- \\
R-squared & 0.178 & 0.175 \\
Observations & 3,149 & 3,149 \\
\addlinespace[0.20cm]
\emph{Panel B:} & \multicolumn{2}{c}{Dependent variable: Horizontal coordination} \\
\cmidrule(lr){2-3}
Rural $\times$ Covered $\times$ Post & 0.653*** & 0.664*** \\
 & (0.188) & (0.190) \\
 & [0.272, 1.034] & [0.280, 1.049] \\
Pre-trend $p$ & --- & --- \\
R-squared & 0.160 & 0.158 \\
Observations & 3,149 & 3,149 \\
\addlinespace[0.20cm]
\emph{Panel C:} & \multicolumn{2}{c}{Dependent variable: Civic participation} \\
\cmidrule(lr){2-3}
Rural $\times$ Covered $\times$ Post & 0.238 & 0.196 \\
 & (0.202) & (0.202) \\
 & [-0.172, 0.647] & [-0.213, 0.605] \\
Pre-trend $p$ & --- & --- \\
R-squared & 0.194 & 0.188 \\
Observations & 3,149 & 3,149 \\
\addlinespace[0.20cm]
\emph{Panel D:} & \multicolumn{2}{c}{Dependent variable: Party linkage} \\
\cmidrule(lr){2-3}
Rural $\times$ Covered $\times$ Post & 0.454* & 0.397 \\
 & (0.238) & (0.243) \\
 & [-0.029, 0.936] & [-0.096, 0.889] \\
Pre-trend $p$ & --- & --- \\
R-squared & 0.115 & 0.113 \\
Observations & 3,149 & 3,149 \\
\addlinespace[0.10cm]\hline\addlinespace[0.15cm]
Years & 2011 and 2023 & 2011 and 2023 \\
Both residence groups at both dates & Y & Y \\
\addlinespace[0.15cm]\hline\hline

\end{tabular}\end{adjustbox}\end{tablecontent}\tablenotesgap
\centering\begin{minipage}{\exhibittablewidth}\tablenotefont\textbf{Notes:} Same municipalities and respondents in both columns; only 2011 and 2023 are used, with both residence groups observed at both dates. Weights in column (2) restore the 2011 municipal sample shares within each coverage-by-residence group, not demographic or settlement composition within municipalities. No pre-period test is possible with these two dates. All regressions include municipality-by-wave and municipality-by-residence fixed effects and respondent sex. Outcomes retain their 2011 rural-comparison standardization. Standard errors are clustered by municipality. *, **, *** denote conventional significance at 10\%, 5\% and 1\%; these tests do not jointly adjust for outcome selection.\end{minipage}\par}\end{table}

\begin{table}[H]{\renewcommand{\arraystretch}{0.95}\setlength{\tabcolsep}{2pt}
\captionsetup{font={normalsize,md}}
\caption{Composition of the Rural Sample}\label{tab:composition}\vspace{-0.3cm}
\begin{tablecontent}\begin{adjustbox}{max width=0.98\linewidth, max totalheight=0.86\textheight}
\begin{tabular}{L{4.5cm}*{4}{C{2.5cm}}}

\hline\hline \addlinespace[0.15cm]
 & (1) & (2) & (3) & (4) \\
 & Rural share & Rural share & Female, rural respondents & Age, rural respondents \\
\addlinespace[0.10cm]\hline\addlinespace[0.15cm]
Covered $\times$ Post & -0.055 & -0.061 & 0.002 & 2.586 \\
 & (0.040) & (0.043) & (0.052) & (1.874) \\
 & [-0.134, 0.025] & [-0.147, 0.025] & [-0.101, 0.105] & [-1.157, 6.330] \\
Mean of dependent variable & 0.240 & 0.240 & 0.496 & 37.696 \\
R-squared & 0.359 & 0.097 & 0.020 & 0.042 \\
Observations & 12,466 & 12,466 & 2,998 & 2,998 \\
\addlinespace[0.10cm]\hline\addlinespace[0.15cm]
Municipality fixed effects & Y & N & Y & Y \\
Sample & All & All & Rural & Rural \\
Waves & 4 & 4 & 4 & 4 \\
\addlinespace[0.15cm]\hline\hline

\end{tabular}\end{adjustbox}\end{tablecontent}\tablenotesgap
\centering\begin{minipage}{\exhibittablewidth}\tablenotefont\textbf{Notes:} Difference-in-differences of sample characteristics, where Post is the 2023 wave. Columns (1) and (2) use all respondents and the rural indicator as outcome, with and without municipality fixed effects, and control for sex; columns (3) and (4) use rural respondents and female and age as outcomes, with municipality fixed effects. Age is measured in completed years in all four waves; the estimate compares 2023 with the pooled earlier waves. It describes observed composition rather than following the same individuals. Socioeconomic stratum is observed for 0 of 828, 21 of 713, 703 of 787 and 0 of 670 rural respondents in 2005, 2011, 2017 and 2023, so it supports no test. Standard errors clustered by municipality in parentheses and 95\% confidence intervals in brackets. *, **, *** denote significance at 10\%, 5\% and 1\%.\end{minipage}\par}\end{table}

\clearpage

\begin{landscape}
\begin{table}[H]{\renewcommand{\arraystretch}{0.95}\setlength{\tabcolsep}{2pt}
\captionsetup{font={normalsize,md}}
\caption{Fixed Effects, Samples and Weights: The Three Constructs}\label{tab:specsc}\vspace{-0.3cm}
\begin{tablecontent}\begin{adjustbox}{max width=0.98\linewidth, max totalheight=0.86\textheight}
\begin{tabular}{L{4.6cm}*{10}{C{1.62cm}}}

\hline\hline \addlinespace[0.15cm]
 &  & \multicolumn{2}{c}{Fixed effects} & \multicolumn{4}{c}{Sample} & \multicolumn{3}{c}{Weights and controls} \\
\cmidrule(lr){3-4}\cmidrule(lr){5-8}\cmidrule(lr){9-11}
 & (1) & (2) & (3) & (4) & (5) & (6) & (7) & (8) & (9) & (10) \\
 & Baseline & No munic.\ $\times$ residence FE & Region $\times$ residence $\times$ wave FE & Strict support & Rural in 2011 and 2023 & Excl.\ Mocoa & Excl.\ Santa Marta & Equal munic.\ weights & Regional reweighting & Age, all waves \\
\addlinespace[0.10cm]\hline\addlinespace[0.15cm]
\emph{Panel A:} & \multicolumn{10}{c}{Dependent variable: Horizontal coordination} \\
\cmidrule(lr){2-11}
Rural $\times$ Covered $\times$ 2023 & 0.445*** & 0.387** & 0.344** & 0.486*** & 0.496*** & 0.458*** & 0.472*** & 0.422** & 0.444** & 0.442*** \\
 & (0.160) & (0.157) & (0.131) & (0.140) & (0.170) & (0.162) & (0.161) & (0.164) & (0.176) & (0.159) \\
Pre-trend $p$ & 0.555 & 0.580 & 0.162 & 0.437 & 0.456 & 0.596 & 0.674 & 0.454 & 0.236 & --- \\
R-squared & 0.127 & 0.120 & 0.131 & 0.162 & 0.163 & 0.127 & 0.128 & 0.151 & 0.140 & 0.127 \\
Observations & 12,466 & 12,466 & 12,466 & 5,032 & 6,312 & 12,406 & 12,226 & 12,466 & 12,466 & 12,466 \\
\addlinespace[0.20cm]
\emph{Panel B:} & \multicolumn{10}{c}{Dependent variable: Civic participation} \\
\cmidrule(lr){2-11}
Rural $\times$ Covered $\times$ 2023 & 0.328* & 0.330** & 0.226 & 0.132 & 0.312 & 0.331* & 0.311* & 0.296 & 0.269 & 0.313* \\
 & (0.178) & (0.164) & (0.153) & (0.168) & (0.186) & (0.183) & (0.178) & (0.184) & (0.187) & (0.179) \\
Pre-trend $p$ & 0.752 & 0.655 & 0.752 & 0.567 & 0.887 & 0.749 & 0.736 & 0.727 & 0.760 & --- \\
R-squared & 0.151 & 0.146 & 0.153 & 0.205 & 0.195 & 0.152 & 0.149 & 0.175 & 0.128 & 0.153 \\
Observations & 12,466 & 12,466 & 12,466 & 5,032 & 6,312 & 12,406 & 12,226 & 12,466 & 12,466 & 12,466 \\
\addlinespace[0.20cm]
\emph{Panel C:} & \multicolumn{10}{c}{Dependent variable: Party linkage} \\
\cmidrule(lr){2-11}
Rural $\times$ Covered $\times$ 2023 & 0.446** & 0.358* & 0.343* & 0.308 & 0.446** & 0.479** & 0.468** & 0.474** & 0.457* & 0.427** \\
 & (0.197) & (0.185) & (0.188) & (0.214) & (0.220) & (0.198) & (0.199) & (0.207) & (0.241) & (0.197) \\
Pre-trend $p$ & 0.775 & 0.838 & 0.603 & 0.792 & 0.863 & 0.728 & 0.740 & 0.864 & 0.595 & --- \\
R-squared & 0.129 & 0.123 & 0.130 & 0.171 & 0.162 & 0.129 & 0.130 & 0.154 & 0.222 & 0.133 \\
Observations & 12,466 & 12,466 & 12,466 & 5,032 & 6,312 & 12,406 & 12,226 & 12,466 & 12,466 & 12,466 \\
\addlinespace[0.10cm]\hline\addlinespace[0.15cm]
Municipality $\times$ residence FE & Y & N & Y & Y & Y & Y & Y & Y & Y & Y \\
Region $\times$ residence $\times$ wave FE & N & N & Y & N & N & N & N & N & N & N \\
\addlinespace[0.15cm]\hline\hline

\end{tabular}\end{adjustbox}\end{tablecontent}\tablenotesgap
\centering\begin{minipage}{\exhibittablewidth}\tablenotefont\textbf{Notes:} Columns (1)--(9) reproduce the specifications of Table~\ref{tab:specs} for each constituent. Column (10) adds age and age-by-rural interactions over all four waves (Appendix~\ref{app:age}). Dropping each covered municipality in turn gives estimates between \BarcasCoordLooMin{} and \BarcasCoordLooMax{} for coordination (largest $p=\BarcasCoordLooP$), 0.293 and 0.345 for civic participation (0.100), and 0.374 and 0.506 for party linkage (0.049). Standard errors clustered by municipality in parentheses. *, **, *** denote significance at 10\%, 5\% and 1\%.\end{minipage}\par}\end{table}
\end{landscape}

\clearpage

\begin{table}[H]{\renewcommand{\arraystretch}{0.95}\setlength{\tabcolsep}{2pt}
\captionsetup{font={normalsize,md}}
\caption{Robustness: Alternative Inference}\label{tab:inference}\vspace{-0.3cm}
\begin{tablecontent}\begin{adjustbox}{max width=0.98\linewidth, max totalheight=0.86\textheight}
\begin{tabular}{L{7.1cm}*{4}{C{1.95cm}}}

\hline\hline \addlinespace[0.15cm]
 & (1) & (2) & (3) & (4) \\
 & Primary index & Coordination & Civic particip. & Party linkage \\
\addlinespace[0.10cm]\hline\addlinespace[0.15cm]
Rural $\times$ Covered $\times$ 2023 & 0.582 & 0.445 & 0.328 & 0.446 \\
\addlinespace[0.10cm]
\emph{Panel A:} & \multicolumn{4}{c}{Standard errors} \\
\cmidrule(lr){2-5}
Clustered by municipality (baseline) & (0.198) & (0.160) & (0.178) & (0.197) \\
Clustered by PDET subregion or department & (0.199) & (0.153) & (0.179) & (0.189) \\
Clustered by department & (0.196) & (0.165) & (0.178) & (0.173) \\
Cluster jackknife over municipalities & (0.203) & (0.163) & (0.182) & (0.201) \\
\addlinespace[0.20cm]
\emph{Panel B:} & \multicolumn{4}{c}{$p$-values} \\
\cmidrule(lr){2-5}
Clustered by municipality (baseline) & 0.005 & 0.007 & 0.070 & 0.027 \\
Clustered by PDET subregion or department & 0.006 & 0.006 & 0.077 & 0.024 \\
Clustered by department & 0.006 & 0.012 & 0.077 & 0.016 \\
Cluster jackknife & 0.005 & 0.008 & 0.077 & 0.030 \\
Wild cluster bootstrap-$t$, municipality & 0.003 & 0.006 & 0.084 & 0.030 \\
Wild cluster bootstrap-$t$, subregion or department & 0.005 & 0.007 & 0.079 & 0.024 \\
Placebo permutation, municipalities & 0.002 & 0.005 & 0.077 & 0.038 \\
Placebo permutation, geographic blocks & 0.010 & 0.005 & 0.085 & 0.042 \\
Romano--Wolf (3), municipal & --- & 0.021 & 0.077 & 0.055 \\
Romano--Wolf (3), blocks & --- & 0.021 & 0.085 & 0.067 \\
Maximum over 84 indices, municipalities & 0.052 & --- & --- & --- \\
Maximum over 84 indices, blocks & 0.057 & --- & --- & --- \\
\addlinespace[0.10cm]\hline\addlinespace[0.15cm]
Municipalities & 66 & 66 & 66 & 66 \\
PDET subregion or department groups & 34 & 34 & 34 & 34 \\
Departments & 27 & 27 & 27 & 27 \\
\addlinespace[0.15cm]\hline\hline

\end{tabular}\end{adjustbox}\end{tablecontent}\tablenotesgap
\centering\begin{minipage}{\exhibittablewidth}\tablenotefont\textbf{Notes:} Standard errors and $p$-values for $\pi_{2023}$ of equation~\eqref{eq:esddd} under the baseline specification. Panel A: cluster-robust standard errors at three levels and the leave-one-municipality-out cluster jackknife, centered on the mean of the 66 leave-one-out estimates, with factor $65/66$. Its $p$-value uses a $t_{65}$ reference. Panel B: conventional $p$-values for the same clusterings; wild cluster bootstrap-$t$ with 9,999 replications; placebo permutation with 1,999 reassignments of coverage within macro-region, among municipalities or among 36 geographic blocks (12 covered); Romano--Wolf step-down over the three constructs from the same draws; and, for the primary index, the reference $p$-value for the maximum over all 84 equal-weight averages of three of the nine families (Appendix~\ref{app:selection}). Monte Carlo uncertainty remains with 1,999 draws; near a $p$-value of 0.05 its standard error is about 0.005. Permutation exercises provide placebo reference distributions, not a randomized assignment mechanism.\end{minipage}\par}\end{table}

\begin{table}[H]{\renewcommand{\arraystretch}{0.95}\setlength{\tabcolsep}{2pt}
\captionsetup{font={normalsize,md}}
\caption{Broader Alternatives to the Primary Index}\label{tab:altindices}\vspace{-0.3cm}
\begin{tablecontent}\begin{adjustbox}{max width=0.98\linewidth, max totalheight=0.86\textheight}
\begin{tabular}{L{4.7cm}*{5}{C{1.95cm}}}

\hline\hline \addlinespace[0.15cm]
 & (1) & (2) & (3) & (4) & (5) \\
 & Primary index & + Formal collective action & + Formal action and membership & All except brokerage & All nine \\
\addlinespace[0.10cm]\hline\addlinespace[0.15cm]
Rural $\times$ Covered $\times$ 2023 & 0.582*** & 0.600** & 0.565** & 0.452** & 0.431* \\
 & (0.198) & (0.232) & (0.263) & (0.215) & (0.221) \\
 & [0.187, 0.976] & [0.136, 1.063] & [0.041, 1.089] & [0.023, 0.881] & [-0.010, 0.872] \\
Rural $\times$ Covered $\times$ 2011 & -0.015 & 0.025 & 0.026 & 0.023 & 0.006 \\
 & (0.120) & (0.127) & (0.133) & (0.118) & (0.118) \\
Rural $\times$ Covered $\times$ 2005 & 0.062 & 0.184 & 0.222 & 0.190 & 0.151 \\
 & (0.169) & (0.176) & (0.181) & (0.172) & (0.170) \\
Pre-trend $p$ & 0.921 & 0.583 & 0.464 & 0.545 & 0.668 \\
R-squared & 0.171 & 0.140 & 0.113 & 0.156 & 0.158 \\
Observations & 12,466 & 12,466 & 12,466 & 12,466 & 12,466 \\
\addlinespace[0.10cm]\hline\addlinespace[0.15cm]
Families averaged & 3 & 4 & 5 & 8 & 9 \\
\addlinespace[0.15cm]\hline\hline

\end{tabular}\end{adjustbox}\end{tablecontent}\tablenotesgap
\centering\begin{minipage}{\exhibittablewidth}\tablenotefont\textbf{Notes:} Each column is equation~\eqref{eq:esddd} under the baseline specification for an equal-weight average of standardized families, re-standardized on the 2011 rural comparison cell: (1) the primary index; (2) adding formal collective action; (3) adding also organizational membership; (4) every family except brokerage and clientelism; (5) all nine families. Standard errors clustered by municipality in parentheses and 95\% confidence intervals in brackets. Outcomes are standardized on the 2011 rural comparison distribution. $X_i$ contains respondent sex. Pre-trend $p$ tests $\pi_{2005}=\pi_{2011}=0$. *, **, *** denote significance at 10\%, 5\% and 1\%.\end{minipage}\par}\end{table}

\begin{table}[H]{\renewcommand{\arraystretch}{0.95}\setlength{\tabcolsep}{2pt}
\captionsetup{font={normalsize,md}}
\caption{Robustness: Multiple-Hypothesis Correction}\label{tab:multiplicity}\vspace{-0.3cm}
\begin{tablecontent}\begin{adjustbox}{max width=0.98\linewidth, max totalheight=0.86\textheight}
\begin{tabular}{L{5.1cm}*{9}{C{1.25cm}}}

\hline\hline \addlinespace[0.15cm]
 & \multicolumn{3}{c}{Constructs} & \multicolumn{2}{c}{Other collective} & \multicolumn{4}{c}{Contrast domains} \\
\cmidrule(lr){2-4}\cmidrule(lr){5-6}\cmidrule(lr){7-10}
 & (1) & (2) & (3) & (4) & (5) & (6) & (7) & (8) & (9) \\
 & Coordination & Civic particip. & Party linkage & Formal coll. action & Org. membership & Electoral particip. & Brokerage & Diffuse relations & Inst. trust \\
\addlinespace[0.10cm]\hline\addlinespace[0.15cm]
Rural $\times$ Covered $\times$ 2023 & 0.445*** & 0.328* & 0.446** & 0.422 & 0.294 & 0.137 & -0.067 & -0.008 & 0.145 \\
 & (0.160) & (0.178) & (0.197) & (0.292) & (0.318) & (0.111) & (0.138) & (0.153) & (0.140) \\
\addlinespace[0.10cm]
\emph{Panel A:} & \multicolumn{9}{c}{Unadjusted $p$-values} \\
\cmidrule(lr){2-10}
Conventional, municipal clusters & 0.007 & 0.070 & 0.027 & 0.154 & 0.357 & 0.223 & 0.631 & 0.959 & 0.304 \\
Permutation, municipalities & 0.005 & 0.077 & 0.038 & 0.094 & 0.300 & 0.315 & 0.586 & 0.955 & 0.320 \\
Permutation, geographic blocks & 0.005 & 0.085 & 0.042 & 0.130 & 0.341 & 0.258 & 0.619 & 0.947 & 0.298 \\
\addlinespace[0.20cm]
\emph{Panel B:} & \multicolumn{9}{c}{Adjusted $p$-values} \\
\cmidrule(lr){2-10}
Romano--Wolf (3), municipal & 0.021 & 0.077 & 0.055 & --- & --- & --- & --- & --- & --- \\
Romano--Wolf (3), blocks & 0.021 & 0.085 & 0.067 & --- & --- & --- & --- & --- & --- \\
Romano--Wolf (5), municipal & 0.022 & 0.111 & 0.064 & 0.126 & 0.300 & --- & --- & --- & --- \\
Romano--Wolf (5), blocks & 0.022 & 0.137 & 0.074 & 0.160 & 0.341 & --- & --- & --- & --- \\
Romano--Wolf (9), municipal & 0.051 & 0.347 & 0.164 & 0.581 & 0.719 & 0.719 & 0.828 & 0.955 & 0.719 \\
Romano--Wolf (9), blocks & 0.055 & 0.328 & 0.159 & 0.527 & 0.700 & 0.670 & 0.842 & 0.947 & 0.700 \\
Holm, all nine & 0.063 & 0.487 & 0.213 & 0.926 & 1.000 & 1.000 & 1.000 & 1.000 & 1.000 \\
Benjamini--Hochberg (9) & 0.063 & 0.209 & 0.120 & 0.347 & 0.459 & 0.402 & 0.710 & 0.959 & 0.457 \\
Bonferroni, all nine & 0.063 & 0.626 & 0.240 & 1.000 & 1.000 & 1.000 & 1.000 & 1.000 & 1.000 \\
\addlinespace[0.15cm]\hline\hline

\end{tabular}\end{adjustbox}\end{tablecontent}\tablenotesgap
\centering\begin{minipage}{\exhibittablewidth}\tablenotefont\textbf{Notes:} Romano--Wolf step-down \citep{romanowolf2005,clarkeromano2020}, re-estimating equation~\eqref{eq:esddd} for all nine families on every permutation draw and taking the maximum $|t|$, so the correction uses the observed correlation among outcomes. The three-family block covers the primary index's constituents; the five-family block adds formal collective action and organizational membership; the nine-family block covers every family. All blocks are reported without a pre-trend screening rule. Draws reassign coverage within macro-region among municipalities or among 36 geographic blocks, 1,999 each. The conventional $p$-values and the Holm, Benjamini--Hochberg and Bonferroni corrections use the cluster-robust $p$-values; Romano--Wolf is permutation-based, so where a family's permutation $p$ lies below its conventional $p$ the Romano--Wolf value can lie below the conventional one. The primary index is examined separately in Appendix~\ref{app:selection}; its fixed-index inference is not automatically exempt from selection sensitivity.\end{minipage}\par}\end{table}

\begin{table}[H]{\renewcommand{\arraystretch}{0.95}\setlength{\tabcolsep}{2pt}
\captionsetup{font={normalsize,md}}
\caption{Initiative Gradients: Pre-Reform Population and Rurality}\label{tab:dosesize}\vspace{-0.3cm}
\begin{tablecontent}\begin{adjustbox}{max width=0.98\linewidth, max totalheight=0.78\textheight}
\begin{tabular}{L{7.0cm}*{2}{C{3.1cm}}}

\hline\hline \addlinespace[0.15cm]
 & (1) & (2) \\
 & \shortstack{Unadjusted} & \shortstack{Population and\\rurality adjusted} \\
\midrule
\multicolumn{3}{l}{\textit{Panel A: Primary index}} \\
Rural $\times$ 2023 $\times$ initiatives & 0.186 & 0.188 \\
 & (0.092) & (0.082) \\
95\% confidence interval & [-0.006, 0.378] & [0.017, 0.358] \\
Conventional $p$-value & 0.057 & 0.032 \\
Wild-bootstrap $p$-value & 0.248 & 0.204 \\
$R^2$ & 0.246 & 0.247 \\
Observations & 3,213 & 3,213 \\
\addlinespace[0.10cm]
\multicolumn{3}{l}{\textit{Panel B: Horizontal coordination}} \\
Rural $\times$ 2023 $\times$ initiatives & 0.066 & 0.059 \\
 & (0.073) & (0.061) \\
95\% confidence interval & [-0.087, 0.218] & [-0.068, 0.187] \\
Conventional $p$-value & 0.381 & 0.345 \\
Wild-bootstrap $p$-value & 0.386 & 0.393 \\
$R^2$ & 0.160 & 0.162 \\
Observations & 3,213 & 3,213 \\
\addlinespace[0.10cm]
\multicolumn{3}{l}{\textit{Panel C: Civic participation}} \\
Rural $\times$ 2023 $\times$ initiatives & 0.064 & 0.077 \\
 & (0.052) & (0.050) \\
95\% confidence interval & [-0.044, 0.171] & [-0.028, 0.181] \\
Conventional $p$-value & 0.231 & 0.142 \\
Wild-bootstrap $p$-value & 0.307 & 0.306 \\
$R^2$ & 0.233 & 0.234 \\
Observations & 3,213 & 3,213 \\
\addlinespace[0.10cm]
\multicolumn{3}{l}{\textit{Panel D: Party linkage}} \\
Rural $\times$ 2023 $\times$ initiatives & 0.260 & 0.257 \\
 & (0.123) & (0.115) \\
95\% confidence interval & [0.005, 0.516] & [0.018, 0.496] \\
Conventional $p$-value & 0.046 & 0.036 \\
Wild-bootstrap $p$-value & 0.251 & 0.215 \\
$R^2$ & 0.195 & 0.196 \\
Observations & 3,213 & 3,213 \\
\addlinespace[0.10cm]
\midrule
Municipalities (clusters) & 22 & 22 \\
Municipality $\times$ wave fixed effects & Yes & Yes \\
Municipality $\times$ residence fixed effects & Yes & Yes \\
Sex adjustment & Yes & Yes \\
\hline\hline

\end{tabular}\end{adjustbox}\end{tablecontent}\tablenotesgap
\centering\begin{minipage}{\exhibittablewidth}\tablenotefont\textbf{Notes:} Covered municipalities, 2005--2023. Each entry estimates the initiative gradient comparing 2023 with pooled 2005, 2011 and 2017, as in Panel A of Table~\ref{tab:dose}. The adjusted model adds rural-by-post interactions with $\log(1+\text{population})$ and the rurality index, both measured in 2015. The two columns use identical respondents; initiative standardization is fixed at that of Table~\ref{tab:dose}. Both models include a rural-by-post interaction, municipality-by-wave and municipality-by-residence effects, and respondent sex. Standard errors clustered by municipality are in parentheses, conventional 95\% cluster-$t$ intervals in brackets. Wild $p$ uses 9,999 municipal-cluster bootstrap replications, seed 20261006.\end{minipage}\par}\end{table}

\begin{table}[H]{\renewcommand{\arraystretch}{0.95}\setlength{\tabcolsep}{2pt}
\captionsetup{font={normalsize,md}}
\caption{Initiative Gradients: Bootstrap, Influence and Excluding 2017}\label{tab:dose_audit}\vspace{-0.3cm}
\begin{tablecontent}\begin{adjustbox}{max width=0.98\linewidth, max totalheight=0.78\textheight}
\begin{tabular}{L{4.4cm}*{6}{C{1.7cm}}}

\hline\hline \addlinespace[0.15cm]
\emph{Panel A:} & \multicolumn{6}{c}{Bootstrap, influence and omitting 2017} \\
 & (1) & (2) & (3) & (4) & (5) & (6) \\
 & All-wave gradient & Wild $p$ & Leave-one-out minimum & Leave-one-out maximum & Omit 2017 & Wild $p$ \\
\addlinespace[0.10cm]\hline\addlinespace[0.15cm]
Primary index & 0.186* & 0.248 & 0.099 & 0.239 & 0.197* & 0.234 \\
Horizontal coordination & 0.066 & 0.386 & 0.021 & 0.105 & 0.045 & 0.556 \\
Civic participation & 0.064 & 0.307 & 0.038 & 0.088 & 0.086 & 0.317 \\
Party linkage & 0.260** & 0.251 & 0.099 & 0.313 & 0.281** & 0.219 \\
Membership, types the statute names & 0.115 & 0.500 & 0.006 & 0.244 & 0.278 & 0.097 \\
\addlinespace[0.15cm]\multicolumn{7}{l}{\emph{Panel B: Initiative-record sensitivities, all-wave gradient}} \\
Rule & Primary index & Coordination & Civic participation & Party linkage & Named types & Clusters \\
Snapshot rows & 0.186 (0.057) & 0.066 (0.381) & 0.064 (0.231) & 0.260 (0.046) & 0.115 (0.489) & 22 \\
Distinct full records & 0.186 (0.057) & 0.066 (0.379) & 0.063 (0.233) & 0.260 (0.046) & 0.112 (0.495) & 22 \\
Municipality, title, description & 0.186 (0.057) & 0.064 (0.390) & 0.064 (0.228) & 0.262 (0.044) & 0.116 (0.484) & 22 \\
\addlinespace[0.15cm]\hline\hline

\end{tabular}\end{adjustbox}\end{tablecontent}\tablenotesgap
\centering\begin{minipage}{\exhibittablewidth}\tablenotefont\textbf{Notes:} Covered municipalities only. All-wave gradients reproduce Panel A of Table~\ref{tab:dose}; the omitted-2017 gradient pools 2005 and 2011 as the pre-period. Wild $p$ uses 9,999 municipality-cluster bootstrap replications, seed 20261006. Leave-one-out extrema omit each of the 22 covered municipalities in turn, holding the initiative scale fixed. Panel B compares snapshot rows with distinct full records and distinct municipality--title--description combinations; cells show the coefficient and conventional $p$-value in parentheses. Each rule is standardized among covered respondents in 2023. Every regression includes municipality-by-wave and municipality-by-residence effects, a rural-by-post interaction and respondent sex. Stars denote conventional cluster-$t$ tests, not bootstrap significance.\end{minipage}\par}\end{table}

\begin{table}[H]{\renewcommand{\arraystretch}{0.95}\setlength{\tabcolsep}{3pt}
\captionsetup{font={normalsize,md}}
\caption{Organizational Membership under Alternative Inactive-Membership Codings}\label{tab:membershipcoding}\vspace{-0.3cm}
\begin{tablecontent}\begin{adjustbox}{max width=0.98\linewidth, max totalheight=0.82\textheight}
\begin{tabular}{L{5.3cm}C{2.6cm}C{0.9cm}C{3.0cm}C{0.9cm}}

\hline\hline \addlinespace[0.15cm]
 & (1) & (2) & (3) & (4) \\
 & 2017 reference & $p$ & 2011 reference, no 2017 & $p$ \\
\addlinespace[0.10cm]\hline\addlinespace[0.15cm]
\emph{Panel A:} & \multicolumn{4}{c}{Named nominating types} \\
\cmidrule(lr){2-5}
Original 0/5/10 coding & 0.672** & 0.032 & 0.498* & 0.096 \\
 & (0.307) &  & (0.295) &  \\
Any membership & 0.399 & 0.165 & 0.349 & 0.225 \\
 & (0.284) &  & (0.284) &  \\
Active membership only & 1.006*** & 0.003 & 0.626** & 0.036 \\
 & (0.320) &  & (0.292) &  \\
\emph{Panel B:} & \multicolumn{4}{c}{Other secular types} \\
\cmidrule(lr){2-5}
Original 0/5/10 coding & 0.206 & 0.538 & 0.155 & 0.586 \\
 & (0.333) &  & (0.283) &  \\
Any membership & 0.055 & 0.876 & 0.110 & 0.710 \\
 & (0.351) &  & (0.293) &  \\
Active membership only & 0.426* & 0.097 & 0.191 & 0.449 \\
 & (0.253) &  & (0.250) &  \\
\emph{Panel C:} & \multicolumn{4}{c}{Named minus other types} \\
\cmidrule(lr){2-5}
Original 0/5/10 coding & 0.466* & 0.099 & 0.343 & 0.196 \\
 & (0.278) &  & (0.263) &  \\
Any membership & 0.345 & 0.182 & 0.239 & 0.259 \\
 & (0.255) &  & (0.210) &  \\
Active membership only & 0.580* & 0.056 & 0.436 & 0.175 \\
 & (0.298) &  & (0.318) &  \\
\addlinespace[0.15cm]\hline\hline

\end{tabular}\end{adjustbox}\end{tablecontent}\tablenotesgap
\centering\begin{minipage}{\exhibittablewidth}\tablenotefont\textbf{Notes:} Event-study post-period coefficients under the indicated reference wave, with municipality-by-wave and municipality-by-residence effects and respondent sex. The 2011-reference regressions exclude 2017 entirely. The original coding is 0/5/10; any membership combines 5 and 10; active-only membership counts 10 alone. Each component and each group index is standardized on the corresponding 2011 rural comparison distribution. Panel C subtracts the two standardized group indices without re-standardizing their difference. All original nonmissing answers are retained; original missing values remain missing. Standard errors clustered by municipality appear in parentheses. Inference is conventional and pointwise. *, **, *** denote 10\%, 5\% and 1\%.\end{minipage}\par}\end{table}

\begin{table}[H]{\renewcommand{\arraystretch}{0.95}\setlength{\tabcolsep}{2pt}
\captionsetup{font={normalsize,md}}
\caption{Common Parallel-Trends Sensitivity Across All Families}\label{tab:allsensitivity}\vspace{-0.3cm}
\begin{tablecontent}\begin{adjustbox}{max width=0.98\linewidth, max totalheight=0.86\textheight}
\begin{tabular}{L{4.0cm}*{10}{C{1.06cm}}}

\hline\hline \addlinespace[0.15cm]
 & (1) & (2) & (3) & (4) & (5) & (6) & (7) & (8) & (9) & (10) \\
 & Primary index & Coordination & Civic particip. & Party linkage & Formal coll. action & Org. membership & Electoral particip. & Brokerage & Diffuse relations & Inst. trust \\
\addlinespace[0.10cm]\hline\addlinespace[0.15cm]
Rural $\times$ Covered $\times$ 2023 & 0.582 & 0.445 & 0.328 & 0.446 & 0.422 & 0.294 & 0.137 & -0.067 & -0.008 & 0.145 \\
 & (0.198) & (0.160) & (0.178) & (0.197) & (0.292) & (0.318) & (0.111) & (0.138) & (0.153) & (0.140) \\
\addlinespace[0.10cm]
Original interval & [0.194, 0.969] & [0.132, 0.758] & [-0.020, 0.676] & [0.061, 0.831] & [-0.152, 0.995] & [-0.328, 0.917] & [-0.081, 0.355] & [-0.338, 0.204] & [-0.307, 0.292] & [-0.129, 0.419] \\
$\bar M=1$ interval & [0.086, 1.166] & [-0.020, 0.864] & [-0.118, 0.854] & [-0.133, 1.001] & [-0.418, 1.273] & [-0.557, 1.146] & [-0.291, 0.542] & [-0.567, 0.428] & [-0.436, 0.446] & [-0.329, 0.550] \\
Breakdown $\bar M$ & 1.50 & 1.00 & 0.00 & 0.50 & 0.00 & 0.00 & 0.00 & 0.00 & 0.00 & 0.00 \\
\addlinespace[0.15cm]\hline\hline

\end{tabular}\end{adjustbox}\end{tablecontent}\tablenotesgap
\centering\begin{minipage}{\exhibittablewidth}\tablenotefont\textbf{Notes:} 95\% sensitivity intervals for $\pi_{2023}$ from the relative-magnitudes restriction of \citet{rambachanroth2023}, implemented by \texttt{honestdid}, C-LF method. The original interval is its asymptotic Gaussian interval, which can differ slightly from the conventional cluster-$t$ interval of Table~\ref{tab:intervals}. Breakdown is the first tested value $\bar M\in\{0,0.25,0.50,0.75,1,1.5,2\}$ whose interval contains zero; $>2$ means zero is excluded on this grid, not that the threshold beyond two has been located. These are pointwise, fixed-outcome intervals.\end{minipage}\par}\end{table}

\begin{table}[H]{\renewcommand{\arraystretch}{0.95}\setlength{\tabcolsep}{1pt}
\captionsetup{font={normalsize,md}}
\caption{Pre-Existing Municipal Characteristics:\\ The Three Constructs}\label{tab:horseracec}\vspace{-0.3cm}
\begin{tablecontent}\begin{adjustbox}{max width=0.98\linewidth, max totalheight=0.86\textheight}
\begin{tabular}{L{4.3cm}*{9}{C{1.24cm}}}

\hline\hline \addlinespace[0.15cm]
 & (1) & (2) & (3) & (4) & (5) & (6) & (7) & (8) & (9) \\
 & Full sample & Same sample & Homicides & Displacement & Basic needs & Rurality & Population & Dist. to capital & All six \\
\addlinespace[0.10cm]\hline\addlinespace[0.15cm]
\emph{Panel A:} & \multicolumn{9}{c}{Dependent variable: Horizontal coordination} \\
\cmidrule(lr){2-10}
Rural $\times$ Covered $\times$ 2023 & 0.445*** & 0.474*** & 0.474*** & 0.511*** & 0.506** & 0.457** & 0.471*** & 0.491*** & 0.561** \\
 & (0.160) & (0.161) & (0.162) & (0.188) & (0.217) & (0.177) & (0.163) & (0.178) & (0.277) \\
Pre-trend $p$ & 0.555 & 0.449 & 0.448 & 0.873 & 0.156 & 0.363 & 0.413 & 0.349 & 0.464 \\
R-squared & 0.127 & 0.122 & 0.122 & 0.122 & 0.122 & 0.122 & 0.122 & 0.122 & 0.123 \\
Observations & 12,466 & 12,105 & 12,105 & 12,105 & 12,105 & 12,105 & 12,105 & 12,105 & 12,105 \\
\addlinespace[0.20cm]
\emph{Panel B:} & \multicolumn{9}{c}{Dependent variable: Civic participation} \\
\cmidrule(lr){2-10}
Rural $\times$ Covered $\times$ 2023 & 0.328* & 0.298 & 0.291 & 0.302 & 0.442* & 0.289 & 0.323* & 0.356* & 0.572** \\
 & (0.178) & (0.182) & (0.181) & (0.230) & (0.240) & (0.226) & (0.180) & (0.207) & (0.278) \\
Pre-trend $p$ & 0.752 & 0.746 & 0.750 & 0.858 & 0.664 & 0.892 & 0.626 & 0.567 & 0.452 \\
R-squared & 0.151 & 0.143 & 0.143 & 0.143 & 0.143 & 0.143 & 0.143 & 0.143 & 0.144 \\
Observations & 12,466 & 12,105 & 12,105 & 12,105 & 12,105 & 12,105 & 12,105 & 12,105 & 12,105 \\
\addlinespace[0.20cm]
\emph{Panel C:} & \multicolumn{9}{c}{Dependent variable: Party linkage} \\
\cmidrule(lr){2-10}
Rural $\times$ Covered $\times$ 2023 & 0.446** & 0.431** & 0.422** & 0.400* & 0.525* & 0.447* & 0.446** & 0.476** & 0.499 \\
 & (0.197) & (0.202) & (0.198) & (0.225) & (0.270) & (0.236) & (0.203) & (0.226) & (0.329) \\
Pre-trend $p$ & 0.775 & 0.810 & 0.806 & 0.824 & 0.918 & 0.951 & 0.791 & 0.672 & 0.354 \\
R-squared & 0.129 & 0.122 & 0.123 & 0.123 & 0.123 & 0.123 & 0.123 & 0.123 & 0.124 \\
Observations & 12,466 & 12,105 & 12,105 & 12,105 & 12,105 & 12,105 & 12,105 & 12,105 & 12,105 \\
\addlinespace[0.10cm]\hline\addlinespace[0.15cm]
Covariate $\times$ rural $\times$ wave & N & N & Y & Y & Y & Y & Y & Y & Y \\
Common covariate sample & N & Y & Y & Y & Y & Y & Y & Y & Y \\
\addlinespace[0.15cm]\hline\hline

\end{tabular}\end{adjustbox}\end{tablecontent}\tablenotesgap
\centering\begin{minipage}{\exhibittablewidth}\tablenotefont\textbf{Notes:} The specifications of Table~\ref{tab:horserace}, estimated for each constituent. Column (1) is the full-sample baseline; columns (2)--(9) use the same complete-covariate sample. Column (2) adds no characteristics, columns (3)--(8) add one and column (9) all six, interacted with rural residence and wave. Characteristics are dated 2015; basic needs are dated 2011. Standard errors clustered by municipality in parentheses. *, **, *** denote 10\%, 5\% and 1\%.\end{minipage}\par}\end{table}

\begin{table}[H]{\renewcommand{\arraystretch}{0.95}\setlength{\tabcolsep}{3pt}
\captionsetup{font={normalsize,md}}
\caption{Age Adjustment on the Four-Wave Sample}\label{tab:agediagnostic}\vspace{-0.3cm}
\begin{tablecontent}\begin{adjustbox}{max width=0.98\linewidth, max totalheight=0.82\textheight}
\begin{tabular}{L{5.3cm}*{2}{C{3.5cm}}}

\hline\hline \addlinespace[0.15cm]
 & (1) & (2) \\
 & Sex adjustment & Sex and age adjustment \\
\addlinespace[0.10cm]\hline\addlinespace[0.15cm]
\emph{Panel A:} & \multicolumn{2}{c}{Dependent variable: Primary index} \\
\cmidrule(lr){2-3}
Rural $\times$ Covered $\times$ Post & 0.582*** & 0.564*** \\
 & (0.198) & (0.198) \\
 & [0.187, 0.976] & [0.169, 0.959] \\
Pre-trend $p$ & 0.921 & 0.957 \\
R-squared & 0.171 & 0.174 \\
Observations & 12,466 & 12,466 \\
\addlinespace[0.20cm]
\emph{Panel B:} & \multicolumn{2}{c}{Dependent variable: Horizontal coordination} \\
\cmidrule(lr){2-3}
Rural $\times$ Covered $\times$ Post & 0.445*** & 0.442*** \\
 & (0.160) & (0.159) \\
 & [0.126, 0.764] & [0.124, 0.759] \\
Pre-trend $p$ & 0.555 & 0.547 \\
R-squared & 0.127 & 0.127 \\
Observations & 12,466 & 12,466 \\
\addlinespace[0.20cm]
\emph{Panel C:} & \multicolumn{2}{c}{Dependent variable: Civic participation} \\
\cmidrule(lr){2-3}
Rural $\times$ Covered $\times$ Post & 0.328* & 0.313* \\
 & (0.178) & (0.179) \\
 & [-0.027, 0.683] & [-0.043, 0.670] \\
Pre-trend $p$ & 0.752 & 0.705 \\
R-squared & 0.151 & 0.153 \\
Observations & 12,466 & 12,466 \\
\addlinespace[0.20cm]
\emph{Panel D:} & \multicolumn{2}{c}{Dependent variable: Party linkage} \\
\cmidrule(lr){2-3}
Rural $\times$ Covered $\times$ Post & 0.446** & 0.427** \\
 & (0.197) & (0.197) \\
 & [0.053, 0.838] & [0.034, 0.821] \\
Pre-trend $p$ & 0.775 & 0.805 \\
R-squared & 0.129 & 0.133 \\
Observations & 12,466 & 12,466 \\
\addlinespace[0.10cm]\hline\addlinespace[0.15cm]
Years & 2005, 2011, 2017, 2023 & 2005, 2011, 2017, 2023 \\
\addlinespace[0.15cm]\hline\hline

\end{tabular}\end{adjustbox}\end{tablecontent}\tablenotesgap
\centering\begin{minipage}{\exhibittablewidth}\tablenotefont\textbf{Notes:} Both columns use the same age-complete sample in 2005, 2011, 2017 and 2023, retaining 2017 as reference. Municipality-by-wave and municipality-by-residence fixed effects and sex adjustment are included throughout; column (2) adds age and its interaction with rural residence. The coefficient is the additional rural--urban change in covered municipalities relative to comparison municipalities. Outcomes retain their 2011 rural-comparison scale. The 2005 and 2011 interactions enter both specifications; the pre-period test assesses them jointly. Age adjustment addresses observed age composition, not unobserved household or settlement turnover. Standard errors clustered by municipality in parentheses; 95\% intervals in brackets. *, **, *** denote conventional significance at 10\%, 5\% and 1\%.\end{minipage}\par}\end{table}

\begin{table}[H]{\renewcommand{\arraystretch}{0.95}\setlength{\tabcolsep}{2pt}
\captionsetup{font={normalsize,md}}
\caption{Evidence on the Identifying Assumption: The Pre-Reform Anchor}\label{tab:anchors}\vspace{-0.3cm}
\begin{tablecontent}\begin{adjustbox}{max width=0.98\linewidth, max totalheight=0.86\textheight}
\begin{tabular}{L{4.3cm}*{6}{C{1.74cm}}}

\hline\hline \addlinespace[0.15cm]
 & (1) & (2) & (3) & (4) & (5) & (6) \\
 & Baseline & Normalized on 2011 & 2023 vs pooled 2005--2017 & Excluding 2017 & Adding 1997 & Adding 1997, excl.\ Santa Marta \\
\addlinespace[0.10cm]\hline\addlinespace[0.15cm]
 & \multicolumn{6}{c}{Dependent variable: Primary index} \\
\cmidrule(lr){2-7}
Rural $\times$ Covered $\times$ Post & 0.582*** & 0.596*** & 0.558*** & 0.587*** & 0.585*** & 0.601*** \\
 & (0.198) & (0.221) & (0.190) & (0.198) & (0.196) & (0.197) \\
 & [0.187, 0.976] & [0.154, 1.038] & [0.178, 0.937] & [0.192, 0.981] & [0.196, 0.974] & [0.208, 0.993] \\
Pre-trend $p$ & 0.921 & --- & --- & --- & 0.389 & 0.533 \\
R-squared & 0.171 & 0.171 & 0.171 & 0.158 & 0.159 & 0.159 \\
Observations & 12,466 & 12,466 & 12,466 & 9,278 & 15,495 & 15,155 \\
\addlinespace[0.10cm]\hline\addlinespace[0.15cm]
Event study & Y & Y & N & N & Y & Y \\
Waves & 2005--23 & 2005--23 & 2005--23 & 2005, 11, 23 & 1997--2023 & 1997--2023 \\
\addlinespace[0.15cm]\hline\hline

\end{tabular}\end{adjustbox}\end{tablecontent}\tablenotesgap
\centering\begin{minipage}{\exhibittablewidth}\tablenotefont\textbf{Notes:} Column (1) is the baseline, equation~\eqref{eq:esddd}. Column (2) normalizes the event study on 2011, so the coefficient is $\pi_{2023}-\pi_{2011}$ of the baseline and the pre-trend test is that of column (1). Columns (3) and (4) estimate the pooled analogue of Section~\ref{sec:spec}, comparing 2023 with the pooled earlier waves, and in column (4) excluding 2017; the pooled analogue has no pre-trend test. Columns (5) and (6) add the 1997 wave with its own interaction; their pre-trend $p$ tests the 1997, 2005 and 2011 coefficients jointly. Column (6) excludes Santa Marta, which supplies 13\% of covered rural respondents in 1997 and none in 2005. Table~\ref{tab:anchorsc} reports the constituents. The first coefficient row is $\pi_{2023}$ in event-study columns and $\pi$ in pooled columns. Standard errors clustered by municipality in parentheses and 95\% confidence intervals in brackets. Outcomes are standardized on the 2011 rural comparison distribution. $X_i$ contains respondent sex. *, **, *** denote significance at 10\%, 5\% and 1\%.\end{minipage}\par}\end{table}

\begin{table}[H]{\renewcommand{\arraystretch}{0.95}\setlength{\tabcolsep}{2pt}
\captionsetup{font={normalsize,md}}
\caption{Pre-Reform Anchor: The Three Constructs}\label{tab:anchorsc}\vspace{-0.3cm}
\begin{tablecontent}\begin{adjustbox}{max width=0.98\linewidth, max totalheight=0.86\textheight}
\begin{tabular}{L{4.3cm}*{6}{C{1.74cm}}}

\hline\hline \addlinespace[0.15cm]
 & (1) & (2) & (3) & (4) & (5) & (6) \\
 & Baseline & Normalized on 2011 & 2023 vs pooled 2005--2017 & Excluding 2017 & Adding 1997 & Adding 1997, excl.\ Santa Marta \\
\addlinespace[0.10cm]\hline\addlinespace[0.15cm]
\emph{Panel A:} & \multicolumn{6}{c}{Dependent variable: Horizontal coordination} \\
\cmidrule(lr){2-7}
Rural $\times$ Covered $\times$ Post & 0.445*** & 0.547*** & 0.503*** & 0.577*** & 0.447*** & 0.478*** \\
 & (0.160) & (0.184) & (0.153) & (0.159) & (0.157) & (0.158) \\
Pre-trend $p$ & 0.555 & --- & --- & --- & 0.399 & 0.083 \\
R-squared & 0.127 & 0.127 & 0.127 & 0.117 & 0.117 & 0.119 \\
Observations & 12,466 & 12,466 & 12,466 & 9,278 & 15,495 & 15,155 \\
\addlinespace[0.20cm]
\emph{Panel B:} & \multicolumn{6}{c}{Dependent variable: Civic participation} \\
\cmidrule(lr){2-7}
Rural $\times$ Covered $\times$ Post & 0.328* & 0.261 & 0.263 & 0.231 & 0.340* & 0.321* \\
 & (0.178) & (0.190) & (0.162) & (0.171) & (0.177) & (0.178) \\
Pre-trend $p$ & 0.752 & --- & --- & --- & 0.645 & 0.776 \\
R-squared & 0.151 & 0.151 & 0.151 & 0.173 & 0.147 & 0.146 \\
Observations & 12,466 & 12,466 & 12,466 & 9,278 & 15,495 & 15,155 \\
\addlinespace[0.20cm]
\emph{Panel C:} & \multicolumn{6}{c}{Dependent variable: Party linkage} \\
\cmidrule(lr){2-7}
Rural $\times$ Covered $\times$ Post & 0.446** & 0.441* & 0.402* & 0.422* & 0.438** & 0.459** \\
 & (0.197) & (0.224) & (0.208) & (0.232) & (0.196) & (0.198) \\
Pre-trend $p$ & 0.775 & --- & --- & --- & 0.512 & 0.667 \\
R-squared & 0.129 & 0.129 & 0.129 & 0.102 & 0.125 & 0.125 \\
Observations & 12,466 & 12,466 & 12,466 & 9,278 & 15,495 & 15,155 \\
\addlinespace[0.10cm]\hline\addlinespace[0.15cm]
Event study & Y & Y & N & N & Y & Y \\
Waves & 2005--23 & 2005--23 & 2005--23 & 2005, 11, 23 & 1997--2023 & 1997--2023 \\
\addlinespace[0.15cm]\hline\hline

\end{tabular}\end{adjustbox}\end{tablecontent}\tablenotesgap
\centering\begin{minipage}{\exhibittablewidth}\tablenotefont\textbf{Notes:} The specifications of Table~\ref{tab:anchors}, estimated for each constituent of the primary index. Standard errors clustered by municipality in parentheses. *, **, *** denote significance at 10\%, 5\% and 1\%.\end{minipage}\par}\end{table}

\begin{table}[H]{\renewcommand{\arraystretch}{0.95}\setlength{\tabcolsep}{2pt}
\captionsetup{font={normalsize,md}}
\caption{Effect of Harmonizing the Non-Comparable Composites}\label{tab:harmonization}\vspace{-0.3cm}
\begin{tablecontent}\begin{adjustbox}{max width=0.98\linewidth, max totalheight=0.86\textheight}
\begin{tabular}{L{4.5cm}*{4}{C{2.5cm}}}

\hline\hline \addlinespace[0.15cm]
 & (1) & (2) & (3) & (4) \\
 & Coordination & Brokerage & Civic particip. & Primary index \\
\addlinespace[0.10cm]\hline\addlinespace[0.15cm]
\emph{Panel A:} & \multicolumn{4}{c}{Unharmonized alternatives} \\
\cmidrule(lr){2-5}
Rural $\times$ Covered $\times$ 2023 & 0.384** & -0.075 & 0.387* & 0.566** \\
 & (0.154) & (0.129) & (0.223) & (0.214) \\
 & [0.076, 0.693] & [-0.332, 0.181] & [-0.058, 0.832] & [0.139, 0.994] \\
Pre-trend $p$ & 0.484 & 0.307 & 0.904 & 0.933 \\
R-squared & 0.170 & 0.089 & 0.186 & 0.193 \\
Observations & 12,466 & 12,466 & 12,466 & 12,466 \\
\addlinespace[0.20cm]
\emph{Panel B:} & \multicolumn{4}{c}{Harmonized outcomes} \\
\cmidrule(lr){2-5}
Rural $\times$ Covered $\times$ 2023 & 0.445*** & -0.067 & 0.328* & 0.582*** \\
 & (0.160) & (0.138) & (0.178) & (0.198) \\
 & [0.126, 0.764] & [-0.343, 0.209] & [-0.027, 0.683] & [0.187, 0.976] \\
Pre-trend $p$ & 0.555 & 0.481 & 0.752 & 0.921 \\
R-squared & 0.127 & 0.079 & 0.151 & 0.171 \\
Observations & 12,466 & 12,466 & 12,466 & 12,466 \\
\addlinespace[0.15cm]\hline\hline

\end{tabular}\end{adjustbox}\end{tablecontent}\tablenotesgap
\centering\begin{minipage}{\exhibittablewidth}\tablenotefont\textbf{Notes:} Estimates of equation~\eqref{eq:esddd} under the baseline specification. Panel A retains mixed-order collective-problems coding and the non-comparable civic composite as diagnostic alternatives; Panel B uses the comparable codings of Appendix~\ref{app:module} and the common-items civic index, as in the paper. Column (4) is the primary index built from the corresponding constructs. Standard errors clustered by municipality in parentheses and 95\% confidence intervals in brackets. Outcomes are standardized on the 2011 rural comparison distribution. $X_i$ contains respondent sex. *, **, *** denote significance at 10\%, 5\% and 1\%.\end{minipage}\par}\end{table}

\begin{table}[H]{\renewcommand{\arraystretch}{0.95}\setlength{\tabcolsep}{2pt}
\captionsetup{font={normalsize,md}}
\caption{The Municipal Difference-in-Differences: Primary Index and Constructs}\label{tab:did1}\vspace{-0.3cm}
\begin{tablecontent}\begin{adjustbox}{max width=0.98\linewidth, max totalheight=0.86\textheight}
\begin{tabular}{L{6.0cm}*{4}{C{2.15cm}}}

\hline\hline \addlinespace[0.15cm]
 & (1) & (2) & (3) & (4) \\
 & Primary index & Coordination & Civic particip. & Party linkage \\
\addlinespace[0.10cm]\hline\addlinespace[0.15cm]
\emph{Panel A:} & \multicolumn{4}{c}{All respondents} \\
\cmidrule(lr){2-5}
Covered $\times$ 2023 & 0.300*** & 0.112 & 0.114 & 0.403*** \\
 & (0.110) & (0.097) & (0.099) & (0.116) \\
 & [0.081, 0.520] & [-0.082, 0.306] & [-0.083, 0.312] & [0.171, 0.635] \\
Covered $\times$ 2011 & 0.027 & 0.131 & -0.188 & 0.113 \\
 & (0.116) & (0.117) & (0.132) & (0.096) \\
Covered $\times$ 2005 & 0.303*** & -0.026 & 0.193 & 0.468*** \\
 & (0.098) & (0.080) & (0.147) & (0.107) \\
Pre-trend $p$ & 0.011 & 0.386 & 0.145 & 0.000 \\
R-squared & 0.081 & 0.040 & 0.057 & 0.049 \\
Observations & 12,466 & 12,466 & 12,466 & 12,466 \\
\addlinespace[0.20cm]
\emph{Panel B:} & \multicolumn{4}{c}{Rural respondents} \\
\cmidrule(lr){2-5}
Covered $\times$ 2023 & 0.529*** & 0.321* & 0.339** & 0.447** \\
 & (0.159) & (0.164) & (0.157) & (0.169) \\
 & [0.211, 0.847] & [-0.006, 0.649] & [0.026, 0.653] & [0.110, 0.784] \\
Covered $\times$ 2011 & -0.185 & 0.067 & -0.255 & -0.200 \\
 & (0.174) & (0.159) & (0.202) & (0.140) \\
Covered $\times$ 2005 & 0.240 & -0.109 & 0.219 & 0.391** \\
 & (0.170) & (0.098) & (0.232) & (0.149) \\
Pre-trend $p$ & 0.224 & 0.483 & 0.332 & 0.006 \\
R-squared & 0.127 & 0.055 & 0.096 & 0.092 \\
Observations & 2,998 & 2,998 & 2,998 & 2,998 \\
\addlinespace[0.20cm]
\emph{Panel C:} & \multicolumn{4}{c}{Urban respondents} \\
\cmidrule(lr){2-5}
Covered $\times$ 2023 & 0.210* & 0.023 & 0.037 & 0.379*** \\
 & (0.114) & (0.102) & (0.103) & (0.116) \\
 & [-0.019, 0.438] & [-0.181, 0.226] & [-0.169, 0.243] & [0.147, 0.612] \\
Covered $\times$ 2011 & 0.168 & 0.188* & -0.068 & 0.231* \\
 & (0.122) & (0.110) & (0.137) & (0.118) \\
Covered $\times$ 2005 & 0.326*** & 0.039 & 0.188 & 0.457*** \\
 & (0.079) & (0.104) & (0.140) & (0.114) \\
Pre-trend $p$ & 0.001 & 0.141 & 0.238 & 0.001 \\
R-squared & 0.072 & 0.040 & 0.048 & 0.039 \\
Observations & 9,468 & 9,468 & 9,468 & 9,468 \\
\addlinespace[0.10cm]\hline\addlinespace[0.15cm]
Wave and macro-region fixed effects & Y & Y & Y & Y \\
Sex controls & Y & Y & Y & Y \\
Municipalities (clusters) & 66 & 66 & 66 & 66 \\
\addlinespace[0.15cm]\hline\hline

\end{tabular}\end{adjustbox}\end{tablecontent}\tablenotesgap
\centering\begin{minipage}{\exhibittablewidth}\tablenotefont\textbf{Notes:} Estimates of equation~\eqref{eq:did}, normalized on 2017, waves 2005--2023, on all respondents (Panel A) and separately on rural (Panel B) and urban (Panel C) respondents. The coefficient on Covered $\times$ 2023 is $\beta_{2023}$. Wave and macro-region fixed effects and respondent sex throughout. Pre-trend $p$ tests $\beta_{2005}=\beta_{2011}=0$. Standard errors clustered by municipality in parentheses and 95\% confidence intervals in brackets. *, **, *** denote significance at 10\%, 5\% and 1\%.\end{minipage}\par}\end{table}

\begin{table}[H]{\renewcommand{\arraystretch}{0.95}\setlength{\tabcolsep}{2pt}
\captionsetup{font={normalsize,md}}
\caption{The Municipal Difference-in-Differences: Other Outcome Families}\label{tab:did2}\vspace{-0.3cm}
\begin{tablecontent}\begin{adjustbox}{max width=0.98\linewidth, max totalheight=0.86\textheight}
\begin{tabular}{L{6.0cm}*{6}{C{1.43cm}}}

\hline\hline \addlinespace[0.15cm]
 & (1) & (2) & (3) & (4) & (5) & (6) \\
 & Formal coll. action & Org. membership & Electoral particip. & Brokerage & Diffuse relations & Inst. trust \\
\addlinespace[0.10cm]\hline\addlinespace[0.15cm]
\emph{Panel A:} & \multicolumn{6}{c}{All respondents} \\
\cmidrule(lr){2-7}
Covered $\times$ 2023 & 0.188 & 0.140 & 0.400*** & -0.180* & 0.080 & 0.101 \\
 & (0.130) & (0.141) & (0.131) & (0.091) & (0.129) & (0.131) \\
 & [-0.073, 0.448] & [-0.142, 0.422] & [0.139, 0.661] & [-0.362, 0.001] & [-0.177, 0.338] & [-0.161, 0.363] \\
Covered $\times$ 2011 & -0.200* & -0.124 & 0.364*** & -0.204** & 0.069 & 0.191 \\
 & (0.106) & (0.090) & (0.117) & (0.095) & (0.129) & (0.130) \\
Covered $\times$ 2005 & 0.123 & 0.100 & 0.427*** & -0.042 & 0.236** & 0.217** \\
 & (0.142) & (0.140) & (0.112) & (0.084) & (0.104) & (0.107) \\
Pre-trend $p$ & 0.104 & 0.145 & 0.001 & 0.105 & 0.076 & 0.109 \\
R-squared & 0.008 & 0.010 & 0.089 & 0.021 & 0.082 & 0.123 \\
Observations & 12,466 & 12,466 & 12,466 & 12,466 & 12,466 & 12,466 \\
\addlinespace[0.20cm]
\emph{Panel B:} & \multicolumn{6}{c}{Rural respondents} \\
\cmidrule(lr){2-7}
Covered $\times$ 2023 & 0.407* & 0.269 & 0.240 & -0.187 & 0.404*** & 0.092 \\
 & (0.220) & (0.238) & (0.164) & (0.115) & (0.149) & (0.188) \\
 & [-0.032, 0.847] & [-0.207, 0.745] & [-0.087, 0.567] & [-0.416, 0.041] & [0.106, 0.703] & [-0.283, 0.468] \\
Covered $\times$ 2011 & -0.220 & -0.194 & 0.014 & -0.223** & 0.181 & 0.107 \\
 & (0.189) & (0.168) & (0.135) & (0.099) & (0.173) & (0.204) \\
Covered $\times$ 2005 & 0.413* & 0.360* & 0.173 & -0.144 & 0.320** & 0.095 \\
 & (0.212) & (0.199) & (0.134) & (0.149) & (0.143) & (0.117) \\
Pre-trend $p$ & 0.068 & 0.043 & 0.433 & 0.056 & 0.079 & 0.716 \\
R-squared & 0.038 & 0.035 & 0.122 & 0.043 & 0.087 & 0.180 \\
Observations & 2,998 & 2,998 & 2,998 & 2,998 & 2,998 & 2,998 \\
\addlinespace[0.20cm]
\emph{Panel C:} & \multicolumn{6}{c}{Urban respondents} \\
\cmidrule(lr){2-7}
Covered $\times$ 2023 & 0.111 & 0.080 & 0.432*** & -0.162 & -0.008 & 0.082 \\
 & (0.133) & (0.143) & (0.144) & (0.105) & (0.150) & (0.134) \\
 & [-0.154, 0.377] & [-0.206, 0.365] & [0.144, 0.720] & [-0.371, 0.047] & [-0.308, 0.291] & [-0.186, 0.350] \\
Covered $\times$ 2011 & -0.126 & -0.054 & 0.506*** & -0.149 & 0.007 & 0.168 \\
 & (0.116) & (0.103) & (0.127) & (0.117) & (0.139) & (0.126) \\
Covered $\times$ 2005 & -0.014 & -0.025 & 0.541*** & 0.033 & 0.090 & 0.204 \\
 & (0.107) & (0.103) & (0.118) & (0.100) & (0.109) & (0.127) \\
Pre-trend $p$ & 0.548 & 0.874 & 0.000 & 0.362 & 0.633 & 0.180 \\
R-squared & 0.004 & 0.006 & 0.082 & 0.017 & 0.075 & 0.107 \\
Observations & 9,468 & 9,468 & 9,468 & 9,468 & 9,468 & 9,468 \\
\addlinespace[0.10cm]\hline\addlinespace[0.15cm]
Wave and macro-region fixed effects & Y & Y & Y & Y & Y & Y \\
Sex controls & Y & Y & Y & Y & Y & Y \\
Municipalities (clusters) & 66 & 66 & 66 & 66 & 66 & 66 \\
\addlinespace[0.15cm]\hline\hline

\end{tabular}\end{adjustbox}\end{tablecontent}\tablenotesgap
\centering\begin{minipage}{\exhibittablewidth}\tablenotefont\textbf{Notes:} As Table~\ref{tab:did1}, for the other six families. *, **, *** denote significance at 10\%, 5\% and 1\%.\end{minipage}\par}\end{table}

\clearpage

\begin{landscape}
\begin{table}[H]{\renewcommand{\arraystretch}{0.95}\setlength{\tabcolsep}{2pt}
\captionsetup{font={normalsize,md}}
\caption{Program Intensity by Pillar}\label{tab:dosepillar}\vspace{-0.3cm}
\begin{tablecontent}\begin{adjustbox}{max width=0.98\linewidth, max totalheight=0.86\textheight}
\begin{tabular}{L{4.5cm}*{9}{C{1.60cm}}}

\hline\hline \addlinespace[0.15cm]
 &  & \multicolumn{8}{c}{Initiatives in pillar} \\
\cmidrule(lr){3-10}
 & (1) & (2) & (3) & (4) & (5) & (6) & (7) & (8) & (9) \\
 & Total & Land tenure & Infra. & Rural health & Rural education & Housing and water & Economic reactivation & Right to food & Reconci\-liation \\
\addlinespace[0.10cm]\hline\addlinespace[0.15cm]
\emph{Panel A:} & \multicolumn{9}{c}{Dependent variable: Primary index} \\
\cmidrule(lr){2-10}
Rural $\times$ Post $\times$ Initiatives & 0.186* & 0.179* & 0.234*** & 0.179** & 0.249*** & 0.105 & 0.029 & 0.152 & 0.167 \\
 & (0.092) & (0.098) & (0.056) & (0.076) & (0.042) & (0.121) & (0.156) & (0.125) & (0.108) \\
R-squared & 0.246 & 0.246 & 0.247 & 0.246 & 0.248 & 0.245 & 0.245 & 0.246 & 0.246 \\
Observations & 3,213 & 3,213 & 3,213 & 3,213 & 3,213 & 3,213 & 3,213 & 3,213 & 3,213 \\
\addlinespace[0.20cm]
\emph{Panel B:} & \multicolumn{9}{c}{Dependent variable: Horizontal coordination} \\
\cmidrule(lr){2-10}
Rural $\times$ Post $\times$ Initiatives & 0.066 & 0.054 & 0.057 & 0.068 & 0.086 & 0.070 & -0.035 & 0.099 & 0.113 \\
 & (0.073) & (0.083) & (0.055) & (0.068) & (0.071) & (0.076) & (0.078) & (0.076) & (0.083) \\
R-squared & 0.160 & 0.160 & 0.160 & 0.160 & 0.160 & 0.160 & 0.160 & 0.160 & 0.160 \\
Observations & 3,213 & 3,213 & 3,213 & 3,213 & 3,213 & 3,213 & 3,213 & 3,213 & 3,213 \\
\addlinespace[0.20cm]
\emph{Panel C:} & \multicolumn{9}{c}{Dependent variable: Civic participation} \\
\cmidrule(lr){2-10}
Rural $\times$ Post $\times$ Initiatives & 0.064 & 0.120** & 0.146*** & 0.043 & 0.124*** & -0.024 & -0.037 & 0.067 & 0.018 \\
 & (0.052) & (0.047) & (0.035) & (0.049) & (0.043) & (0.067) & (0.078) & (0.072) & (0.067) \\
R-squared & 0.233 & 0.234 & 0.234 & 0.233 & 0.234 & 0.233 & 0.233 & 0.233 & 0.233 \\
Observations & 3,213 & 3,213 & 3,213 & 3,213 & 3,213 & 3,213 & 3,213 & 3,213 & 3,213 \\
\addlinespace[0.20cm]
\emph{Panel D:} & \multicolumn{9}{c}{Dependent variable: Party linkage} \\
\cmidrule(lr){2-10}
Rural $\times$ Post $\times$ Initiatives & 0.260** & 0.202 & 0.286*** & 0.265** & 0.312*** & 0.175 & 0.133 & 0.151 & 0.219 \\
 & (0.123) & (0.132) & (0.097) & (0.106) & (0.084) & (0.156) & (0.215) & (0.153) & (0.130) \\
R-squared & 0.195 & 0.194 & 0.196 & 0.196 & 0.197 & 0.194 & 0.193 & 0.193 & 0.194 \\
Observations & 3,213 & 3,213 & 3,213 & 3,213 & 3,213 & 3,213 & 3,213 & 3,213 & 3,213 \\
\addlinespace[0.20cm]
\emph{Panel E:} & \multicolumn{9}{c}{Dependent variable: Membership, types the statute names} \\
\cmidrule(lr){2-10}
Rural $\times$ Post $\times$ Initiatives & 0.115 & -0.009 & 0.091 & --- & 0.055 & --- & --- & --- & 0.186 \\
 & (0.163) & (0.153) & (0.205) &  & (0.183) &  &  &  & (0.151) \\
R-squared & 0.248 & 0.248 & 0.248 & --- & 0.248 & --- & --- & --- & 0.249 \\
Observations & 3,213 & 3,213 & 3,213 & --- & 3,213 & --- & --- & --- & 3,213 \\
\addlinespace[0.15cm]\hline\hline

\end{tabular}\end{adjustbox}\end{tablecontent}\tablenotesgap
\centering\begin{minipage}{\exhibittablewidth}\tablenotefont\textbf{Notes:} Covered municipalities only; Infra. denotes infrastructure. Each cell is a separate regression of the outcome on the interaction of rural residence, the post-reform wave and the number of initiatives in the municipality's action plan, in total (column 1) or in one pillar (columns 2 to 9), standardized, with a rural-by-post interaction, municipality-by-wave and municipality-by-residence effects, and $X_i$. Column (1) reproduces Panel A of Table~\ref{tab:dose}. Panel E, membership in the types the statute names as nominators, is estimated for total initiatives and pillars 1, 2, 4 and 8 only. Standard errors clustered by municipality in parentheses. *, **, *** denote significance at 10\%, 5\% and 1\%.\end{minipage}\par}\end{table}
\end{landscape}

\clearpage

\begin{table}[H]{\renewcommand{\arraystretch}{0.95}\setlength{\tabcolsep}{2pt}
\captionsetup{font={normalsize,md}}
\caption{Which \texttt{area} Code Is the Rural One, Wave by Wave}\label{tab:areaval}\vspace{-0.3cm}
\begin{tablecontent}
\begin{tabular}{lcccccccc}

\hline\hline \addlinespace[0.15cm]
 & (1) & (2) & (3) & (4) & (5) & (6) & (7) & (8) \\
 & \multicolumn{2}{c}{Bogot\'a: \% coded rural} & \multicolumn{2}{c}{Corr.\ with CEDE index} & \multicolumn{2}{c}{Mean CEDE index} & \multicolumn{2}{c}{Mean NBI} \\
\cmidrule(lr){2-3}\cmidrule(lr){4-5}\cmidrule(lr){6-7}\cmidrule(lr){8-9}
Wave & Documented & Used & Documented & Used & $area=1$ & $area=2$ & $area=1$ & $area=2$ \\
\addlinespace[0.15cm]\hline\addlinespace[0.15cm]
1997 & 100.0 &   0.0 & -0.863 & 0.863 & 0.316 & 0.070 & 44.6 & 19.9 \\
2005 &  99.7 &   0.3 & -0.893 & 0.893 & 0.469 & 0.128 & 47.7 & 25.1 \\
2011 & 100.0 &   0.0 & -0.975 & 0.975 & 0.499 & 0.139 & 50.1 & 25.9 \\
2017 & 100.0 &   0.0 & -0.948 & 0.948 & 0.486 & 0.151 & 51.1 & 27.2 \\
2023 &   0.6 &   0.6 & 0.912 & 0.912 & 0.165 & 0.492 & 27.9 & 52.2 \\
\addlinespace[0.15cm]\hline\hline

\end{tabular}\end{tablecontent}\tablenotesgap
\centering\begin{minipage}{\exhibittablewidth}\tablenotefont\textbf{Notes:} ``Documented'' is the coding the five BARCAS dictionaries state, \texttt{area}~$=2$ is rural; ``Used'' is the coding adopted here: \texttt{area}~$=1$ is rural before 2023 and \texttt{area}~$=2$ in 2023. Columns 2--3 give the share of the Bogot\'a sample classified rural under each. Columns 4--5 give the correlation, across the municipalities sampled in that wave, between the municipal share classified rural and the 2017 CEDE rurality index. Columns 6--9 give the mean of the CEDE index and of 2011 unsatisfied basic needs (NBI, percent) for respondents carrying each \texttt{area} code, without reference to either coding.\end{minipage}\par}
\end{table}

\begin{table}[H]{\renewcommand{\arraystretch}{0.95}\setlength{\tabcolsep}{2pt}
\captionsetup{font={normalsize,md}}
\caption{Diagnostics of the Distributed Item Files}\label{tab:sourceaudit}\vspace{-0.3cm}
\begin{tablecontent}\begin{adjustbox}{max width=0.98\linewidth, max totalheight=0.78\textheight}
\begin{tabular}{L{1.8cm}*{5}{C{2.55cm}}}

\hline\hline \addlinespace[0.15cm]
 & (1) & (2) & (3) & (4) & (5) \\
 & Respondents & Common-item repeated profiles & Expanded-item repeated profiles & Expanded items & Constant-indicator municipalities \\
\addlinespace[0.10cm]\hline\addlinespace[0.15cm]
2005 & 3025 & 0 & 0 & 116 & 0/57 \\
2011 & 3028 & 4 & 0 & 116 & 0/57 \\
2017 & 3188 & 1219 & 1004 & 361 & 41/62 \\
2023 & 3225 & 0 & 0 & 323 & 0/62 \\
\addlinespace[0.15cm]\hline\hline

\end{tabular}\end{adjustbox}\end{tablecontent}\tablenotesgap
\centering\begin{minipage}{\exhibittablewidth}\tablenotefont\textbf{Notes:} Common profiles use the same 98 numeric items with over 90\% nonmissing observations and more than one value in every wave, after excluding identifiers and demographics. Expanded profiles use wave-specific, sufficiently observed numeric fields with between two and half as many distinct values as respondents; item counts therefore differ across waves. A repeated row shares its entire selected-item vector with at least one other row; this is a response-profile diagnostic, not proof of duplicated persons. The last column counts municipalities where the any-mention joint-resolution indicator has no within-municipality variation. Raw-file SHA-256 hashes and exact item selection are recorded in the replication code and source manifest.\end{minipage}\par}\end{table}

\begin{table}[H]{\renewcommand{\arraystretch}{0.95}\setlength{\tabcolsep}{2pt}
\captionsetup{font={normalsize,md}}
\caption{Cross-Residence Profile Repetition and Municipal Engagement Gaps}\label{tab:sourcecross}\vspace{-0.3cm}
\begin{tablecontent}\begin{adjustbox}{max width=0.98\linewidth, max totalheight=0.78\textheight}
\begin{tabular}{L{1.8cm}*{5}{C{2.55cm}}}

\hline\hline \addlinespace[0.15cm]
 & (1) & (2) & (3) & (4) & (5) \\
 & Shared raw profiles (percent) & Shared family profiles (percent) & Mixed-residence municipalities & Zero primary-index gaps & Median absolute gap \\
\addlinespace[0.10cm]\hline\addlinespace[0.15cm]
2005 & 0.0 & 0.3 & 49 & 0 & 0.299 \\
2011 & 0.0 & 1.2 & 51 & 0 & 0.239 \\
2017 & 15.8 & 33.5 & 46 & 0 & 0.074 \\
2023 & 0.0 & 0.9 & 49 & 0 & 0.416 \\
\addlinespace[0.15cm]\hline\hline

\end{tabular}\end{adjustbox}\end{tablecontent}\tablenotesgap
\centering\begin{minipage}{\exhibittablewidth}\tablenotefont\textbf{Notes:} Columns (1)--(2) report percentages of all respondents in each wave whose exact profile is observed in the other residence group of the same municipality. Column (1) uses the 98 raw items of Table~\ref{tab:sourceaudit}; column (2) uses the three harmonized family indices entering the primary index. Missing item values match only missing values. Residence and municipal identifiers use the harmonization described in Appendix~\ref{app:data}. Columns (3)--(5) concern municipalities with both rural and urban respondents: their count, the number of mean primary-index gaps within $10^{-10}$ of zero, and the median absolute rural-minus-urban gap. Means are unweighted and all waves use the 2011 rural-comparison outcome scale; municipal support varies. These diagnostics do not identify the reason for repeated responses.\end{minipage}\par}\end{table}

\begin{table}[H]{\renewcommand{\arraystretch}{0.95}\setlength{\tabcolsep}{2pt}
\captionsetup{font={normalsize,md}}
\caption{Changing the Reference Wave and Excluding 2017}\label{tab:reference2011}\vspace{-0.3cm}
\begin{tablecontent}\begin{adjustbox}{max width=0.98\linewidth, max totalheight=0.78\textheight}
\begin{tabular}{L{5.2cm}*{3}{C{3.1cm}}}

\hline\hline \addlinespace[0.15cm]
 & (1) & (2) & (3) \\
 & 2011 reference, all waves & 2011 reference, omit 2017 & Pooled 2005--2011 reference \\
\addlinespace[0.10cm]\hline\addlinespace[0.15cm]
\emph{Panel A:} & \multicolumn{3}{c}{Dependent variable: Primary index} \\
\cmidrule(lr){2-4}
Rural $\times$ Covered $\times$ Post & 0.596*** & 0.625*** & 0.587*** \\
 & (0.221) & (0.227) & (0.198) \\
 & [0.154, 1.038] & [0.172, 1.078] & [0.192, 0.981] \\
Pre-trend $p$ & --- & 0.686 & --- \\
R-squared & 0.171 & 0.158 & 0.158 \\
Observations & 12,466 & 9,278 & 9,278 \\
\addlinespace[0.20cm]
\emph{Panel B:} & \multicolumn{3}{c}{Dependent variable: Horizontal coordination} \\
\cmidrule(lr){2-4}
Rural $\times$ Covered $\times$ Post & 0.547*** & 0.582*** & 0.577*** \\
 & (0.184) & (0.188) & (0.159) \\
 & [0.180, 0.914] & [0.208, 0.957] & [0.258, 0.895] \\
Pre-trend $p$ & --- & 0.994 & --- \\
R-squared & 0.127 & 0.117 & 0.117 \\
Observations & 12,466 & 9,278 & 9,278 \\
\addlinespace[0.20cm]
\emph{Panel C:} & \multicolumn{3}{c}{Dependent variable: Civic participation} \\
\cmidrule(lr){2-4}
Rural $\times$ Covered $\times$ Post & 0.261 & 0.253 & 0.231 \\
 & (0.190) & (0.192) & (0.171) \\
 & [-0.118, 0.641] & [-0.131, 0.636] & [-0.111, 0.573] \\
Pre-trend $p$ & --- & 0.764 & --- \\
R-squared & 0.151 & 0.173 & 0.173 \\
Observations & 12,466 & 9,278 & 9,278 \\
\addlinespace[0.20cm]
\emph{Panel D:} & \multicolumn{3}{c}{Dependent variable: Party linkage} \\
\cmidrule(lr){2-4}
Rural $\times$ Covered $\times$ Post & 0.441* & 0.474** & 0.422* \\
 & (0.224) & (0.230) & (0.232) \\
 & [-0.007, 0.889] & [0.014, 0.935] & [-0.041, 0.885] \\
Pre-trend $p$ & --- & 0.543 & --- \\
R-squared & 0.129 & 0.102 & 0.102 \\
Observations & 12,466 & 9,278 & 9,278 \\
\addlinespace[0.10cm]\hline\addlinespace[0.15cm]
2017 observations & Included & Excluded & Excluded \\
\addlinespace[0.15cm]\hline\hline

\end{tabular}\end{adjustbox}\end{tablecontent}\tablenotesgap
\centering\begin{minipage}{\exhibittablewidth}\tablenotefont\textbf{Notes:} Column (1) retains all four analysis waves and omits 2011. Column (2) omits 2011 and excludes every 2017 record. Column (3) compares 2023 with pooled 2005 and 2011, also excluding 2017. Pre-period tests in column (2) concern the 2005 coefficient alone; pooled columns have no pre-period test. All regressions include municipality-by-wave and municipality-by-residence fixed effects and respondent sex. Outcomes retain their 2011 rural-comparison standardization. Standard errors are clustered by municipality. *, **, *** denote conventional significance at 10\%, 5\% and 1\%; these tests do not jointly adjust for outcome selection.\end{minipage}\par}\end{table}

\begin{table}[H]{\renewcommand{\arraystretch}{0.95}\setlength{\tabcolsep}{2pt}
\captionsetup{font={normalsize,md}}
\caption{Parallel-Trends Sensitivity with 2011 as Reference}\label{tab:clean_honest}\vspace{-0.3cm}
\begin{tablecontent}\begin{adjustbox}{max width=0.98\linewidth, max totalheight=0.78\textheight}
\begin{tabular}{L{5.3cm}*{2}{C{4.5cm}}}

\hline\hline \addlinespace[0.15cm]
 & (1) & (2) \\
 & Lower bound & Upper bound \\
\addlinespace[0.10cm]\hline\addlinespace[0.15cm]
$\bar M=0.0$ & 0.176 & 1.074 \\
$\bar M=0.5$ & 0.135 & 1.170 \\
$\bar M=1.0$ & 0.075 & 1.322 \\
$\bar M=1.5$ & -0.023 & 1.515 \\
$\bar M=2.0$ & -0.132 & 1.710 \\
$\bar M=2.5$ & -0.318 & 1.909 \\
$\bar M=3.0$ & -0.508 & 2.112 \\
\addlinespace[0.15cm]\hline\hline

\end{tabular}\end{adjustbox}\end{tablecontent}\tablenotesgap
\centering\begin{minipage}{\exhibittablewidth}\tablenotefont\textbf{Notes:} 95\% C-LF sensitivity intervals for the primary index from the no-2017 regression, with the 2005 coefficient as the single pre-period differential and 2023 as the post-period coefficient. Relative-magnitudes restrictions bound post-period deviations by $\bar M$ times the pre-period change. The intervals incorporate coefficient uncertainty, but are conditional on the fixed index definition. Wave intervals have unequal calendar length, as explained in the text.\end{minipage}\par}\end{table}

\begin{table}[H]{\renewcommand{\arraystretch}{0.95}\setlength{\tabcolsep}{3pt}
\captionsetup{font={normalsize,md}}
\caption{Concurrent Rural Exposure: Baseline Reference}\label{tab:shocks2017}\vspace{-0.3cm}
\begin{tablecontent}\begin{adjustbox}{max width=0.98\linewidth, max totalheight=0.82\textheight}
\begin{tabular}{L{5.3cm}*{4}{C{2.3cm}}}

\hline\hline \addlinespace[0.15cm]
 & (1) & (2) & (3) & (4) \\
 & Unadjusted & Exclude early ETCR & Coca reporting status & Recorded coca and reporting status \\
\addlinespace[0.10cm]\hline\addlinespace[0.15cm]
\emph{Panel A:} & \multicolumn{4}{c}{Dependent variable: Primary index} \\
\cmidrule(lr){2-5}
Rural $\times$ Covered $\times$ Post & 0.582*** & 0.510** & 0.580** & 0.581** \\
 & (0.198) & (0.201) & (0.237) & (0.241) \\
 & [0.187, 0.976] & [0.107, 0.913] & [0.106, 1.054] & [0.100, 1.061] \\
Pre-trend $p$ & 0.921 & 0.908 & 0.654 & 0.638 \\
R-squared & 0.171 & 0.163 & 0.171 & 0.171 \\
Observations & 12,466 & 11,466 & 12,466 & 12,466 \\
\addlinespace[0.20cm]
\emph{Panel B:} & \multicolumn{4}{c}{Dependent variable: Horizontal coordination} \\
\cmidrule(lr){2-5}
Rural $\times$ Covered $\times$ Post & 0.445*** & 0.361** & 0.444*** & 0.402** \\
 & (0.160) & (0.170) & (0.155) & (0.160) \\
 & [0.126, 0.764] & [0.021, 0.702] & [0.135, 0.752] & [0.083, 0.722] \\
Pre-trend $p$ & 0.555 & 0.666 & 0.657 & 0.747 \\
R-squared & 0.127 & 0.122 & 0.127 & 0.128 \\
Observations & 12,466 & 11,466 & 12,466 & 12,466 \\
\addlinespace[0.20cm]
\emph{Panel C:} & \multicolumn{4}{c}{Dependent variable: Civic participation} \\
\cmidrule(lr){2-5}
Rural $\times$ Covered $\times$ Post & 0.328* & 0.331* & 0.331* & 0.361* \\
 & (0.178) & (0.186) & (0.191) & (0.191) \\
 & [-0.027, 0.683] & [-0.040, 0.703] & [-0.050, 0.712] & [-0.020, 0.742] \\
Pre-trend $p$ & 0.752 & 0.749 & 0.967 & 0.947 \\
R-squared & 0.151 & 0.144 & 0.151 & 0.151 \\
Observations & 12,466 & 11,466 & 12,466 & 12,466 \\
\addlinespace[0.20cm]
\emph{Panel D:} & \multicolumn{4}{c}{Dependent variable: Party linkage} \\
\cmidrule(lr){2-5}
Rural $\times$ Covered $\times$ Post & 0.446** & 0.376** & 0.441 & 0.454 \\
 & (0.197) & (0.182) & (0.284) & (0.287) \\
 & [0.053, 0.838] & [0.011, 0.740] & [-0.127, 1.008] & [-0.119, 1.027] \\
Pre-trend $p$ & 0.775 & 0.902 & 0.344 & 0.311 \\
R-squared & 0.129 & 0.124 & 0.130 & 0.130 \\
Observations & 12,466 & 11,466 & 12,466 & 12,466 \\
\addlinespace[0.10cm]\hline\addlinespace[0.15cm]
\addlinespace[0.15cm]\hline\hline

\end{tabular}\end{adjustbox}\end{tablecontent}\tablenotesgap
\centering\begin{minipage}{\exhibittablewidth}\tablenotefont\textbf{Notes:} Estimates of equation~\eqref{eq:esddd}, normalized on 2017, using 2005--2023. Municipality-by-wave and municipality-by-residence effects and respondent sex throughout. Column (1) is unadjusted; column (2) excludes six municipalities on the 2018 ETCR list; columns (3)--(4) add the pre-2016 coca interactions described in the text. Missing coca observations remain a distinct reporting category, not observed zeros. Pre-trend $p$ tests the 2005 and 2011 differentials jointly. Standard errors clustered by municipality in parentheses; 95\% intervals in brackets. Outcomes use the 2011 rural comparison scale. *, **, *** denote 10\%, 5\% and 1\%.\end{minipage}\par}\end{table}

\begin{table}[H]{\renewcommand{\arraystretch}{0.95}\setlength{\tabcolsep}{3pt}
\captionsetup{font={normalsize,md}}
\caption{Concurrent Rural Exposure: 2011 Reference without 2017}\label{tab:shocks2011}\vspace{-0.3cm}
\begin{tablecontent}\begin{adjustbox}{max width=0.98\linewidth, max totalheight=0.82\textheight}
\begin{tabular}{L{5.3cm}*{4}{C{2.3cm}}}

\hline\hline \addlinespace[0.15cm]
 & (1) & (2) & (3) & (4) \\
 & Unadjusted & Exclude early ETCR & Coca reporting status & Recorded coca and reporting status \\
\addlinespace[0.10cm]\hline\addlinespace[0.15cm]
\emph{Panel A:} & \multicolumn{4}{c}{Dependent variable: Primary index} \\
\cmidrule(lr){2-5}
Rural $\times$ Covered $\times$ Post & 0.625*** & 0.565** & 0.691*** & 0.694*** \\
 & (0.227) & (0.254) & (0.237) & (0.240) \\
 & [0.172, 1.078] & [0.057, 1.073] & [0.217, 1.166] & [0.214, 1.174] \\
Pre-trend $p$ & 0.686 & 0.665 & 0.491 & 0.455 \\
R-squared & 0.158 & 0.150 & 0.158 & 0.158 \\
Observations & 9,278 & 8,528 & 9,278 & 9,278 \\
\addlinespace[0.20cm]
\emph{Panel B:} & \multicolumn{4}{c}{Dependent variable: Horizontal coordination} \\
\cmidrule(lr){2-5}
Rural $\times$ Covered $\times$ Post & 0.582*** & 0.519*** & 0.422*** & 0.388** \\
 & (0.188) & (0.190) & (0.147) & (0.153) \\
 & [0.208, 0.957] & [0.140, 0.899] & [0.129, 0.715] & [0.083, 0.694] \\
Pre-trend $p$ & 0.994 & 0.777 & 0.401 & 0.499 \\
R-squared & 0.117 & 0.113 & 0.117 & 0.118 \\
Observations & 9,278 & 8,528 & 9,278 & 9,278 \\
\addlinespace[0.20cm]
\emph{Panel C:} & \multicolumn{4}{c}{Dependent variable: Civic participation} \\
\cmidrule(lr){2-5}
Rural $\times$ Covered $\times$ Post & 0.253 & 0.289 & 0.366* & 0.405** \\
 & (0.192) & (0.213) & (0.189) & (0.192) \\
 & [-0.131, 0.636] & [-0.137, 0.715] & [-0.011, 0.743] & [0.020, 0.789] \\
Pre-trend $p$ & 0.764 & 0.599 & 0.903 & 0.834 \\
R-squared & 0.173 & 0.164 & 0.173 & 0.174 \\
Observations & 9,278 & 8,528 & 9,278 & 9,278 \\
\addlinespace[0.20cm]
\emph{Panel D:} & \multicolumn{4}{c}{Dependent variable: Party linkage} \\
\cmidrule(lr){2-5}
Rural $\times$ Covered $\times$ Post & 0.474** & 0.375 & 0.660** & 0.662** \\
 & (0.230) & (0.243) & (0.312) & (0.315) \\
 & [0.014, 0.935] & [-0.111, 0.861] & [0.036, 1.284] & [0.031, 1.292] \\
Pre-trend $p$ & 0.543 & 0.780 & 0.401 & 0.310 \\
R-squared & 0.102 & 0.099 & 0.103 & 0.103 \\
Observations & 9,278 & 8,528 & 9,278 & 9,278 \\
\addlinespace[0.10cm]\hline\addlinespace[0.15cm]
\addlinespace[0.15cm]\hline\hline

\end{tabular}\end{adjustbox}\end{tablecontent}\tablenotesgap
\centering\begin{minipage}{\exhibittablewidth}\tablenotefont\textbf{Notes:} Estimates of equation~\eqref{eq:esddd}, normalized on 2011, using 2005, 2011 and 2023 and excluding 2017. Municipality-by-wave and municipality-by-residence effects and respondent sex throughout. Column (1) is unadjusted; column (2) excludes six municipalities on the 2018 ETCR list; columns (3)--(4) add the pre-2016 coca interactions described in the text. Missing coca observations remain a distinct reporting category, not observed zeros. Pre-trend $p$ tests the 2005 differential. Standard errors clustered by municipality in parentheses; 95\% intervals in brackets. Outcomes use the 2011 rural comparison scale. *, **, *** denote 10\%, 5\% and 1\%.\end{minipage}\par}\end{table}

\begin{table}[H]{\renewcommand{\arraystretch}{0.95}\setlength{\tabcolsep}{3pt}
\captionsetup{font={normalsize,md}}
\caption{Magnitude Sensitivity to Bilateral Municipal Trimming}\label{tab:trimming}\vspace{-0.3cm}
\begin{tablecontent}\begin{adjustbox}{max width=0.98\linewidth, max totalheight=0.82\textheight}
\begin{tabular}{L{4.5cm}*{4}{C{2.55cm}}}

\hline\hline \addlinespace[0.15cm]
 & (1) & (2) & (3) & (4) \\
 & Full baseline & Endpoint support & One per tail and group & Two per tail and group \\
\addlinespace[0.10cm]\hline\addlinespace[0.15cm]
\emph{Panel A:} & \multicolumn{4}{c}{Dependent variable: Primary index} \\
\cmidrule(lr){2-5}
Rural $\times$ Covered $\times$ Post & 0.582*** & 0.541** & 0.446*** & 0.434*** \\
 & (0.198) & (0.217) & (0.148) & (0.136) \\
 & [0.187, 0.976] & [0.102, 0.980] & [0.146, 0.746] & [0.156, 0.711] \\
Pre-trend $p$ & 0.921 & 0.925 & 0.563 & 0.766 \\
R-squared & 0.171 & 0.212 & 0.210 & 0.217 \\
Observations & 12,466 & 6,116 & 5,568 & 4,949 \\
\addlinespace[0.20cm]
\emph{Panel B:} & \multicolumn{4}{c}{Dependent variable: Horizontal coordination} \\
\cmidrule(lr){2-5}
Rural $\times$ Covered $\times$ Post & 0.445*** & 0.462*** & 0.357** & 0.345** \\
 & (0.160) & (0.169) & (0.143) & (0.150) \\
 & [0.126, 0.764] & [0.122, 0.803] & [0.067, 0.647] & [0.040, 0.651] \\
Pre-trend $p$ & 0.555 & 0.272 & 0.026 & 0.059 \\
R-squared & 0.127 & 0.156 & 0.147 & 0.149 \\
Observations & 12,466 & 6,116 & 5,568 & 4,949 \\
\addlinespace[0.20cm]
\emph{Panel C:} & \multicolumn{4}{c}{Dependent variable: Civic participation} \\
\cmidrule(lr){2-5}
Rural $\times$ Covered $\times$ Post & 0.328* & 0.280 & 0.299** & 0.361** \\
 & (0.178) & (0.193) & (0.139) & (0.142) \\
 & [-0.027, 0.683] & [-0.111, 0.670] & [0.018, 0.580] & [0.072, 0.651] \\
Pre-trend $p$ & 0.752 & 0.548 & 0.611 & 0.705 \\
R-squared & 0.151 & 0.198 & 0.198 & 0.203 \\
Observations & 12,466 & 6,116 & 5,568 & 4,949 \\
\addlinespace[0.20cm]
\emph{Panel D:} & \multicolumn{4}{c}{Dependent variable: Party linkage} \\
\cmidrule(lr){2-5}
Rural $\times$ Covered $\times$ Post & 0.446** & 0.392* & 0.278 & 0.202 \\
 & (0.197) & (0.213) & (0.180) & (0.165) \\
 & [0.053, 0.838] & [-0.038, 0.822] & [-0.087, 0.643] & [-0.135, 0.538] \\
Pre-trend $p$ & 0.775 & 0.804 & 0.764 & 0.818 \\
R-squared & 0.129 & 0.168 & 0.162 & 0.166 \\
Observations & 12,466 & 6,116 & 5,568 & 4,949 \\
\addlinespace[0.10cm]\hline\addlinespace[0.15cm]
Municipalities (clusters) & 66 & 42 & 38 & 34 \\
Covered municipalities & 22 & 22 & 20 & 18 \\
\addlinespace[0.15cm]\hline\hline

\end{tabular}\end{adjustbox}\end{tablecontent}\tablenotesgap
\centering\begin{minipage}{\exhibittablewidth}\tablenotefont\textbf{Notes:} Equation~\eqref{eq:esddd} with 2017 as reference, municipality-by-wave and municipality-by-residence effects, and sex adjustment. Column (2) requires both residence groups at both endpoints; columns (3)--(4) remove four and eight municipalities, respectively, from this common support. Ranking is based on the primary-index gap change separately within covered and comparison groups. All waves of retained municipalities enter. Outcomes are not re-standardized after selection. Standard errors clustered by municipality in parentheses; conventional 95\% intervals in brackets. Stars and pre-period tests condition on the selected sample and do not adjust for outcome-based trimming. *, **, *** denote 10\%, 5\% and 1\%.\end{minipage}\par}\end{table}

\begin{table}[H]{\renewcommand{\arraystretch}{0.95}\setlength{\tabcolsep}{3pt}
\captionsetup{font={normalsize,md}}
\caption{Reconstruction Following the Original BARCAS Dimension Taxonomy}\label{tab:barcasdimensionmap}\vspace{-0.3cm}
\begin{tablecontent}\begin{adjustbox}{max width=0.98\linewidth, max totalheight=0.82\textheight}
\begin{tabular}{L{3.8cm}L{5.7cm}L{4.7cm}}

\hline\hline
 & (1) & (2) \\
Dimension & Retained raw-scale rule & Common-component restriction \\
\midrule
Solidarity/mutuality & Solidarity + reciprocity + conflict resolution $-$ atomization $-$ opportunism & Common opportunism items; distributed atomization composite. \\
\addlinespace[0.10cm]
Horizontal relations & Horizontal help + work and neighborhood activities + joint problem-solving & External-agent component (AGETEX) omitted throughout; fixed either-mention coding. \\
\addlinespace[0.10cm]
Vertical articulation & Organizational belonging + hierarchical help + vertical problem-solving + church, guild, union and party mediation & Vertical resolution counts representatives positively, individuals and brokers negatively. \\
\addlinespace[0.10cm]
Civic participation & Secular membership + local, civic and media activities + volunteering & Membership restricted to 16 common organization types. \\
\addlinespace[0.10cm]
Political participation & Voting + legislative and executive linkage + political activity + party component & Political skills and mechanism knowledge omitted; common party items only. \\
\addlinespace[0.10cm]
Institutional trust & Law, media, group, legal, political, movement and government trust + education, police and large-company trust $-$ corruption & Separate military, church and social-media trust items omitted; five common government items. \\
\addlinespace[0.10cm]
Social control & Trust in control institutions + accountability & Eight common trust fields; changing control-mechanism component omitted. \\
\addlinespace[0.10cm]
Media & Media trust + media activity + explanations of public problems + newspaper use & Common item fields and published composites. \\
\addlinespace[0.10cm]
Information/transparency & Information from state, citizens, media and government $-$ no source + knowledge of public resources & Six common items, with the producer's signs. \\
\addlinespace[0.10cm]
Civic republicanism & Political education + citizen--politics connection $-$ particularism + public responsibility $-$ clientelism & Five published composites and original signs. \\
\addlinespace[0.10cm]
\hline\hline

\end{tabular}\end{adjustbox}\end{tablecontent}\tablenotesgap
\centering\begin{minipage}{\exhibittablewidth}\tablenotefont\textbf{Notes:} Rules follow the producer's documentation on constructing variables and dimensions. Components are added on their raw scales with the displayed signs; each resulting score is standardized on the 2011 rural comparison distribution. The table summarizes fixed common-component reconstructions, not released official scores or factor estimates. The replication code identifies the exact source sheets, cells, variables, harmonizations and omissions. Simple composites and political-activity and trust components are reconstructed with common content across waves. Other retained published composites inherit their response-coding conventions. Social media is excluded because unavailable in the early waves. Both reference-period specifications use the same reconstructed outcomes.\end{minipage}\par}\end{table}

\begin{table}[H]{\renewcommand{\arraystretch}{0.95}\setlength{\tabcolsep}{2pt}
\captionsetup{font={normalsize,md}}
\caption{Triple Differences Using Reconstructed BARCAS Dimensions}\label{tab:barcasdimensions}\vspace{-0.3cm}
\begin{tablecontent}\begin{adjustbox}{max width=0.98\linewidth, max totalheight=0.82\textheight}
\begin{tabular}{L{3.8cm}C{0.95cm}C{2.3cm}C{0.75cm}C{0.85cm}C{1.15cm}C{1.1cm}C{2.1cm}}

\hline\hline
 & (1) & (2) & (3) & (4) & (5) & (6) & (7) \\
Dimension & $\pi_{2023}$ & 95\% interval & $p$ & \shortstack{Holm\\$p$} & \shortstack{Pre-test\\$p$} & $N$ & \shortstack{Primary index\\$\pi$ ($p$)} \\
\midrule
\multicolumn{8}{l}{\textit{Panel A: 2017 reference (baseline)}} \\
\addlinespace[0.08cm]
Solidarity/mutuality & 0.084 & [-0.188, 0.357] & 0.539 & 1.000 & 0.302 & 12,332 & 0.595 (0.003) \\
Horizontal relations & 0.070 & [-0.275, 0.415] & 0.687 & 1.000 & 0.866 & 12,466 & 0.582 (0.005) \\
Vertical articulation & 0.240 & [-0.033, 0.512] & 0.084 & 0.836 & 0.630 & 12,466 & 0.582 (0.005) \\
Civic participation & 0.401 & [-0.175, 0.977] & 0.169 & 1.000 & 0.246 & 12,466 & 0.582 (0.005) \\
Political participation & 0.142 & [-0.075, 0.359] & 0.196 & 1.000 & 0.944 & 12,466 & 0.582 (0.005) \\
Institutional trust & 0.197 & [-0.117, 0.510] & 0.214 & 1.000 & 0.681 & 12,155 & 0.572 (0.005) \\
Social control & 0.104 & [-0.210, 0.418] & 0.509 & 1.000 & 0.853 & 11,969 & 0.614 (0.004) \\
Media & 0.186 & [-0.049, 0.421] & 0.118 & 1.000 & 0.950 & 12,329 & 0.558 (0.005) \\
Information/transparency & 0.038 & [-0.201, 0.277] & 0.753 & 1.000 & 0.223 & 12,055 & 0.539 (0.003) \\
Civic republicanism & 0.192 & [-0.062, 0.447] & 0.137 & 1.000 & 0.560 & 12,466 & 0.582 (0.005) \\
\addlinespace[0.12cm]
\midrule
\multicolumn{8}{l}{\textit{Panel B: 2011 reference, excluding 2017 (supplementary)}} \\
\addlinespace[0.08cm]
Solidarity/mutuality & -0.030 & [-0.328, 0.268] & 0.842 & 1.000 & 0.963 & 9,144 & 0.643 (0.005) \\
Horizontal relations & 0.117 & [-0.251, 0.484] & 0.528 & 1.000 & 0.522 & 9,278 & 0.625 (0.008) \\
Vertical articulation & 0.158 & [-0.255, 0.570] & 0.448 & 1.000 & 0.996 & 9,278 & 0.625 (0.008) \\
Civic participation & 0.361 & [-0.156, 0.878] & 0.168 & 1.000 & 0.093 & 9,278 & 0.625 (0.008) \\
Political participation & 0.198 & [-0.157, 0.552] & 0.270 & 1.000 & 0.976 & 9,278 & 0.625 (0.008) \\
Institutional trust & 0.341 & [-0.062, 0.744] & 0.096 & 0.958 & 0.431 & 8,967 & 0.629 (0.007) \\
Social control & 0.237 & [-0.148, 0.622] & 0.224 & 1.000 & 0.669 & 8,781 & 0.680 (0.005) \\
Media & 0.215 & [-0.081, 0.511] & 0.152 & 1.000 & 0.901 & 9,141 & 0.605 (0.009) \\
Information/transparency & -0.152 & [-0.524, 0.220] & 0.416 & 1.000 & 0.643 & 8,867 & 0.596 (0.007) \\
Civic republicanism & 0.027 & [-0.355, 0.410] & 0.887 & 1.000 & 0.768 & 9,278 & 0.625 (0.008) \\
\hline\hline

\end{tabular}\end{adjustbox}\end{tablecontent}\tablenotesgap
\centering\begin{minipage}{\exhibittablewidth}\tablenotefont\textbf{Notes:} Each row estimates equation~\eqref{eq:esddd} for the indicated reconstructed dimension, with municipality-by-wave and municipality-by-residence fixed effects and sex adjustment. Panel A retains 2017 as reference; Panel B uses 2011 and excludes all 2017 observations. Coefficients and intervals use each outcome's 2011 rural comparison scale. Conventional intervals and $p$-values use municipality-clustered standard errors (66 clusters in A, 64 in B). Holm adjustment covers all ten dimensions separately within each panel. Pre-period $p$ tests the 2005 and 2011 differentials jointly in A and the 2005 differential in B. Column (7) reports the unchanged primary index on the exact sample used in that row, with its conventional $p$-value; its scale remains the full-sample 2011 rural comparison scale. Missing retained score components are not replaced by zero. No dimension is selected or excluded using its estimated coefficient or pre-period test.\end{minipage}\par}\end{table}

\begin{table}[H]{\renewcommand{\arraystretch}{0.95}\setlength{\tabcolsep}{3pt}
\captionsetup{font={normalsize,md}}
\caption{Aggregate Engagement under Alternative Outcome Constructions}\label{tab:aggregatecomparison}\vspace{-0.3cm}
\begin{tablecontent}\begin{adjustbox}{max width=0.98\linewidth, max totalheight=0.82\textheight}
\begin{tabular}{L{6.2cm}*{4}{C{2.2cm}}}

\hline\hline
 & (1) & (2) & (3) & (4) \\
 & \shortstack{Primary\\index} & \shortstack{Eleven\\measures} & \shortstack{Fourteen\\measures} & \shortstack{Original\\dimensions} \\
\midrule
Rural $\times$ Covered $\times$ 2023 & 0.582 & 0.533 & 0.347 & 0.282 \\
 & (0.198) & (0.196) & (0.187) & (0.187) \\
95\% confidence interval & [0.187, 0.976] & [0.141, 0.925] & [-0.026, 0.720] & [-0.092, 0.656] \\
Common-scale coefficient & 0.582 & 0.541 & 0.350 & 0.284 \\
95\% interval, common scale & [0.187, 0.976] & [0.143, 0.939] & [-0.026, 0.727] & [-0.093, 0.661] \\
Conventional $p$-value & 0.005 & 0.008 & 0.068 & 0.137 \\
Joint pre-period $p$-value & 0.921 & 0.858 & 0.744 & 0.614 \\
$R^2$ & 0.171 & 0.168 & 0.190 & 0.193 \\
Observations & 12,466 & 12,466 & 12,466 & 12,466 \\
Municipalities (clusters) & 66 & 66 & 66 & 66 \\
\midrule
Equal weighting across & Families & Measures & Measures & Dimensions \\
Distinct component measures & 11 & 11 & 14 & 14 \\
Municipality $\times$ wave fixed effects & Yes & Yes & Yes & Yes \\
Municipality $\times$ residence fixed effects & Yes & Yes & Yes & Yes \\
Sex adjustment & Yes & Yes & Yes & Yes \\
\hline\hline

\end{tabular}\end{adjustbox}\end{tablecontent}\tablenotesgap
\centering\begin{minipage}{\exhibittablewidth}\tablenotefont\textbf{Notes:} All columns estimate equation~\eqref{eq:esddd}, with 2017 as reference. Columns (2) and (3) apply the same equal-component rule to different component sets; columns (3) and (4) hold fourteen measures fixed while changing weights. For common scaling, write each raw aggregate as a linear combination of component $z$-scores and normalize its weights to sum to one. Let $Q_k$ denote that weighted mean and $s_0$ the primary weighted mean's standard deviation among 2011 rural comparison respondents. Common-scale outcomes are $(Q_k-\bar Q_{k0})/s_0$; own-scale outcomes use their own reference standard deviations. This fixes calibration, not substantive meaning. Counts refer to constructed measures, not independent questions. Municipality-clustered standard errors appear in parentheses and conventional 95\% intervals in brackets. Scaling leaves $p$-values unchanged. Pre-period $p$ tests both earlier differentials jointly. Inference is pointwise and conditional on the definition, not selection-adjusted.\end{minipage}\par}\end{table}

\begin{table}[H]{\renewcommand{\arraystretch}{0.95}\setlength{\tabcolsep}{3pt}
\captionsetup{font={normalsize,md}}
\caption{Sensitivity to Excluding Outcome Families and Components}\label{tab:outcomeomissions}\vspace{-0.3cm}
\begin{tablecontent}\begin{adjustbox}{max width=0.98\linewidth, max totalheight=0.82\textheight}
\begin{tabular}{L{6.2cm}C{1.45cm}C{1.7cm}C{2.7cm}C{1.0cm}C{1.6cm}}

\hline\hline
 & (1) & (2) & (3) & (4) & (5) \\
Definition / omitted measure & \shortstack{Own SD\\$\pi_{2023}$} & \shortstack{Common\\$\pi_{2023}$} & \shortstack{95\% interval,\\common scale} & $p$ & $N$ \\
\midrule
\multicolumn{6}{l}{\textit{Panel A: Family omitted}} \\
Original index & 0.582 & 0.582 & [0.187, 0.976] & 0.005 & 12,466 \\
Coordination & 0.445 & 0.525 & [0.073, 0.976] & 0.024 & 12,466 \\
Civic participation & 0.604 & 0.747 & [0.262, 1.231] & 0.003 & 12,466 \\
Party linkage & 0.521 & 0.508 & [0.155, 0.860] & 0.005 & 12,466 \\
\midrule
\multicolumn{6}{l}{\textit{Panel B: Constructed component omitted}} \\
Joint resolution & 0.548 & 0.584 & [0.177, 0.991] & 0.006 & 12,466 \\
Conflict-resolution orientation & 0.530 & 0.560 & [0.148, 0.973] & 0.009 & 12,466 \\
Petition disposition & 0.571 & 0.579 & [0.184, 0.973] & 0.005 & 12,466 \\
Local participation & 0.584 & 0.573 & [0.201, 0.945] & 0.003 & 12,466 \\
Media engagement & 0.555 & 0.561 & [0.180, 0.942] & 0.005 & 12,466 \\
Volunteering & 0.588 & 0.599 & [0.182, 1.017] & 0.006 & 12,466 \\
Citizen--politics connection & 0.592 & 0.607 & [0.204, 1.009] & 0.004 & 12,466 \\
Political education & 0.579 & 0.597 & [0.172, 1.022] & 0.007 & 12,466 \\
Membership, trust and activity & 0.581 & 0.565 & [0.182, 0.949] & 0.004 & 12,466 \\
Membership and trust & 0.567 & 0.582 & [0.189, 0.976] & 0.004 & 12,466 \\
Party activity & 0.581 & 0.588 & [0.185, 0.991] & 0.005 & 12,466 \\
\midrule
\multicolumn{6}{l}{\textit{Panel C: Underlying party item omitted: complete-item sample}} \\
Original index, matched sample & 0.575 & 0.575 & [0.178, 0.973] & 0.005 & 12,356 \\
Each party item once & 0.587 & 0.543 & [0.158, 0.929] & 0.006 & 12,356 \\
Party membership & 0.546 & 0.518 & [0.171, 0.865] & 0.004 & 12,356 \\
Trust in parties & 0.579 & 0.564 & [0.138, 0.991] & 0.010 & 12,356 \\
Party activity & 0.579 & 0.537 & [0.150, 0.925] & 0.007 & 12,356 \\
\hline\hline

\end{tabular}\end{adjustbox}\end{tablecontent}\tablenotesgap
\centering\begin{minipage}{\exhibittablewidth}\tablenotefont\textbf{Notes:} Equation~\eqref{eq:esddd}, with unchanged controls, fixed effects and municipality clustering (66 clusters). Panel A averages the two retained families without changing their definitions. Panel B preserves the original component $z$-scores, the two unaffected families and equal family weights; only the affected family is reconstructed and re-standardized. Panel C replaces party composites with their underlying items counted once, then excludes membership, trust or activity. Every Panel C regression uses the same 12,356 complete-item respondents, including the matched original-index benchmark. Reference moments retain the original 2011 rural comparison population and available-item convention; the linear-weight comparison is evaluated on complete-item estimation rows. Column (1) uses each outcome's own reference scale; columns (2) and (3) use the common calibration of Table~\ref{tab:aggregatecomparison}. No variant is screened on its coefficient, significance or pre-trend test. All intervals and $p$-values are conventional and pointwise, not simultaneous or selection-adjusted.\end{minipage}\par}\end{table}

\clearpage
\section{Re-estimation with 2011 as the Reference}\label{app:reference2011}

This appendix distinguishes changing the omitted wave from removing the 2017 observations altogether. Table~\ref{tab:reference2011} reports both comparisons and a pooled 2005--2011 reference. The specification excluding 2017 estimates equation~\eqref{eq:esddd} on 2005, 2011 and 2023, omitting 2011; its only pre-reform differential is therefore 2005 relative to 2011. This is a new regression, not simply subtracting a baseline coefficient. The main article retains 2017 as its baseline.

The primary index remains positive (\BarcasMainCleanB{}, conventional $p=\BarcasMainCleanP$, wild cluster bootstrap $p=\BarcasMainCleanWildP$). Coordination gives \BarcasCoordCleanB{} ($p=\BarcasCoordCleanP$), civic participation \BarcasCivicCleanB{} ($p=\BarcasCivicCleanP$), and party linkage \BarcasPartyCleanB{} ($p=\BarcasPartyCleanP$). Party linkage loses conventional 5 percent significance when the pre-period is pooled instead, and individual party activity is less precise. Table~\ref{tab:clean2011} reports all families, nominating-type comparisons and the relevant individual items. Thus the aggregate and coordination findings are more stable than the stronger behavioral and mechanism interpretations.

With rural and urban observations required in 2023 and at least one retained pre-wave, 19 covered and 24 comparison municipalities provide this direct longitudinal support, compared with 22 and 25 when 2017 is available. Regression cluster counts also include municipalities observed in other cells and should not be read as this identifying support. Table~\ref{tab:clean_cells} uses the same 33-municipality sample as the main descriptive decomposition, so changing the reference does not change its municipalities.

\subsection{Party-item construction}

The baseline party family averages standardized composites with overlapping membership, trust and activity content. Table~\ref{tab:party_once} constructs an alternative family by standardizing \texttt{ovgptido}, \texttt{cfptido} and \texttt{actptid} individually on the same reference group, averaging them once each, and re-standardizing. It then replaces the party family in the primary index. Available-item averaging is accompanied by a common complete-item sample comparison. The primary finding survives this alternative; the individual items are less precisely estimated and should not be described as three separately established responses.

\subsection{Observed sample allocation and municipal support}

Table~\ref{tab:allocation} retains municipalities with both residence groups in both 2011 and 2023. Within each coverage-by-residence group, it weights 2023 respondents by the ratio of their municipality's 2011 sample share to its 2023 share; 2011 observations retain weight one. The unweighted regression uses the identical endpoint sample. The primary estimate changes from \BarcasAllocUnweightedB{} to \BarcasAllocB{}, so observed municipal allocation alone does not account for the contrast. These are sensitivity weights, not producer sampling weights, and do not adjust selection within municipalities.

Table~\ref{tab:balance} describes pre-existing municipal differences. Table~\ref{tab:support} estimates the baseline on the complete-covariate sample and on three restrictions chosen without using outcomes: the overlap in estimated coverage propensity scores, exclusion of departmental capitals, and a pre-agreement rural population share of at least 0.30. These changes test sensitivity to the comparison set; none ensures balance on unobserved counterfactual rural trends.

\subsection{Program-activity inference and influence}

Table~\ref{tab:dose_audit} supplements the main initiative gradients with wild cluster bootstrap inference, leave-one-covered-municipality-out ranges, and a comparison pooling only 2005 and 2011 against 2023. The initiative normalization remains fixed when municipalities are omitted, so the ranges compare like-scaled gradients. Bootstrap tests leave the primary and party gradients imprecise. These checks support reporting positive associations while avoiding independent causal attribution to either mechanism.

\subsection{Parallel-trends sensitivity without 2017}

Table~\ref{tab:clean_honest} recomputes the relative-magnitudes bounds of \citet{rambachanroth2023} from the no-2017 regression and its covariance matrix. The interval remains positive at $\bar M=1$ but includes zero at $\bar M=1.5$. There is only one observed pre-reform change, from 2005 to 2011, and the post comparison spans twelve rather than six years. The restriction is on changes between observed waves, not annual slopes: $\bar M=2$ permits twice the pre-period change over the twice-as-long post interval, and its interval includes zero. These bounds qualify robustness to counterfactual trends; they neither estimate actual violations nor correct sampling changes or index selection.

\section{Reported Activities and Petition Coding}\label{app:completed}

The complementary behavioral index combines memberships and activities, including a petition item coded 0 (would never sign), 5 (might sign) and 10 (has signed). To distinguish reported participation from hypothetical willingness, Table~\ref{tab:behavioral} retains that original index and adds three variants. The first omits petitions. The second replaces the item with 10 for having signed and 0 for either other response, preserving missing values. The third uses only the four completed activities and excludes memberships. Each averages component $z$-scores and is re-standardized on the 2011 rural comparison distribution. These variants complement the primary index and nine-family analysis without changing their definitions. All three estimates remain positive, but none excludes zero at 5 percent, leaving the evidence for broad realized collective activity inconclusive.

\section{Engagement Gaps, Membership and Rural Exposure}\label{app:submissionchecks}

\subsection{Regression decomposition of the rural--urban gap}

Table~\ref{tab:regressiongaps} decomposes the triple difference within the estimated model, preserving its sample, fixed effects, sex adjustment and regression weights. Column (1) reports the comparison-group change in its rural--urban gap, $\rho_{2023}$; column (2) the corresponding covered-group change, $\rho_{2023}+\pi_{2023}$; and column (3) their difference, $\pi_{2023}$. Confidence intervals for the covered-group change use the estimated covariance between the two coefficients.

Against 2011, excluding 2017, the primary gap improves by \BarcasGapCoveredB{} in covered municipalities ($p=\BarcasGapCoveredP$) and deteriorates by \BarcasGapComparisonB{} in comparison municipalities ($p=\BarcasGapComparisonP$). Their difference is \BarcasMainCleanB{}, exactly the primary regression estimate. The latter decline is imprecise at the 5 percent level; this does not conflict with a precisely estimated difference between the two changes. Against 2017, the covered-gap improvement is smaller and imprecise. The decomposition clarifies the relative finding, but neither group-specific change identifies an absolute rural treatment effect: both can reflect rural and urban movements and other changes within their municipalities.

\subsection{Inactive membership and organizational contrasts}

The distributed membership items code nonmembership as zero, inactive membership as five and active membership as ten. Among respondents in the four estimation waves, the share reporting inactive membership in at least one of the nine organization types used in the nominating-category comparison is 16.6 percent in 2005, 12.4 percent in 2011, 38.6 percent in 2017 and 22.8 percent in 2023. This change in the use of the intermediate category motivates a coding sensitivity, not the deletion of respondents on the basis of their answers.

Table~\ref{tab:membershipcoding} compares the original grading with two binary codings applied consistently in every wave: any membership (both five and ten) and active membership only (ten). All components and family indices are standardized anew on their respective 2011 rural comparison distributions, so magnitudes across codings are not measured in identical raw units. The named-types estimate remains positive but its precision changes. In the no-2017 comparison, the active-only estimate rejects at 5 percent while the any-membership estimate does not; the direct named-minus-other contrast does not reject at 5 percent under either coding. The organizational findings therefore do not establish a differential response specific to explicitly named nominating types. The survey also lacks a comparable separate item for victims' organizations, another statutory nominating category.

\subsection{Exposure to reincorporation sites and illicit crops}

Tables~\ref{tab:shocks2017} and~\ref{tab:shocks2011} assess two specific sources of concurrent rural change, using the 2017 baseline and the 2011 reference excluding 2017, respectively. The historical list of Territorial Spaces for Training and Reincorporation (ETCR) in \citet[Annex 1]{minsalud2018etcr} contains 26 sites in 25 municipalities. Six appear in the estimation sample, all covered: Dabeiba, Vig\'ia del Fuerte, Tierralta, Tumaco, Tib\'u and San Jos\'e del Guaviare. Column (2) excludes them. The primary estimate is \BarcasEtcrBaseB{} ($p=\BarcasEtcrBaseP$) with the baseline reference and \BarcasEtcrEarlyB{} ($p=\BarcasEtcrEarlyP$) without 2017. This check limits dependence on the early sites' immediate municipal environments, not on demobilization more broadly or subsequent relocations.

Columns (3) and (4) use the CEDE panel's recorded coca hectares in 2015, before the agreement. The field is nonmissing in 16 covered and four comparison municipalities; missing observations are not interpreted as zero coca cultivation. Column (3) interacts an indicator for missing coca data with rural residence and each non-reference wave. Column (4) additionally interacts $\log(1+\text{recorded hectares})$ with residence and wave. For computation this logarithm is set to zero where the field is missing, with the missing-data interactions included throughout. Thus the specification allows a separate path for municipalities without recorded hectares and variation by quantity among those with records.

The primary estimate in column (4) is \BarcasCocaBaseB{} ($p=\BarcasCocaBaseP$) with the baseline reference and \BarcasCocaEarlyB{} ($p=\BarcasCocaEarlyP$) excluding 2017. Coordination remains positive and conventionally significant throughout; significance for civic participation and party linkage varies across specifications and reference periods. These controls assess sensitivity to measured pre-agreement exposure and reporting status, not the effect of crop substitution. They do not control for post-reform PNIS participation, and cannot rule out later coca-market, security or political shocks.

\section{Index Definition and Robustness of Inference}\label{app:selection}

This appendix assesses how the evidence varies across index definitions and inference benchmarks. It considers all $\binom{9}{3}=84$ equally weighted averages of three of the nine standardized outcome families, including combinations drawn entirely from contrast domains. The baseline specification, sample and reference scaling stay fixed. Conventional inference conditions on the primary index's component set; the broader benchmark allows for choosing among all 84 trios. Re-standardizing an average does not change its $t$-statistic.

Each reassignment fits the baseline model with both sets of fixed effects and reconstructs the municipality-clustered covariance across the nine family coefficients. Fixed-index inference compares the primary index's observed $|t|$ with its own permutation distribution. The index-choice benchmark compares the same statistic with the largest $|t|$ among all 84 averages in each draw. Both use $p=(1+e)/(B+1)$, where $e$ counts draws at least as extreme as the observed statistic, with $B=1{,}999$ and seed 20260915. The primary index has the largest observed $|t|$ in the baseline candidate set. Figure~\ref{fig:indexspec} compares the full set under both time references. The no-2017 panel displays conventional estimates and intervals.

Municipal reassignment holds the covered count fixed within each macro-region. Block reassignment groups covered municipalities by PDET subregion and comparison municipalities by department, intersecting each group with macro-region. The resulting 36 homogeneous groups include 12 covered groups, and reassignment preserves their number within each region. These differ from the 34 unsplit PDET-subregion/department groups used for clustered standard errors. Because actual coverage was not randomized, both schemes rely on conditional exchangeability. All draws have finite statistics; Table~\ref{tab:inference} reports the fixed-index $p$-values.

The 84-trio benchmark gives reference $p$-values of \BarcasSelectionUnit{} under municipal and \BarcasSelectionBlock{} under block reassignment. Both lie below 10 percent and slightly above 5 percent, indicating weaker evidence than fixed-index inference while retaining support at the 10 percent threshold under this broad candidate set. The primary estimate itself is unchanged: the difference comes from comparing its statistic with the most extreme statistic across the 84 alternatives. These calculations depend on the specified candidate set and the design and exchangeability assumptions; they adjust the reference test, not the point estimate, for index choice. Full-model statistics agree with independent Stata regressions.

Broader outcome definitions provide complementary evidence on the positive differential. Table~\ref{tab:altindices} expands the primary index in order of the outcomes' substantive connection to the mechanisms: formal collective action, then organizational membership, then every family except brokerage and clientelism, whose orientation is ambiguous, and finally all nine families. Estimates remain positive throughout, with smaller magnitudes and lower precision as contrast domains enter. These indices retain the primary trio, and formal action and membership also overlap with each other and with civic participation, so they assess sensitivity to outcome coverage rather than provide independent tests of index choice.

Family-level corrections in Table~\ref{tab:multiplicity} likewise show where the evidence is strongest. Coordination remains significant at 5 percent within both the three-family and five-family collective-engagement blocks under either reassignment scheme. The correction across all nine families is more demanding, with no family reaching the 5 percent threshold. Together, the exercises distinguish a stable positive engagement pattern and comparatively strong coordination evidence from claims about all domains of social capital.

\section{Additional Evidence on the Mechanisms}\label{app:attribution}

The main text relates engagement to adopted program initiatives and compares membership between organization types explicitly named as constituency nominators and other secular types. This appendix examines variation across program pillars and membership components, sensitivity to municipal coverage, and an additional comparison of legislative and executive linkage.

The initiative measure counts records of adopted proposals in the ART release. The source identifier is rounded. Removing exact duplicate records gives a primary gradient of \BarcasDedupB{} ($p=\BarcasDedupP$), similar to the original count (Table~\ref{tab:dose_audit}). The replication documentation records the deduplication rules. Organizational capacity may affect both proposals and engagement. Municipalities that adopted many proposals can also differ in electoral mobilization or implementation \citep{bachmann2025}. Table~\ref{tab:dosepillar} reports pillar comparisons.

Larger or more rural municipalities may adopt more initiatives independently of participatory mobilization. Adding rural-by-post interactions with population and rurality measured in 2015 leaves the primary gradient nearly unchanged, from \BarcasMainDoseB{} to \BarcasDoseSizePrimaryB{}, as well as the party gradient, from \BarcasPartyDoseB{} to \BarcasDoseSizePartyB{}. All \BarcasDoseSizeClusters{} covered municipalities and \BarcasDoseSizeN{} respondents remain in both models; initiative normalization is unchanged. The association thus survives these observed size controls, with bootstrap $p$-values of \BarcasDoseSizePrimaryWildP{} and \BarcasDoseSizePartyWildP{}.

Relating membership in the statutory nominating types to total initiatives gives a gradient of 0.115 ($p=0.489$; Table~\ref{tab:dose}, column 5). Its wide interval leaves the relationship with program activity unresolved, so this comparison provides little leverage for separating the two instruments.

The membership estimate is concentrated in the ethnic-organization membership component, which gives 0.985, compared with 0.210 for peasant associations and 0.299 for women's organizations. The latter two alone give 0.326 ($p=0.182$). The direct named-minus-other-types difference is 0.466, with conventional $p=0.099$ and permutation $p=0.113$. Requiring rural respondents in both 2011 and 2023 reduces the named-types estimate to 0.296 ($p=0.261$); this restriction is not a balanced four-wave panel. Excluding departments with concentrated ethnic authorities gives 0.442 ($p=0.167$). The 2011-reference comparison without 2017 also weakens the evidence (Table~\ref{tab:clean2011}). Because Indigenous and Afro-Colombian organizations also participate in program assemblies, their membership can respond to either instrument.

The legislative-minus-executive linkage contrast is 0.237 ($p=0.053$), with a pre-period differential ($p=0.018$). Pooling earlier waves gives 0.048 ($p=0.683$). These results do not support an attribution to legislative linkage.

\section{Parallel-Trends Sensitivity Across All Families}\label{app:allsensitivity}

The sensitivity exercise in Table~\ref{tab:allsensitivity} bounds post-reform departures from parallel trends relative to the largest pre-reform wave-to-wave change in each outcome's differential. It applies this relative-magnitudes restriction to every family and the primary index, without screening on pre-trend tests. Each interval conditions on its outcome definition; the exercise does not jointly adjust for index selection and departures from parallel trends.

\section{Age and Municipal-Extreme Diagnostics}

\subsection{Age adjustment as a composition diagnostic}\label{app:age}

Age is recorded in completed years under different names across releases: \texttt{edad} in the earlier waves and \texttt{a\~nos}/\texttt{annos} in the later ones. Table~\ref{tab:agediagnostic} compares sex adjustment alone with the addition of age and its interaction with rural residence over all four analysis waves, holding the sample fixed and retaining 2017 as reference. The primary estimate changes from \BarcasAgeSexB{} to \BarcasAgeAdjustedB{} ($p=\BarcasAgeFullP$); the joint pre-period test gives $p=\BarcasAgeFullPreP$. The similar magnitude and positive sign support stability to observed age composition.

The median municipality has only \BarcasAgeDistinctSeventeen{} distinct recorded ages in 2017, compared with \BarcasAgeDistinctTwentyThree{} in 2023. The adjustment uses the recorded ages without inferring demographic continuity from this concentration. It addresses observed age differences but cannot remove unobserved household or settlement turnover. No municipalities are classified as having a particular questionnaire version without producer documentation.

\subsection{Sensitivity to bilateral municipal trimming}\label{app:trimming}

Table~\ref{tab:trimming} assesses the influence of extreme municipal changes using the current outcome construction and the baseline reference of 2017. Column (1) reproduces the full baseline. Column (2) retains the 42 municipalities, including all 22 covered municipalities, with both rural and urban respondents in both 2017 and 2023. Within each coverage group, municipalities are ranked by the change in their unadjusted mean primary-index rural--urban gap between those dates. Column (3) excludes one municipality from each tail in each group; column (4) excludes two. Ties are broken by municipal code. Each retained municipality contributes all its available waves, and the same municipality selection is used for all four outcomes. The outcome scale remains the full-sample 2011 rural comparison scale.

The primary estimate is \BarcasTrimSupportB{} on the endpoint-support sample, \BarcasTrimOneB{} after the first trim and \BarcasTrimTwoB{} after the second. Thus the positive aggregate differential is not confined to the most extreme changes, although its magnitude falls under these restrictions. These are outcome-selected populations, not estimates of an unchanged target population under a superior estimator. The reported standard errors and tests condition on the selected sample and do not account for the trimming rule. In particular, the coordination pre-period test rejects at 5 percent after the first trim, and party linkage becomes imprecise. The exercise therefore diagnoses sensitivity rather than strengthening all of the identification evidence.

\section{The Municipal Difference-in-Differences}\label{app:did}

A municipal difference-in-differences compares changes in engagement between covered and comparison municipalities, either pooling residence groups or estimating them separately. Unlike the baseline triple difference, it does not remove municipal shocks shared by rural and urban residents. Its event-study specification, normalized on 2017, is
\begin{equation}\label{eq:did}
Y_{imt}=\sum_{\tau\neq 2017}\beta_{\tau}\bigl(T_m\times\mathbf 1\{t=\tau\}\bigr)+\lambda T_m+\mu_t+\kappa_{r(m)}+X_i'\theta+u_{imt},
\end{equation}
where $T_m$ indicates constituency coverage, $\mu_t$ are wave effects, $\kappa_{r(m)}$ are effects for municipality $m$'s macro-region, and $X_i$ contains respondent sex. The coefficient $\lambda$ captures a persistent coverage-group difference, while $\beta_\tau$ measures the differential change relative to 2017; $u_{imt}$ is the error term. Tables~\ref{tab:did1} and \ref{tab:did2} estimate the model on all respondents and separately on rural and urban respondents.

Pooling respondents, the primary index and party linkage show differential pre-reform trajectories (joint $p=0.011$ and $p<0.001$), raising concern about the parallel municipal trends required for causal interpretation. The primary-index post-reform coefficient is 0.300 in the pooled sample, 0.529 among rural respondents and 0.210 among urban respondents. Separate estimates therefore reveal different residence-group changes, but their causal interpretation still requires parallel municipal trends within each group; the urban primary index also shows differential pre-reform movement ($p=0.001$). These comparisons motivate removing common municipal changes, rather than claiming that the triple difference is preferable merely because its coefficient is larger.

\clearpage
\section{Data Appendix}\label{app:data}

Combining the survey waves requires harmonizing municipal identifiers, rural--urban coding and the collective-problems module. This appendix documents these corrections and their validation, distinguishes unresolved source-release and sampling questions from corrected coding, and explains the treatment of the sparsely sampled 1997 wave and macro-region codes.

\subsection{The municipal identifier changes format}\label{app:mnp}
Municipal identifiers use two formats across releases. From 2017 onward, \texttt{mnp} holds the five-digit DANE code; in 1997--2011, it holds the final three digits, with the department recorded in \texttt{dpto}. Concatenating the earlier fields recovers the code and identifies covered municipalities in the historical waves. The recovered codes are consistent with the sampling frame: 98 percent of municipalities sampled in 2005 and 100 percent of those in 2011 reappear in 2017 or 2023 (96 and 98 percent reappear in 2017 alone).

\subsection{The urban--rural code reverses in 2023}\label{app:area}
All five dictionaries document \texttt{area} as 1 = \emph{cabecera} (urban), 2 = \emph{centro poblado} (rural). The data follow that only in 2023. Under the documented coding, 100.0, 99.7, 100.0 and 100.0 percent of the Bogot\'a sample is rural in 1997, 2005, 2011 and 2017, against 0.6 percent in 2023, and respondents coded rural live in municipalities that are \emph{less} rural on the CEDE index than those coded urban. Under the coding used here---\texttt{area}~$=1$ before 2023, \texttt{area}~$=2$ in 2023---Bogot\'a is 0.0 to 0.6 percent rural in every wave, the CEDE index orders correctly (0.32--0.50 rural against 0.07--0.16 urban), and the municipality-level share classified rural correlates with that index at 0.863, 0.893, 0.975, 0.948 and 0.912. Taking the documentation at face value would select an urban sample before the reform and a rural one after it.

Municipal poverty data from 2011 provide a further check on residence coding. Unsatisfied basic needs range from 44.6 to 51.1 for respondents with \texttt{area}~$=1$, compared with 19.9 to 27.2 for those with \texttt{area}~$=2$ in 1997--2017; the ordering reverses in 2023, to 27.9 against 52.2. These differences align with the rurality check and the predominantly urban Bogot\'a sample under the adopted coding. Table~\ref{tab:areaval} reports the wave-specific diagnostics.

These checks support the adopted wave-level coding, which defines the rural--urban comparison throughout the paper. They are less informative about consistency within each wave: some fieldwork teams could have applied a different convention. The municipality-level correlations with CEDE rurality, ranging from 0.89 to 0.98 in the analysis waves, provide additional evidence of aggregate consistency but do not bound or rule out within-wave miscoding.

\subsection{Comparable measures and collective-problem responses}\label{app:module}
Political activity is reconstructed from discussion, demonstrations and seeking or holding elective office, excluding the ever-voted item added in later releases. Media trust retains press and television; legal trust retains the legal system and prosecutor; control-institution trust retains eight common items. Radio, courts and social networks added to the published aggregates in later waves are excluded from these fixed-content measures. These reconstructions also enter the alternative family and producer-taxonomy comparisons.

Nine simple composites are reconstructed from available survey items. The 2017 source comparison flags five exceptional records. Excluding these records gives a primary estimate of \BarcasSourceExcludedB{}. The compound files lack individual identifiers, which limits reconstruction of the original record linkage. The replication documentation provides the component-level calculations and release checks.

The collective-problems module records multiple alternatives, but their storage changes across waves. In the early releases, several alternatives can simultaneously receive 10; a first-mentioned alternative cannot be recovered consistently. The 2017 release adds suffixed mention fields and signed base scores, while 2023 distinguishes mention slots and sometimes codes the second mention as 5. The harmonized indicator therefore records whether an alternative is mentioned at all, using every available positive mention field and recoding it to 10. It does not assign an order to early-wave alternatives. This rule is used for joint resolution, recourse to representatives and personal connections; the last also enters opportunism. The harmonized indicators use the suffix \texttt{\_f} to denote any mention.

Table~\ref{tab:harmonization} contrasts the harmonized families with diagnostic alternatives that retain mixed mention coding or non-comparable producer composites. They are not alternative estimates of an identically measured outcome. Coordination under the harmonized construction is \BarcasCoordB{} in the baseline and \BarcasCoordCleanB{} with 2011 as the reference and all 2017 observations removed. The corresponding primary-index estimates are \BarcasMainB{} and \BarcasMainCleanB{}. Thus the positive aggregate finding does not depend on either interpreting early responses as first mentions or retaining 2017.

Figure~\ref{fig:cells_allwaves} displays low levels in 2017 across all four coverage-by-residence groups. These cannot be interpreted as a substantive decline and recovery without resolving that release's processing history. Municipality-by-wave effects remove measurement shifts common to rural and urban respondents, but not differential shifts associated with coverage. Regional rural paths address shared regional differences, not every municipality-specific error. The source audit below and the re-estimation in Appendix~\ref{app:reference2011} are therefore distinct from routine covariate adjustment.

\subsection{Unresolved features of the 2017 source release}\label{app:sourceaudit}
The source-item files, before any merge or index construction, contain unusually repeated response profiles in 2017. Table~\ref{tab:sourceaudit} compares the same 98 sufficiently observed, varying numeric items across the four analysis waves, excluding geographic and demographic identifiers. In 2017, 1,219 of 3,188 rows belong to repeated profiles; the corresponding counts are zero in 2005, four in 2011 and zero in 2023. Using a broader, wave-specific set of 361 items still gives 1,004 repeated rows in 2017. These counts include every row belonging to a repeated profile, not only copies beyond the first, and do not establish that the same person was interviewed twice.

The harmonized joint-resolution indicator is also constant within 41 of the 62 sampled municipalities in 2017, including 18 of the 22 covered municipalities; no municipality has a constant indicator in the other analysis waves. Repeated profiles and low within-municipality variation warrant clarification of the distributed file's processing history. They do not by themselves distinguish imputation, recoding, anonymization, aggregation or another explanation. The code does not delete observations by response similarity or claim to reconstruct undistributed original responses. Instead, it authenticates the releases and reports estimates excluding the entire 2017 wave.

The rural--urban comparison makes cross-residence repetition especially relevant. Table~\ref{tab:sourcecross} requires an exact profile match in the other residence group within the same municipality. Using the common raw items, \BarcasCrossRawN{} respondents in 2017 (\BarcasCrossRawPct{} percent of that wave) satisfy this criterion, compared with none in the other analysis waves. Using instead the vector of the three current outcome families gives \BarcasCrossFamilyPct{} percent in 2017, compared with at most 1.2 percent elsewhere. These statistics count respondents, not distinct profiles, and the denominators include the entire wave rather than a selected group of municipalities. Matching a lower-dimensional family vector is less demanding than matching the original item vector.

The same table distinguishes repeated responses from identical group means. Among municipalities with both residence groups, the median absolute primary-index gap is lower in 2017, but none of its municipal gaps is zero. The outcome scale is held fixed across waves; the municipalities observed in both residence groups are not identical across waves. These are descriptive diagnostics, not estimates of compression caused by a particular processing procedure. They support caution about the baseline's apparent stability without implying that its earlier-wave coefficients must be zero by construction.

\subsection{The 2023 national and CITREP samples}\label{app:sampling2023}
The producer describes a dedicated CITREP sample for the first time in 2023: 1,125 interviews, comprising 616 urban and 509 rural interviews. Of these, 700 (420 urban and 280 rural) belong to the national sample, and 425 are additional \citep[pp.~17--19, Table~11]{barcascontexto}. The distributed file used here has 3,225 records, matching the documented national-sample total; all are retained. Applying statutory municipal coverage identifies 856 respondents in 22 covered municipalities, including 338 rural respondents, rather than the 700 interviews classified as national-and-CITREP in the producer's table. The available identifiers do not reconcile these counts. They establish neither inclusion of the 425 additional interviews nor their exclusion through a pre-period municipal-support restriction; no such restriction removes 2023 records.

The available identifiers do not resolve membership in the overlapping national and dedicated samples, continuity of sampled settlements, inclusion probabilities or replacement protocols. A producer crosswalk and fieldwork documentation are needed to settle these questions. The analysis therefore does not remove respondents based on an inferred frame assignment. Instead, Table~\ref{tab:allocation} holds observed municipal allocation constant within coverage and residence groups. This checks a specific source of composition change without ruling out treatment-correlated selection within those cells.

The source files contain education fields in all four analysis waves, but their variables and coding change: \texttt{educaci\'on} in 2005--2011 and \texttt{Educ}, \texttt{Nivel1} and \texttt{Nivel2} in 2017--2023. A demographic ethnicity field, \texttt{etnia}, appears only in the latter two waves. Each dictionary labels \texttt{Lcd}/\texttt{lcd} as \emph{Localidad}, without establishing a stable cross-wave settlement or primary-sampling-unit identifier. These fields offer possible additional composition checks, but their availability alone does not establish comparable education categories or continuity of rural sampling points. Further harmonization requires the producer's documentation.

\vspace{-\baselineskip}
\subsection{The 1997 wave}\label{app:1997}
The main analysis begins in 2005 because the 1997 wave covers substantially fewer constituency municipalities (Section~\ref{sec:wave1997}). Including it, with its own interaction, gives a primary-index estimate of \BarcasWaveNinetySevenB{} rather than \BarcasMainB{} (Table~\ref{tab:anchors}, columns 5 and 6). The longer pre-period changes the visible earlier path without materially changing the post-reform estimate, but cannot distinguish coverage and measurement differences from substantive change.

\vspace{-\baselineskip}
\subsection{Macro-region codes}\label{app:region}
Seven municipalities have inconsistent respondent-level macro-region codes; six discrepancies involve only 1997. Quibd\'o changes from 6 in 2005 and 2011 to 2 in 2017 and 2023. Code 2 matches the other Choc\'o municipalities sampled after 1997; code 6 otherwise appears only in Putumayo and Guaviare after that wave. I assign each municipality its latest observed code. This affects permutation strata and blocks, municipal DiD controls, reweighting and regional rural paths. All reported estimates and permutation distributions use this common assignment.

\section{Sensitivity to the Original BARCAS Dimensions}\label{app:barcasdimensions}

The paper's nine outcome families organize measures around its conceptual question; they are not the producer's original dimensions. This appendix follows the BARCAS classification \citep{sudarsky2022methodological,barcascontexto}. The producer's documentation lists the component hierarchy and signs for ten original dimensions, with social media added later. The available releases do not permit an identical full reconstruction in every wave, so Table~\ref{tab:barcasdimensionmap} defines fixed common-component reconstructions. Social media is excluded because unavailable before 2017. These are neither released official dimension scores nor re-estimated factor scores.

Retained components are added on their raw scales with the producer's signs. Published composites are retained where comparable, with harmonizations for collective-problem resolution, secular membership, party items, political activity and institutional trust. Nine simple composites are rebuilt directly from their questionnaire items. Government and control-institution trust, information items and the remaining media items are recovered from the original item files. The source crosswalk records omissions and substitutions. The exercise inherits published composites' response coding, without independently validating every underlying item. Each score is standardized on its 2011 rural comparison distribution. All retained score components must be observed; missing is not replaced by zero.

Table~\ref{tab:barcasdimensions}, Panel A, retains 2017 as reference and the baseline specification, reporting all ten dimensions with conventional and Holm-adjusted inference. Missing components reduce some samples; the last column re-estimates the unchanged primary index on each dimension's exact sample. The supplementary Panel B uses 2011 as reference and excludes 2017, without changing outcome definitions or scaling.

The baseline differentials are positive throughout, but none rejects zero at 5 percent, even before multiplicity adjustment. Reconstructed civic participation is \BarcasDimCivicB{} ($p=\BarcasDimCivicP$), political participation \BarcasDimPoliticalB{} ($p=\BarcasDimPoliticalP$), and horizontal relationships \BarcasDimHorizontalB{} ($p=\BarcasDimHorizontalP$). These constructs are broader than the paper's similarly named families: civic participation includes membership, political participation includes voting and legislator links, and horizontal relationships includes mutual help and workplace and neighborhood activities. The primary estimate remains positive and conventionally significant on every matching sample, ranging from \BarcasDimPrimaryMin{} to \BarcasDimPrimaryMax{}, so sample loss alone does not account for the contrast. The supplementary estimates likewise do not establish effects at 5 percent. Because these outcomes change content as well as weights, the aggregate comparison below examines both changes directly.

\subsection{Comparing aggregate indices}\label{app:aggregatecomparison}

Table~\ref{tab:aggregatecomparison} compares four aggregates on the identical baseline sample and specification, with 2017 as reference. Columns (1) and (2) retain the primary index's eleven measures, weighting families or measures equally. Column (3) equally averages the fourteen standardized measures underlying the reconstructed horizontal, civic and political dimensions; column (4) averages those three standardized dimensions. The eleven- and fourteen-measure sets are not nested: six constructed measures are common, and party content overlaps through different composites. Own-scale rows standardize each aggregate separately; common-scale rows hold the calibration fixed as described in the notes.

Direct averaging preserves the eleven-measure result (\BarcasAggregateEqualElevenB{}, $p=\BarcasAggregateEqualElevenP$). With uniform measure weights, the fourteen-measure estimate is \BarcasAggregateEqualFourteenB{} ($p=\BarcasAggregateEqualFourteenP$); dimension weighting gives \BarcasAggregateOriginalTrioB{} ($p=\BarcasAggregateOriginalTrioP$). On the common scale, changing the component set moves the point estimate from \BarcasAggregateEqualElevenCommonB{} to \BarcasAggregateEqualFourteenCommonB{}, and dimension weighting moves it to \BarcasAggregateOriginalTrioCommonB{}. Thus both content and weights matter for these point estimates; the comparison neither establishes statistically different effects across constructs nor ranks their validity.

\subsection{Excluding families and components in turn}\label{app:outcomeomissions}

Table~\ref{tab:outcomeomissions} excludes each family (Panel A) and component (Panel B) in turn, retaining 2017 as reference. Every exclusion retains a positive conventional interval: common-scale estimates range from \BarcasOmitFamilyMin{} to \BarcasOmitFamilyMax{} across families and from \BarcasOmitComponentMin{} to \BarcasOmitComponentMax{} across components. The underlying party-item exclusions in Panel C likewise retain positive intervals. Within these perturbations, the finding does not depend on a single family or listed component; inference remains conditional on each definition.

\clearpage
\section{Survey Response and Measurement}\label{app:novmeasurement}

Table~\ref{tab:novmeasurement} uses both reference choices to distinguish changes in particular responses from changes in survey response patterns. The total number of alternatives named in the collective-problems module changes little, while the estimate for alternatives other than joint resolution becomes more negative when 2017 is excluded. Joint resolution is mentioned more often in the relative rural comparison, with some substitution away from other answers.

The diagnostics use eight institutional-trust items and four conflict-resolution items with comparable response categories across waves. The questionnaire files group \emph{do not know} and nonresponse under the same missing code. The released earlier waves record no missing answers for these common batteries. For trust, the estimate for missing responses is imprecise. For conflict resolution, the missing share falls by about four percentage points in the relative rural comparison. Estimates for giving identical answers to all items in a battery are imprecise.

The conflict-resolution component follows the released scoring convention, which assigns zero to 111 responses recorded as missing in 2023. Restricting to observed responses moves the baseline standardized differential from \NovConflictItemBase{} to \NovConflictObservedBase{}, and the 2011-reference differential from \NovConflictItemEarlier{} to \NovConflictObservedEarlier{}. Differences in earlier releases' handling of item nonresponse limit what can be inferred about response behavior across waves.

\begin{table}[H]{\renewcommand{\arraystretch}{0.90}\setlength{\tabcolsep}{4pt}
\captionsetup{font={normalsize,md}}
\caption{Response Patterns and Survey Measurement}\label{tab:novmeasurement}\vspace{-0.3cm}
\begin{tablecontent}\begin{adjustbox}{max width=0.98\linewidth, max totalheight=0.80\textheight}
\begin{tabular}{L{4.8cm}C{2.96cm}C{0.86cm}C{2.96cm}C{0.86cm}}

\toprule
Outcome & \multicolumn{2}{c}{2017 reference} & \multicolumn{2}{c}{2011 reference, omitting 2017} \\
\cmidrule(lr){2-3}\cmidrule(lr){4-5}
 & Estimate [95\% CI] & \(p\) & Estimate [95\% CI] & \(p\) \\
\midrule
\multicolumn{5}{l}{\emph{A. Collective-problems module}} \\
Total alternatives mentioned & -0.006 [-0.179, 0.166] & 0.940 & -0.094 [-0.281, 0.093] & 0.320 \\
Other alternatives mentioned & -0.167 [-0.379, 0.044] & 0.118 & -0.276 [-0.499, -0.053] & 0.016 \\
\addlinespace \multicolumn{5}{l}{\emph{B. Selected binary outcomes}} \\
Joint action mentioned (pp) & 16.099 [0.710, 31.488] & 0.041 & 18.164 [-3.363, 39.691] & 0.097 \\
Party work reported (pp) & 9.578 [0.965, 18.191] & 0.030 & 9.064 [-1.198, 19.327] & 0.082 \\
Signed a petition (pp) & 7.670 [-2.405, 17.746] & 0.133 & 3.002 [-9.292, 15.296] & 0.627 \\
\addlinespace \multicolumn{5}{l}{\emph{C. Common questionnaire batteries}} \\
Trust battery, NS/NR (pp) & 1.307 [-2.022, 4.637] & 0.436 & 1.319 [-1.921, 4.560] & 0.419 \\
Conflict battery, NS/NR (pp) & -4.027 [-8.090, 0.036] & 0.052 & -4.532 [-8.494, -0.569] & 0.026 \\
Trust battery, items answered & -0.105 [-0.371, 0.162] & 0.436 & -0.106 [-0.365, 0.154] & 0.419 \\
Conflict battery, items answered & 0.161 [-0.001, 0.324] & 0.052 & 0.181 [0.023, 0.340] & 0.026 \\
Trust battery, uniform responses (pp) & -3.866 [-17.051, 9.318] & 0.560 & -3.511 [-16.052, 9.030] & 0.578 \\
Conflict battery, uniform responses (pp) & -9.268 [-23.717, 5.182] & 0.205 & 2.832 [-14.523, 20.187] & 0.745 \\
\addlinespace \multicolumn{5}{l}{\emph{D. Conflict resolution item and missingness}} \\
Conflict resolution item, baseline coding (SD) & 0.368 [0.070, 0.666] & 0.016 & 0.540 [0.179, 0.901] & 0.004 \\
Conflict resolution item, observed responses (SD) & 0.350 [0.050, 0.650] & 0.023 & 0.522 [0.152, 0.893] & 0.006 \\
\bottomrule

\end{tabular}\end{adjustbox}\end{tablecontent}\tablenotesgap
\centering\begin{minipage}{\exhibittablewidth}\tablenotefont
\textbf{Notes:} NS/NR combines \emph{do not know} with item nonresponse. Triple differences with municipality-by-wave and municipality-by-residence fixed effects and municipality-clustered 95\% confidence intervals. The 2011-reference specification excludes 2017. Indicator estimates are in percentage points, mention counts in units of answers, and conflict-resolution component estimates in 2011 rural comparison standard deviations.
\end{minipage}\par}
\end{table}

\section{Municipal Support across Time Comparisons}\label{app:novsupport}

Both reference specifications can be estimated on the same set of \NovCommonMunicipalities{} municipalities (\NovCommonCovered{} covered) with rural and urban observations in 2005, 2011 and 2023. This holds the set of municipalities fixed while using the observations available in each specification. The estimate using 2017 as reference is \NovCommonBaseline{} ($p=\NovCommonBaselineP$) and that using 2011 is \NovCommonEarlier{} ($p=\NovCommonEarlierP$).

Figure~\ref{fig:municipalchanges} uses the narrower set of 33 municipalities from Figure~\ref{fig:cells}, imposing the latter's fixed longitudinal support and comparing the municipal means at the endpoints. The mean rural change in comparison municipalities is negative when each municipality is dropped in turn, ranging from \NovRuralCompLooMin{} to \NovRuralCompLooMax{} standard deviations. Requiring more than five rural respondents at each endpoint leaves \NovRuralCompNonthinN{} comparison municipalities and an average change of \NovRuralCompNonthin{}.

Decree 1650 of 2017 identifies areas most affected by conflict for the ZOMAC tax program. Of the comparison municipalities, \NovZomacMunicipalities{} are listed as ZOMAC and are outside PDET. Only \NovZomacEarlierEndpoints{} have rural and urban respondents in both 2011 and 2023, and \NovZomacBaselineEndpoints{} do in both 2017 and 2023. In the first set, average baseline unmet basic needs are similar to those in covered municipalities on the same endpoints (55.9 and 55.4 percent), while rural population shares are lower (21 and 48 percent). The small number of comparison municipalities, remaining differences in their characteristics, and their eligibility for a separate tax incentive limit what this comparison can establish.

\clearpage
\section{Political and Violence Diagnostics}\label{app:novpolitics}

Table~\ref{tab:novpolitics} uses a matched sample and allows differential rural trajectories by presidential vote share or violence. Vote shares use Petro's votes over valid votes, including blank ballots, in the second round of the 2018 and 2022 presidential elections. The violence measure is the log of one plus the cumulative count of intentional homicides recorded by the National Police from 2016 to 2022. One older survey municipality code lacks a match. Each proxy interacts with rural residence and all nonreference waves in its own column.

The primary-index and coordination estimates remain positive, close to their matched-sample estimates, and conventionally distinguishable from zero. Party linkage remains positive, with intervals that include zero when allowing differential rural paths by the 2022 vote share, and with the homicide adjustment under the 2011-reference specification. The proxies measure municipal conditions during the reform period and can themselves respond to implementation.

\begin{table}[H]{\renewcommand{\arraystretch}{0.90}\setlength{\tabcolsep}{4pt}
\captionsetup{font={normalsize,md}}
\caption{Political and Violence Diagnostics}\label{tab:novpolitics}\vspace{-0.3cm}
\begin{tablecontent}\begin{adjustbox}{max width=0.98\linewidth, max totalheight=0.80\textheight}
\begin{tabular}{L{4.1cm}*{4}{C{2.2cm}}}

\toprule
 & Matched sample & Petro 2018 & Petro 2022 & Homicides 2016--22 \\
\midrule
\multicolumn{5}{l}{\emph{2017 reference}} \\
Primary index &  0.582 &  0.588 &  0.554 &  0.564 \\
  & [ 0.187,  0.977] & [ 0.173,  1.002] & [ 0.153,  0.956] & [ 0.150,  0.978] \\
Coordination &  0.445 &  0.468 &  0.452 &  0.462 \\
  & [ 0.126,  0.764] & [ 0.132,  0.803] & [ 0.121,  0.784] & [ 0.129,  0.795] \\
Party linkage &  0.446 &  0.445 &  0.401 &  0.417 \\
  & [ 0.053,  0.838] & [ 0.029,  0.862] & [-0.011,  0.814] & [ 0.030,  0.804] \\
Respondents / municipalities & \multicolumn{4}{c}{12431 / 65} \\
\midrule
\multicolumn{5}{l}{\emph{2011 reference, excluding 2017}} \\
Primary index &  0.625 &  0.628 &  0.617 &  0.633 \\
  & [ 0.172,  1.078] & [ 0.184,  1.073] & [ 0.171,  1.063] & [ 0.163,  1.102] \\
Coordination &  0.582 &  0.565 &  0.576 &  0.581 \\
  & [ 0.207,  0.957] & [ 0.183,  0.947] & [ 0.187,  0.965] & [ 0.187,  0.975] \\
Party linkage &  0.474 &  0.489 &  0.463 &  0.456 \\
  & [ 0.013,  0.935] & [ 0.021,  0.956] & [-0.010,  0.937] & [-0.028,  0.939] \\
Respondents / municipalities & \multicolumn{4}{c}{9243 / 63} \\
\midrule
\bottomrule

\end{tabular}\end{adjustbox}\end{tablecontent}\tablenotesgap
\centering\begin{minipage}{\exhibittablewidth}\tablenotefont
\textbf{Notes:} Estimates of the 2023 triple difference in standard deviations with 95\% municipality-clustered intervals in brackets. All columns in each panel use the same matched respondent sample. The adjustments add each proxy separately, interacted with rural residence and indicators for all nonreference waves.
\end{minipage}\par}
\end{table}

\end{document}